\documentclass[]{aa}
\usepackage{graphicx}

\usepackage[varg]{txfonts}
\usepackage{cases}
\usepackage{underscore}
\usepackage{hyperref}
\usepackage{amsmath}

\usepackage[dvipsnames]{xcolor}           

\begin{document}
      \title{On the Constraints and Observational Manifestations of Failed Solar Eruptions in Toroidal Magnetic Cage}
      
      \author{J. H. Guo\inst{1,2}, Y. Guo\inst{1}, H. Wu\inst{1}, B. Schmieder\inst{2,3,4}, P. D\'emoulin\inst{3}, Y. W. Ni\inst{1}, C. Wang\inst{1}, S. Poedts\inst{2,5}, T. Li\inst{6}, Wensi Wang\inst{7}, Y. H. Zhou\inst{1}, P. F. Chen\inst{1}}
      
      \institute{School of Astronomy and Space Science and Key Laboratory of Modern Astronomy and Astrophysics, Nanjing University, Nanjing 210023, China \\
      	\email{jinhan.guo@nju.edu.cn}
 	    \and 
 	    Centre for Mathematical Plasma Astrophysics, Department of Mathematics, KU Leuven, Celestijnenlaan 200B, B-3001 Leuven, Belgium 
        \and
        LIRA, Observatoire de Paris, CNRS, UPMC, Universit\'{e} Paris Diderot, 5 place Jules Janssen, 92190 Meudon, France
        \and
        LUNEX EMMESI Institut, SBIC, Kapteyn straat 1, Noordwijk2201 BB Netherlands
        \and
        Institute of Physics, University of Maria Curie-Skłodowska, ul.\ Radziszewskiego 10, 20-031 Lublin, Poland
        \and 
        State Key Laboratory of Solar Activity and Space Weather, National Space Science Center, Chinese Academy of Sciences, Beijing 100190, China
        \and 
        CAS Key Laboratory of Geospace Environment, Department of Geophysics and Planetary Sciences, University of Science and Technology of China, 230026, Hefei, China
 	    }
    \titlerunning{Confined solar eruption in magnetic cage}
 \authorrunning{Guo et al.}
\date{}

\abstract 
{Observations show that many solar eruptions remain confined in the magnetic configuration of strong overlying magnetic fields, which is commonly referred to as ``magnetic cage".}
{Confined eruptions under strong poloidal overlying fields have been widely studied. 
In contrast, the confined eruption scenario under strong external toroidal fields remains unrevealed.} 
{We use three-dimensional magnetohydrodynamic simulations to systematically study confined eruptions in a toroidal magnetic cage, focusing on the roles of the Lorentz force and magnetic reconnection, as well as their observational manifestations, such as flare ribbons and loops. We further apply the test particle method with the guiding-centre approximation to 
synthesise hard X-ray sources, enabling comparison between thermal and non-thermal responses.}
{Our results show that overlying toroidal magnetic fields are crucial in confining eruptions. They generate strong return currents that produce a significant downward Lorentz force, suppressing the rise of the flux rope. Simultaneously, they drive the large-angle rotation of the rope, triggering reconnection with the overlying fields and ultimately causing its break-up. The synthesised EUV images display multi-ribbon flare structures with highly sheared loops with a global “cowboy-hat-like” shape. Additionally, comparisons with hard X-ray sources reveal that thermal and non-thermal responses are not co-spatial, in which return current is a major accelerator to energetic electrons.}
{The simulations clarify how the magnetic cage constrains solar eruptions. First, the downward Lorentz force related to return current effectively suppresses eruptions, explaining why confined flares tend to occur in electric-current neutralised active regions. Second, we demonstrate that toroidal-field-induced force 
is the key driver for rotation and confinement of the flux rope. This explains why many filament eruptions with rotation 
fail despite being torus-unstable. Finally, we suggest that the global morphology of flare loops (“cowboy-hat–like” or “saddle-like”) and the shearing degree of flare loops can serve as useful diagnostics to distinguish confined from eruptive flares.}
\keywords{Sun: corona -- Sun: solar flares -- Sun: magnetic fields --methods: numerical -- Magnetohydrodynamical (MHD)}

\maketitle
\nolinenumbers

\section{Introduction}\label{introduction} 

Solar eruptions, such as solar flares and coronal mass ejections (CMEs), are widely recognised as the most intense activities in the solar system \citep{Chen2011}. Research in this field is constantly a priority in both astrophysics and solar-terrestrial space physics. On the one hand, the extensively observed solar eruptions encompass numerous fundamental astrophysical processes \citep{Tsurutani2023}. This suggests that novel insights into the explanation of other high-energy astrophysical phenomena may be obtained by analogy with solar eruptions \citep{Dai2006, Meng2014}. On the other hand, solar eruptions can expel a substantial amount of magnetised plasma into interplanetary space, thereby giving rise to disturbances in the solar-terrestrial space environment. As a result, exploring the dominant underlying physical mechanisms of solar eruptions is of great importance both for forecasting adverse space weather events and enhancing our understanding of the universe.

Although almost all solar eruptions can release substantial electromagnetic radiation, namely, solar flares, not all solar eruptions can produce CMEs. If the eruptive core manages to break through the constraints of the overlying fields and escapes into interplanetary space, it will evolve into a CME and later on into an interplanetary CME (ICME). These successful solar eruptions, or eruptive flares, have the potential to induce geomagnetic storms and thus jeopardise human satellite operations. However, in certain events, eruptive cores fail to escape from the solar corona \citep{Ji2003, Nindos2015, Thalmann2015,Zuccarello2017}. They are often referred to as confined flares or failed eruptions. Exactly how to predict eruptive and confined flares is still crucial and challenging for space weather research.

Previous studies have shown that the success or confinement of solar eruptions is strongly influenced by their magnetic environment, particularly the eruptive core and the overlying background fields. For instance, \citet{Wang2007} reported that confined flares preferentially occur near the centre of active regions, highlighting the role of overlying fields in producing confinement. Similarly, \citet{Lit2021} implied that magnetic flux is a decisive quantity to distinguish eruptive and confined flares. Further studies revealed that eruptions are more likely to fail if the critical height (e.g., where the decay index $n(r)=1.5$) is high \citep{Wang2017, Lit2022} or if the $n(r)$ profile exhibits a saddle-like profile with a dip \citep{Guo2010}, in which the decay index first rises, then dips, and finally increases again with height (see Figure~\ref{fig2} for the decay index derived from the total horizontal field). From laboratory plasma experiments, \citet{Myers2015} proposed a combined parameter of the decay index ($n$) and the safety factor ($1/T_{w}$) to better differentiate confined and eruptive flares, a criterion later tested in solar plasma environments by \citet{Jing2018} and \citet{Duan2019}. Hereafter, \citet{Lit2022} proposed a new parameter considering the force-free factor ($\alpha$) of core fields and magnetic flux to distinguish two types of flares. Recently, \citet{Teraoka2025} demonstrated that the amount of twist and the height of field lines connecting flare kernels can distinguish eruptive from confined flares.


\begin{figure}
  \includegraphics[width=8cm,clip]{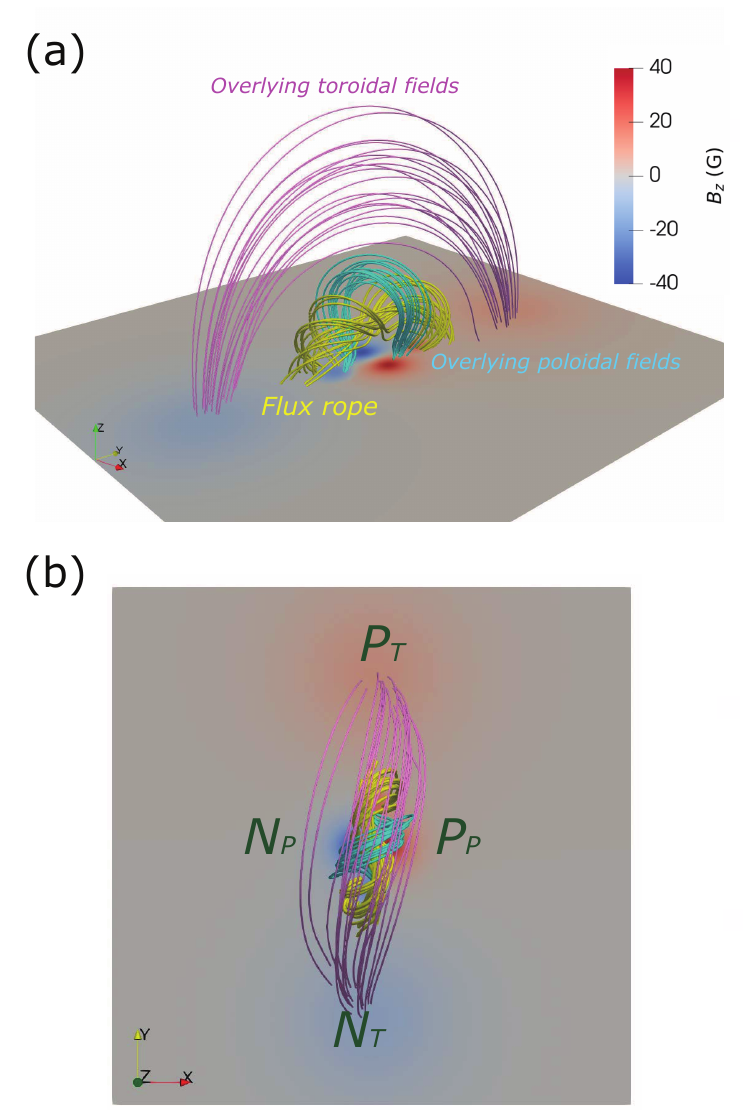}
  \centering
  \caption{Visualisation of the initial magnetic fields viewed from the (a) side and (b) top. The yellow, cyan and magenta tubes represent the twisted flux rope, overlying poloidal and toroidal magnetic fields, respectively. 
  $N_{\rm _{P}}$/$N_{\rm _{T}}$ and $P_{\rm _{P}}$/$P_{\rm _{T}}$ labels the negative and positive sub-photosphere magnetic charges to build the poloidal/toroidal magnetic fields, respectively. 
  \label{fig1}}
\end{figure}

Although these statistical findings reveal key discriminants between confined and eruptive flares, the underlying physical mechanisms remain poorly understood. For example, numerous failed filament eruptions are torus-unstable ($n>1.5$) and exhibit large-angle rotation \citep{Ji2003, Zhou2019,Guo2023b}. Additionally, counterexamples in \citet{Jing2018} and \citet{Duan2019} may reflect the possibility that external magnetic reconnection sometimes plays a negative role in forming CMEs. In this scenario, the toroidal-field induced tension force \citep{Myers2015, Zhang2024, Guo2024,Guo2024tor}, non-axisymmetry induced force \citep{Zhong2021} and external magnetic reconnection involving the flux rope \citep{Jiang2023, Chen2023} may have some effects in resulting in failed eruptions. Recently, \citet{Liu2017} and \citet{Liu2024} pointed out that the amount of non-neutralised electric current, measured by the ratio of photospheric direct current (DC) to return current (RC), is a reliable proxy for assessing the success or not of an eruption. Direct current (DC) is generally associated with the twist of a flux rope and is often concentrated near the core of the flux tube. The return current forms a surrounding surface current that envelopes the DC, effectively isolating the flux tube from its ambient magnetic environment. In flare loops, the role of the shear and induced electric current was demonstrated \citep{Aulanier2012}. However, why active regions with neutralised electric current prefer to form confined flares still remains debated.

To better distinguish between confined and eruptive flares, it is essential to establish a basic framework for different magnetic configurations. Existing numerical MHD models of confined flares have largely focused on cases with strong overlying poloidal fields \citep{Torok2005, Jiang2023}, or on magnetic configurations involving a null point where the flux from outer polarities exceeds that of the core fields \citep{Chen2023}. However, so far, the scenario of strong overlying toroidal fields, the so-called toroidal magnetic cage, remains less explored because the eruptive flux rope easily escapes from the overlying toroidal magnetic fields intuitively, as shown in \citet{Fan2007}. Actually, the downward force induced by the toroidal field scales with the square of the electric current intensity, whereas the downward force induced by the poloidal field scales linearly with the current. This implies that toroidal fields can also play a significant role in suppressing the eruption. In addition, the toroidal field can drive a large-angle rotation of the flux rope \citep{Kliem2012, Zhou2023, Zhang2024,Guo2023b}, which alters the distribution of the resulting Lorentz force and may also generate an external Lorentz force. These indicate that the role of overlying toroidal fields cannot be neglected.

To this end, we carry out a series of MHD simulations and conduct a systematic analysis encompassing (1) the evolution of magnetic fields, (2) electric currents, (3) Lorentz force, (4) magnetic reconnection, and (5) forward modellings (thermal and non-thermal). The organisation of this paper is as follows: Section~\ref{sec:met} introduces the numerical setup, followed by Section~\ref{sec:res1} presenting the global evolution of simulation results. Subsequent sections analyse the confined mechanisms (Section~\ref{sec:res2}) and forward modelling results (Section~\ref{sec:res3}). We conclude with the discussion and summary in Sections~\ref{sec:dis} and \ref{sec:sum}.

\begin{figure}
  \includegraphics[width=9cm,clip]{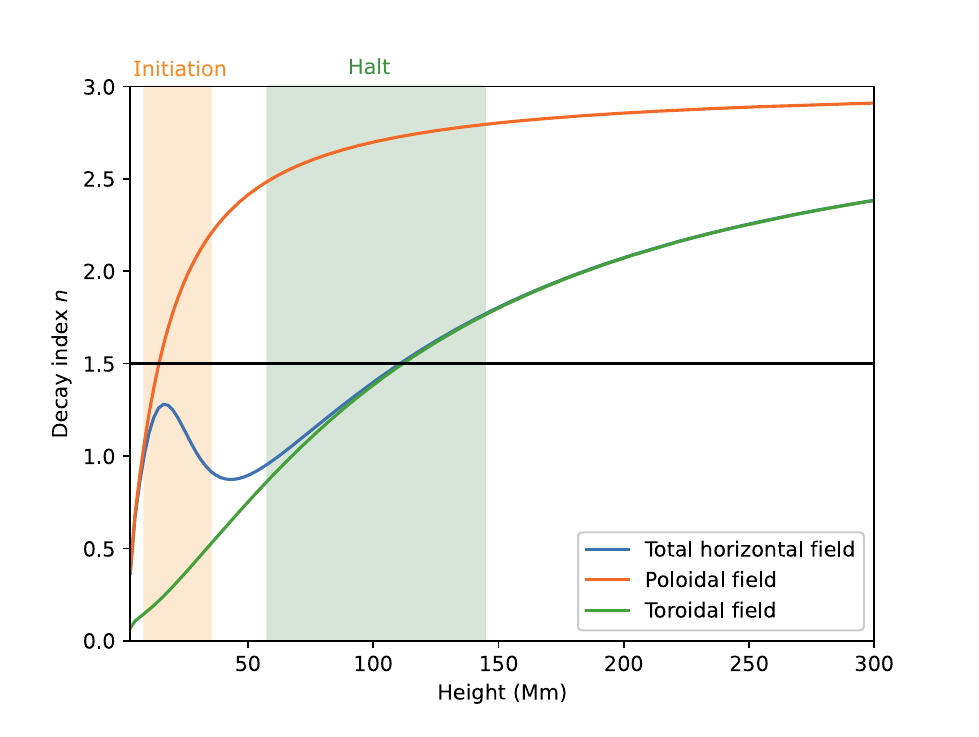}
  \centering
  \caption{Height profiles of the decay index computed from external total horizontal magnetic fields ($B_{h}$, blue line), poloidal fields ($B_{\rm p}$, orange line) and toroidal fields ($B_{\rm T}$, green line). The orange and green bands indicate regions of the flux rope at initial and stopping moments, respectively. 
  }\label{fig2}   
\end{figure}

\begin{figure*}
  \includegraphics[width=16cm,clip]{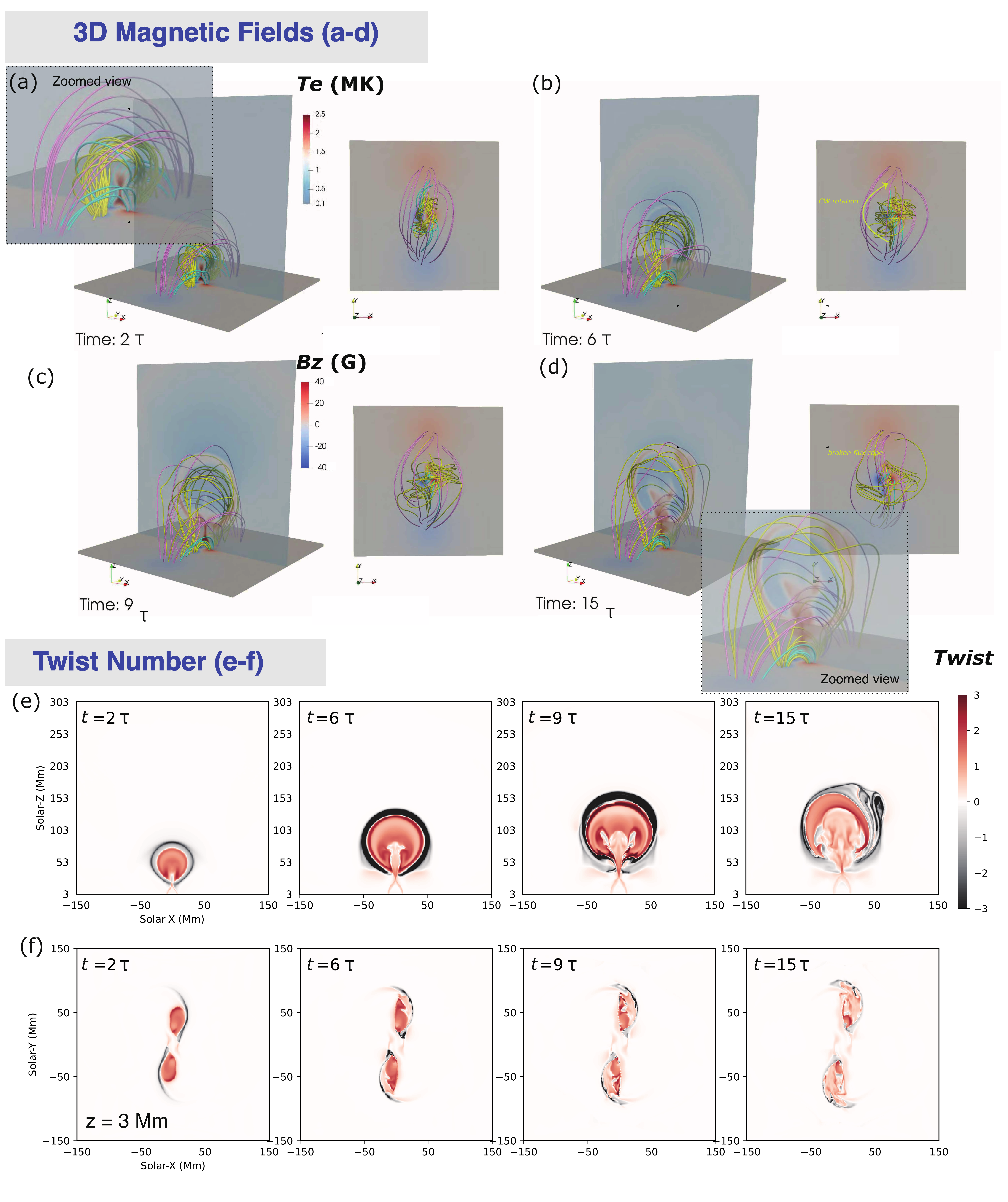}
  \centering
  \caption{Temporal evolution of (a--d) the magnetic field and (e, f) the twist number $T_{w}$ distributions in the eruption process. The yellow, cyan and pink tubes are traced from the $N_{\rm _{FR}}$, $N_{\rm _{P}}$ and $N_{\rm _{T}}$ polarities, respectively. The colours in the bottom and vertical planes exhibit the distributions of the $B_{z}$ and $T_{e}$, respectively. Panels (e) and (f) display the $T_{w}$ distributions on the vertical and bottom planes, respectively.}
  \label{fig3} 
\end{figure*}

\section{Numerical setup}\label{sec:met}

\subsection{Magnetic Configuration}

We build the initial magnetic configuration similarly to the one introduced in \citet{Titov1999} and \citet{Titov2014}. The external poloidal field ($B_{\rm p}$) is defined by two sub-photosphere magnetic charges, of strength $q_{\rm _{P}}$, positioned along the $x$-axis at depth $d_{\rm _{P}}$, and separated by $2L_{\rm _{P}}$. Next, to introduce toroidal fields ($B_{\rm T}$) aligned with the flux-rope axis ($y$-axis), two additional magnetic charges, of strength $q_{\rm _{T}}$, are placed on the $y$-axis at $y=\pm L_{\rm _{T}}$ with depth $q_{\rm _{T}}$. Both $B_{\rm p}$, $B_{\rm T}$ and their combinations are in accordance with the potential field model without electric currents. The core fields are then constructed by embedding a force-free toroidal flux rope ($B_{\rm FR}$) along the $y$-axis.  $B_{\rm FR}$ is characterised by its major radius $R$ and its minor radius $a$. It is implemented with the regularised Biot-Savart laws \citep{Titov2018}. The equilibrium current in the flux rope is governed by Shafranov's stability criterion \citep[Equation 7 in][]{Titov2014}).We call the above model FR-PT, while model FR-P when only the external poloidal fields are included.

To better reach a force-free state in the initial magnetic configuration, we employ a magneto-frictional relaxation on the superimposed fields \citep{Guo2016b}. Figure~\ref{fig1} displays the magnetic configuration with the following parameters: $R=30$~Mm, $a=12$~Mm, $q_{\rm _{P}}=150$~T~Mm$^{2}$, $q_{\rm _{T}}=600$~T~Mm$^{2}$, $d_{\rm _{P}}=10$~Mm, $d_{\rm _{T}}=50$~Mm, $L_{\rm _{P}}=10$~Mm and $L_{\rm _{T}}=100$~Mm. In this figure, we can see  a twisted flux rope (yellow lines), overlying poloidal field lines connecting $P_{\rm _{P}}$ and $N_{\rm _{P}}$ polarities (cyan lines), and overlying large-scale toroidal field lines connecting $P_{\rm_{T}}$ and $N_{\rm _{T}}$ polarities (pink lines). 

Figure~\ref{fig2} shows the profiles of the decay index $n_{p}$ (orange line), $n_{t}$ (green line) and $n_{h}$ (blue line). They are respectively computed from $B_{\rm p}$, $B_{\rm T}$ and $B_{h}=B_{\rm T}+B_{\rm p}$ components, with the following formula:
\begin{eqnarray}
n(r)=-\frac{d\ \rm{log}(B_{\rm ex})}{d\ \rm{log}(r)}=-\frac{r}{B_{\rm ex}} \frac{dB_{\rm ex}}{dr}
\end{eqnarray}
where $B_{\rm ex}$ is the selected horizontal magnetic-field component. The decay index $n_{h}$ calculated from $B_{h}$ exhibits a saddle-like profile, while the profiles of $n_{t}$ and $n_{p}$ monotonically increase. The flux rope is positioned near the typical threshold ($n_{p}=1.5$) of torus instability \citep{Aulanier2010}, enabling triggering of an eruption despite the initial equilibrium. It should be noted that the flux rope occupied areas (red and green bands) are determined by the Lorentz force ($J_{\rm FR} \times B_{\rm T}$) in Figure~\ref{fig9}d.

\subsection{MHD modeling}

We employ a thermodynamic MHD model considering the thermal conduction to simulate the eruption process. The governing equations are as follows:

\begin{align}
    \frac{\partial\rho}{\partial t} + \nabla \cdot (\rho \boldsymbol{v}) &= 0 ,\\
    \frac{\partial(\rho\boldsymbol{v})}{\partial t} + 
    \nabla \cdot [\rho\boldsymbol{v}\boldsymbol{v} + (p+\frac{\boldsymbol{B}^{2}}{2\mu_{0}})\boldsymbol{I} - \frac{\boldsymbol{BB}}{\mu_0}] 
    &= \rho \boldsymbol{g},\\
    \frac{\partial e_{int}}{\partial t} + \nabla \cdot (\boldsymbol{v}e_{int}) &= - p \nabla \cdot
    \boldsymbol{v} + \eta J^{2} + \\
    \nabla \cdot(\boldsymbol{\kappa} \cdot \nabla T) \notag,\\
    \frac{\partial \boldsymbol{B}}{\partial t} + 
    \nabla \cdot (\boldsymbol{vB - Bv}) &=  -\nabla \times(\eta J)
\end{align}
where $\boldsymbol{\kappa}=\kappa_{\parallel}\boldsymbol{\hat{b}\hat{b}}$ is field-aligned thermal conduction, $\kappa_{\parallel} =10^{-6}\ T^{\frac{5}{2}}\ \rm erg\ cm^{-1}\ s^{-1}\ K^{-1}$ is the Spitzer heat conductivity, $\boldsymbol{g}=-k\,g_{\odot}\,r_{\odot}^2/(r_{\odot}+z)^2\boldsymbol{e_{z}}$ represents the gravitational acceleration, and $g_{\odot}= \rm 274\ m\ s^{-2}$ is the gravitational acceleration at the solar surface, $r_{\odot}$ is the solar radius, $\eta=4.65 \times 10^{11}\ \rm cm^{2}\ s^{-1}$ is the uniform resistivity, and other variables have their general meanings. The imposed resistivity $\eta$ is comparable to that used by \citet{Kilpua2021} and has two main effects. First, it improves numerical stability when a high-order limiter is employed. Second, it accounts for Joule heating in regions of strong electric current. Following \citet{Jiang2021}, we scale gravitational acceleration by a factor $k=1.8$. This operation can effectively reduce the scale height, and ensures that plasma $\beta<1$ even in the overlying weak magnetic fields. The environment of $\beta<1$ throughout the computational domain ensures that the eruption dynamics are dominated by magnetic fields in solar active regions rather than thermodynamic effects. The plasma density is initialised using an isothermal hydrostatic atmospheric model with uniform temperature (1 MK).

The three-dimensional (3D) MHD equations are numerically solved with the Message Passing Interface Adaptive Mesh Refinement Versatile Advection Code \citep[MPI-AMRVAC,][]{Xia2018, Keppens2023}. We use Cartesian coordinates, with a domain of $[x_{\rm min},x_{\rm max}]\ \times [y_{\rm min},y_{\rm max}]\ \times [z_{\rm min},z_{\rm max}] = [-150, 150]\ \times [-150,150]\ \times[0,300]\;\rm Mm^{3}$. We use a four-level adaptive mesh refinement with a basic mesh grid of $40 \times 40 \times 40$ points. To mitigate the divergence of the magnetic fields during the numerical calculation, we employ the constrained-transport (CT) method on staggered grids \citep{Gardiner2005}. This method keeps the divergence of magnetic fields unchanged up to machine accuracy. This can be evaluated by the  $<|f_{i}|> =(\sum_i |f_i| \Delta V_i)/(\sum_i \Delta V_i>) ~\sim 10^{-16}$, 
where $f_i=(\nabla \cdot \mathbf{B})_i \Delta V_i/B_i A_i$ denotes the fractional flux increase in a small discrete volume about grid point $i$, $V_i$, $A_i$ and $B_i$ represent the cell volume, cell surface area, and the magnitude of the magnetic field at point $i$, respectively. The validity of this approach has been discussed in \citet{Schmieder2024}.

Regarding the numerical scheme, we adopt an HLL flux scheme, a three-step Runge-Kutta time discretisation approach, and the fifth-order WENO-limited reconstruction. For the bottom boundary conditions, we implement a line-tied condition as follows. First, we set the horizontal electric field ($E_{h}$) on the lower bottom plane to zero, ensuring that the normal magnetic-field component remains unchanged over time. Second, the plasma velocities are set to zero in the bottom plane between the inner ghost cells and the physical domains. The lateral and top boundaries are prescribed as open with zero-gradient extrapolation.

\begin{figure*}
  \includegraphics[width=\textwidth,clip]{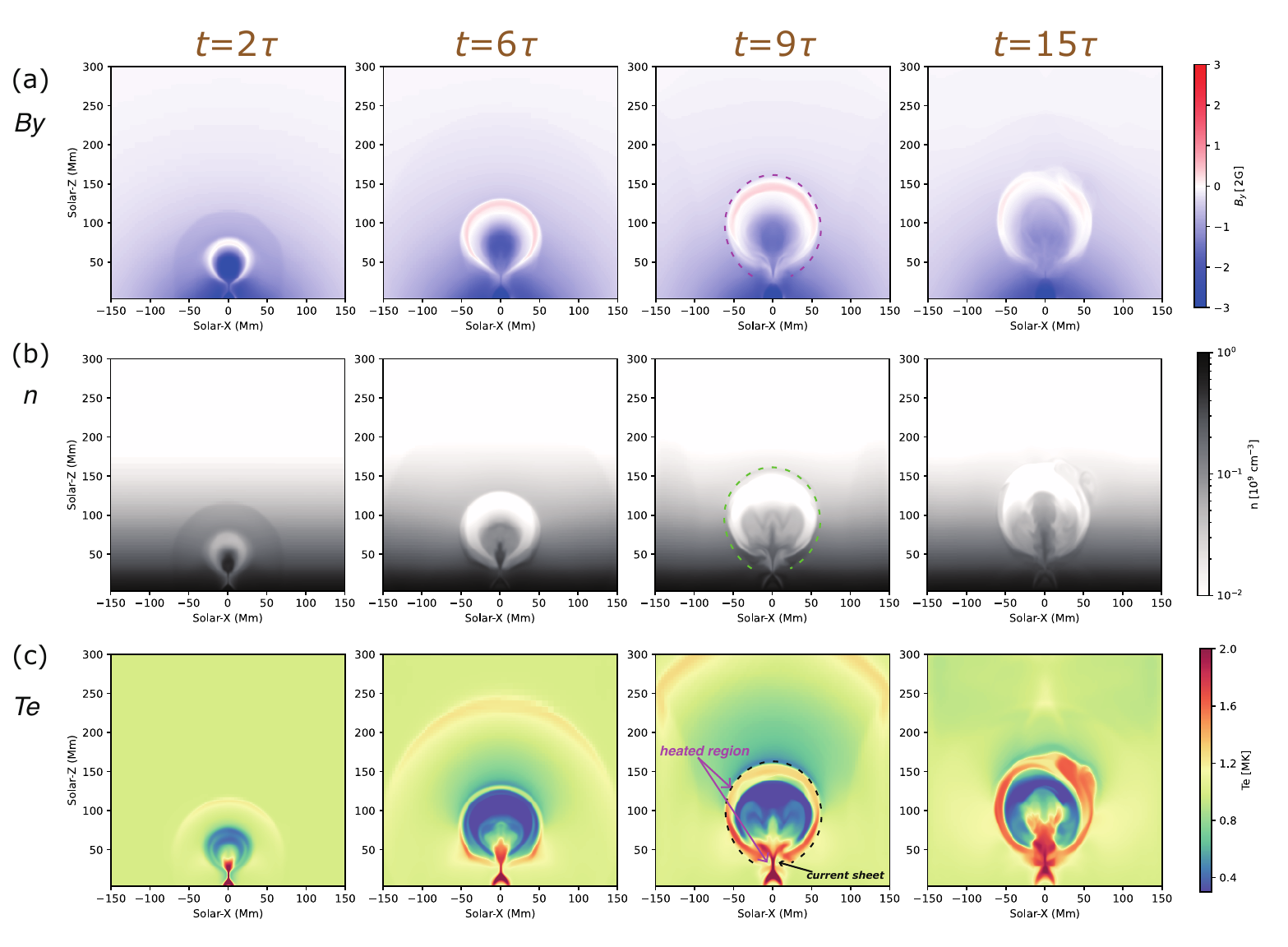}
  \centering
  \caption{Temporal evolution of (a) the axial magnetic field component $B_{y}$, (b)  the number density $n$ and (c) the temperature $T_{e}$ at $t=2, 6, 9, 15\tau$. The dashed line in the $t=9$ column outlines the border of the flux rope (with added reconnected flux).
  \label{fig4}}
\end{figure*}

\begin{figure}
  \includegraphics[width=9cm,clip]{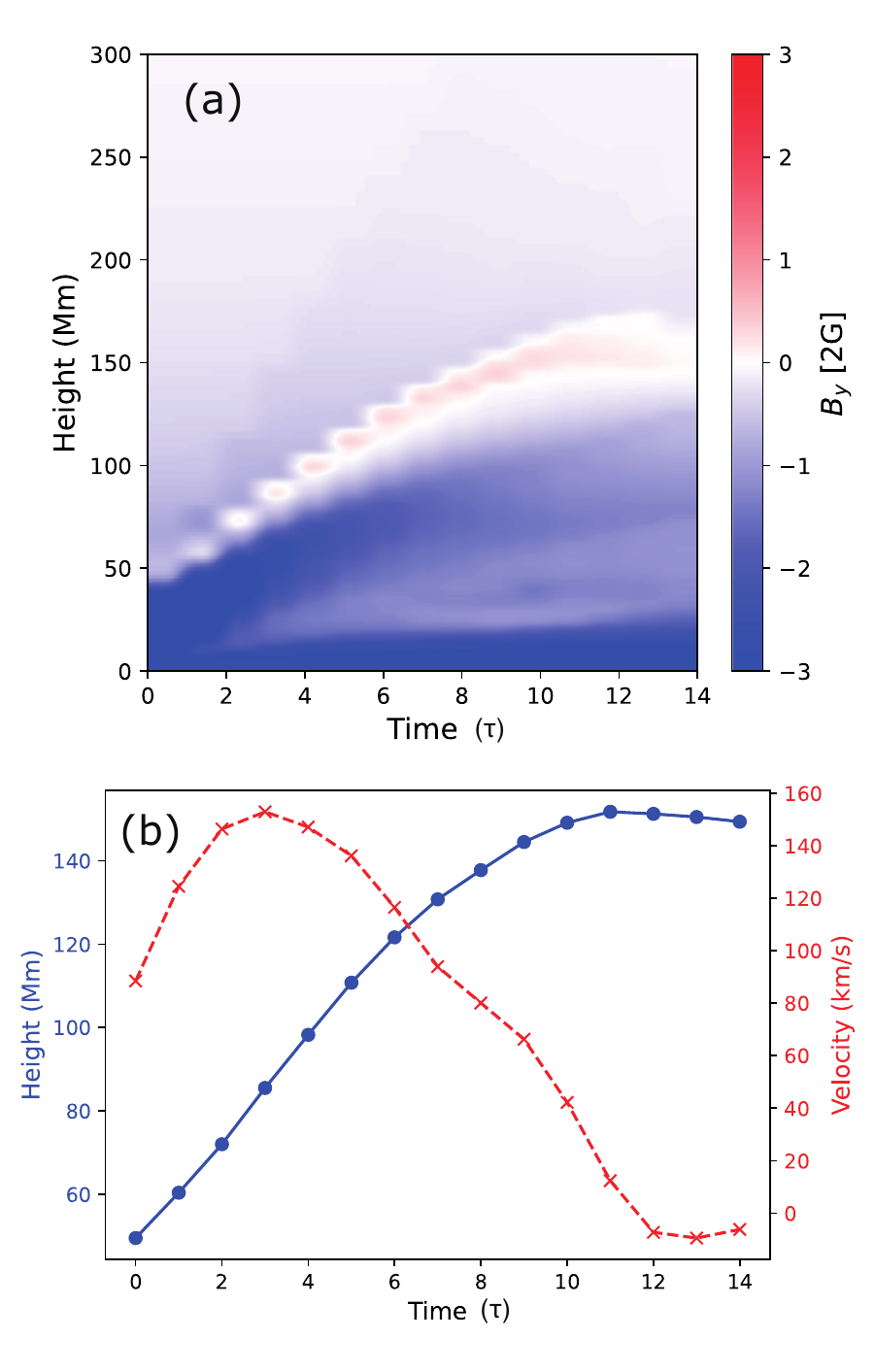}
  \centering
  \caption{Kinematics of the eruptive flux rope. Panel (a) shows the time-distance diagram of the $B_{y}$ component along the $z$-axis. In panel (b) the blue dots and red crosses show the evolution of height and velocity during the eruption, which are measured with the positive $B_{y}$ front in panel (a).}
  \label{fig5}
\end{figure}

\section{Global Evolution during Eruption}  \label{sec:res1}

Figure~\ref{fig3} illustrates the 3D dynamic evolution of magnetic fields during the eruption. The flux rope erupts directly due to torus instability. As it rises, a current sheet forms beneath it, where magnetic reconnection generates highly sheared loops. Meanwhile, the pre-existing flux rope (yellow lines) undergoes a clockwise rotation, reaching 90$^{\circ}$ by about $t=6\tau$ ($\tau=85.87$ s is the normalisation unit in the simulation). This rotation direction is consistent with observations showing that sinistral filaments (positive helicity) tend to rotate clockwise \citep{Green2007}. It should be noted that the asymmetry in the simulation results may be associated with the nonlinear WENO5 reconstruction and the rotation of the flux rope. As the underlying magnetic reconnection persists, the flux rope undergoes a gradual, decelerating rise. Eventually, the structure reverses its motion and begins to descend, with the newly reconnected field lines anchored to the northern toroidal polarity ($N_{T}$). This implies the reconnection between the eruptive flux rope and the overlying toroidal fields.

To further elucidate the topology evolution during the eruption, we compute the twist number ($T_{w}$) distributions using the parallel electric current ($J_{\parallel}=\textbf{J} \cdot \textbf{B}/B$) integration method \citep{Berger&Prior2006, Liu2016}, as follows:
\begin{eqnarray}
&& T_{w} = \int \frac{\mu_{0}J_{\parallel}}{4\pi B}dl
\end{eqnarray}

Figures~\ref{fig3}e and \ref{fig3}f show the distribution of $T_{w}$ distribution on the side and bottom planes, respectively. The initial extension of the flux-rope footpoints, accompanied by an increase in $T_{w}$, reflects magnetic reconnection in the ambient arcades (Figure~\ref{fig3}c). Subsequently, $T_{w}$ decreases as the twist is partially converted into writhe (manifested as the rotation of the flux rope) and further reduced by reconnection between the flux rope and the overlying toroidal fields. Notably, a shell of opposite-sign twist develops around the flux rope, corresponding to return current (RC).

Figure~\ref{fig4} shows the temporal evolution of the toroidal magnetic field ($B_{y}$), number density ($n$), and temperature ($T_{e}$) in the $x$–$z$ plane. Similar to the twist distribution in Figure~\ref{fig3}e, the outer shell of the eruptive core exhibits a $+B_{y}$ component opposite to that of the central region, where the plasma is heated to twice the background temperature. In contrast, the temperature and density inside the circular dome decrease owing to expansion. The reversal of the $B_{y}$ component is due to the flux rope's rotation, with the rotation angle increasing with height, as shown in Figure~\ref{fig3}. Consequently, regions with positive, zero, and negative $B_{y}$ correspond to rotation angles greater than, equal to, and less than $90^{\circ}$, respectively. This is also validated in Figure~\ref{fig10}d, corresponding to strong $-B_{x}$ component and positive $B_{y}$ component. In addition, the heated current sheet beneath the flux rope is consistent with the standard flare model. A heated and dense plasma blob also appears in the upper-right portion of the flux rope at $t=15\tau$, suggesting external magnetic reconnection involving the flux rope. To further quantify the flux rope kinematics, we measure the height of the outer $+B_{y}$ shell to derive its velocity. As shown in Figure~\ref{fig5}, the flux rope accelerates to 150 km s$^{-1}$ at $t=3\tau$, then decelerates to –10 km s$^{-1}$ at $t=12\tau$, indicating a failed eruption.

\begin{figure*}
  \includegraphics[width=\textwidth,clip]{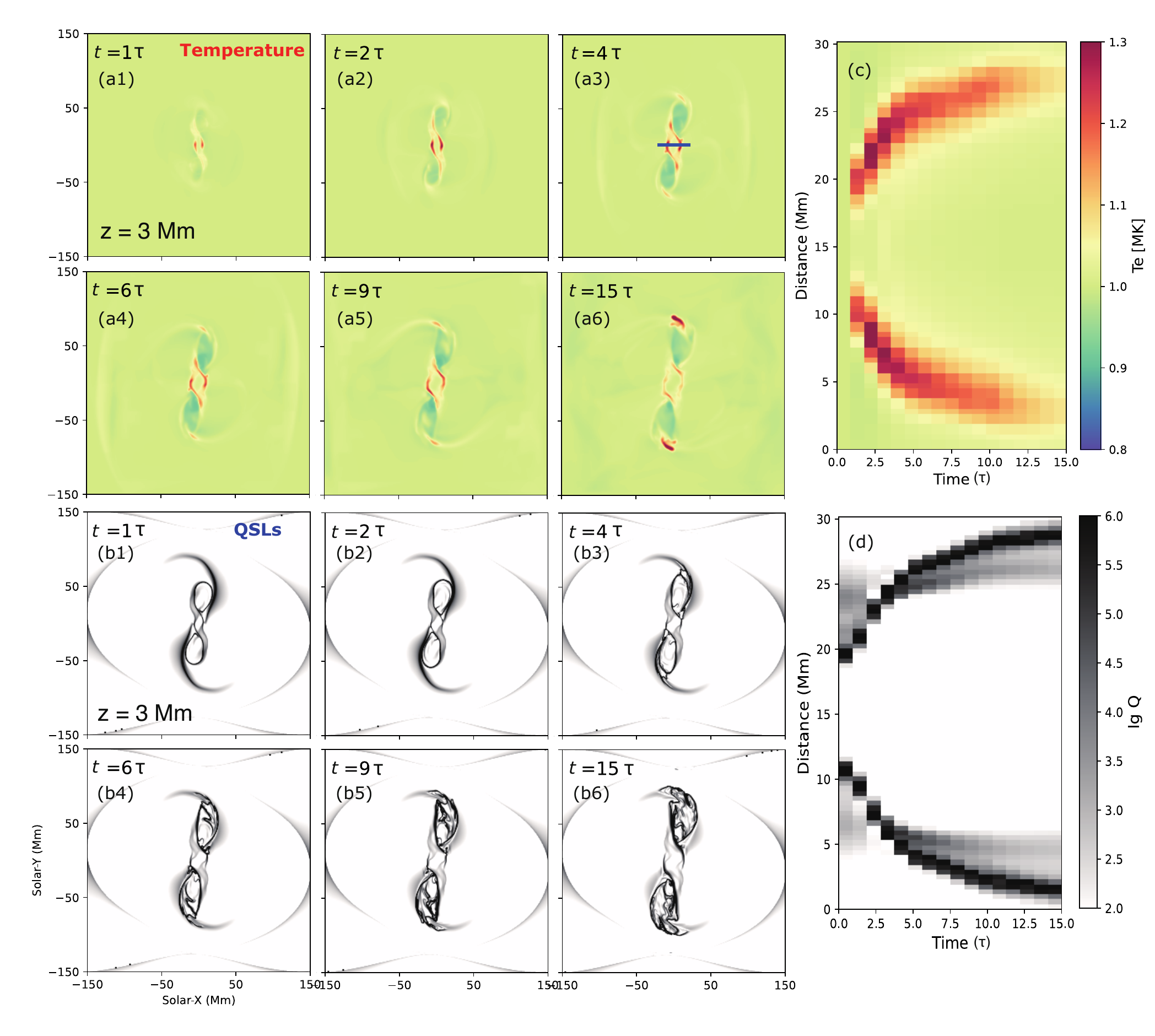}
  \centering
  \caption{Evolution of flare ribbons depicted from the (a, c) temperature and (b, d) QSLs distributions. Temporal evolution of the (a) temperature and (b) squashing degree $Q$ of QSLs on the bottom plane. Panels (c) and (d) show the time-distance diagram of temperature and $Q$ along the blue line in Panel (a3), respectively. Two wave-like fronts in Panels (a3) and (a4) correspond to the shock in Figure~\ref{fig4}c. The artifacts of $Q$ near the side boundaries arise from the finite size of the computational domain. However, they do not give rise to significant heating fronts, electric currents, or Lorentz forces, and thus have a small impact on the derived conclusions. The box size is equal to that in plots (e) and (f) of Figure~\ref{fig3}.}
  \label{fig6}
\end{figure*}

The temperature distribution on the bottom plane, shown in Figures~\ref{fig6}(a1)--(a6), reflects the evolution of flare ribbons to a great extent. As seen in the top panels, the heated regions form elongated two-ribbon structures, closely resembling the separated flare ribbons in observations. Away from the central portion, the ribbons show hook-like shapes, corresponding to the footpoints of the flux rope. Initially, the ribbons extend along the PIL and subsequently separate perpendicular to it, consistent with the 3D standard flare model \citep{Janvier2014}. After $t=10\tau$, two additional heated spots appear at the ribbon endpoints, as shown in Figure~\ref{fig6}(a6). To further analyse the magnetic topology, we compute the distribution of the squashing degree $Q$ \citep{Priest1995, Demoulin1996}, shown in Figures~\ref{fig6}(b1)–(b6). Quasi-separatrix layers (QSLs) are areas where magnetic connectivity changes drastically, and the squashing factor ($Q$) is larger than 2. They are the favourite places for magnetic reconnection \citep{Priest1995, Demoulin1996, Titov2002}. The closed circular QSLs outline the footpoints of the twisted flux rope, while the separated straight QSLs correspond to two flare ribbons. During the eruption, the circular QSLs drift along the $y$–axis, expand significantly, and eventually merge with the outer hooked QSLs, as predicted by \citet{chen12}. Fine structures at the hook endpoints are also evident, coinciding with the additional heated regions in Figure~\ref{fig4}b and the outer twist shell with opposite helicity (Figure~\ref{fig3}f). To quantify the ribbon dynamics, we extract a slit along the separation direction of the ribbons in Figure~\ref{fig6}(a3). The time–distance diagrams of temperature ($T_{e}$) and QSLs, shown in Figures~\ref{fig6}(c) and \ref{fig6}(d), reveal nearly identical behaviours: the ribbons initially separate at $\sim 30$ km s$^{-1}$, then gradually decelerate, and eventually stop near $x=\pm 30$ Mm. This demonstrates that confined flares can also exhibit ribbon separation motions.

\begin{figure*}
  \includegraphics[width=18 cm,clip]{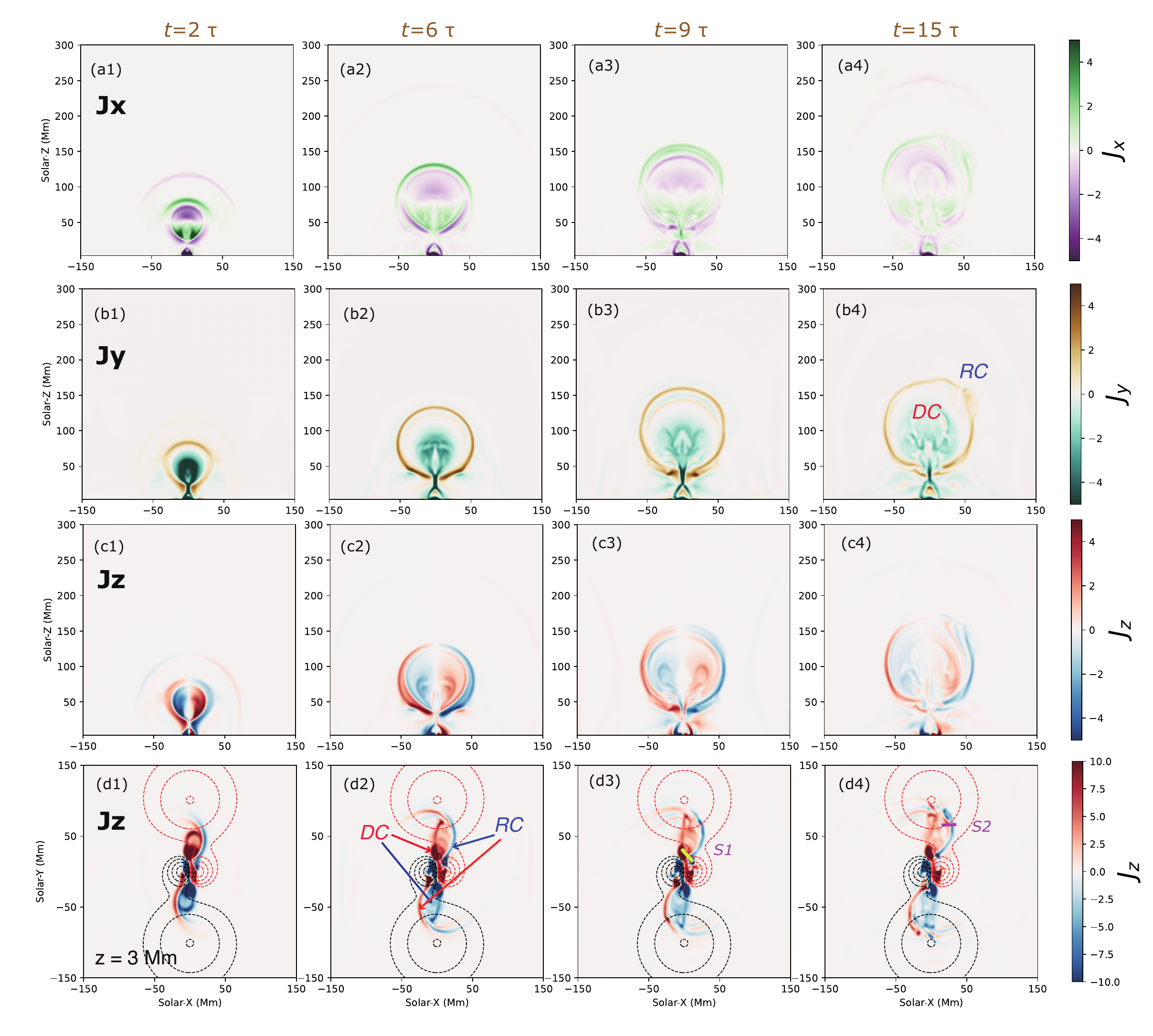}
  \centering
  \caption{Evolution of the electric current during eruption.} The (a1--a4) first, (b1--b4) second and (c1--c4) third rows display the distributions of $J_{x}$, $J_{y}$ and $J_{z}$, respectively, in the central vertical plane. 
  The (d1--d4) bottom row displays the distributions of $J_{z}$ in the bottom horizontal plane. The red (positive $B_{z}$) and black (negative $B_{z}$) dash lines represent the contour lines of $B_{z}=\pm \rm 40,\ 20,\ 10,\ 5$ G. In regions of positive/negative $B_z$ (red/black contours), positive/negative currents (red/blue) represent the direct-current (DC) system, whereas negative/positive currents (blue/red) form the return-current (RC) system. The 1D slits S1 and S2 are defined in panels (d3) and (d4) by yellow and pink segments, respectively, showing the . Some averaged quantities at $z=0$ within these slits are shown in Figure \ref{fig8}.
  \label{fig7}
\end{figure*}

\begin{figure}
  \includegraphics[width=7 cm,clip]{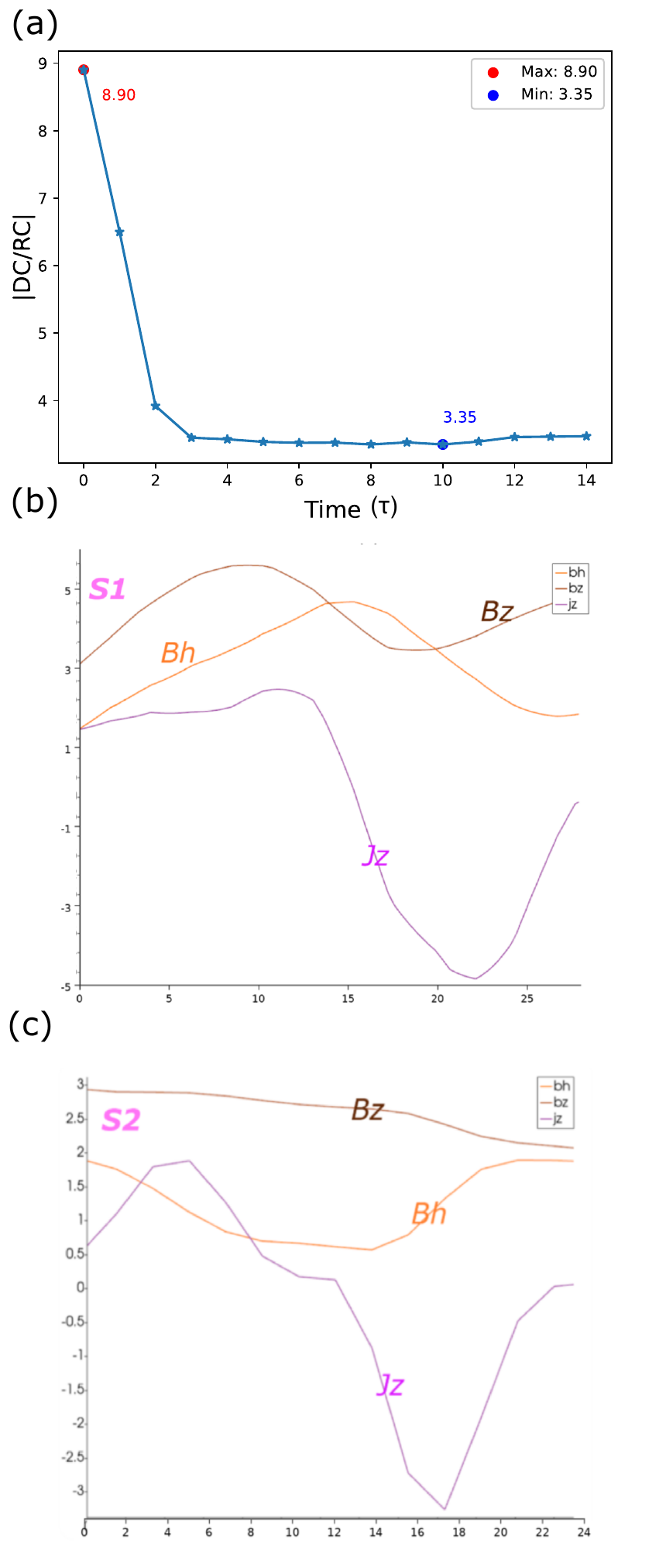}
  \centering
  \caption{Panel (a) shows the temporal evolution of |DC/RC| in the eruption process. Panels (b) and (c) display the average magnetic properties of the direct-return currents near the poloidal polarity ($S1$, Figure~\ref{fig7}d3) and the toroidal polarity ($S2$, Figure~\ref{fig7}d3). The violet, orange, and brown lines represent the profiles of normal electric current ($J_{z}$), horizontal magnetic field ($B_{h}$), and normal magnetic field ($B_{z}$), respectively.
  \label{fig8}}
\end{figure}

Figure~\ref{fig7} shows the distribution of electric current. Panels in rows (a)--(c) display the $J_{x}$, $J_{y}$ and $J_{z}$ components in the $x-z$ plane, respectively. An outer electric current shell with an opposite sign relative to the central core is evident, corresponding to the return current (RC). The return-current regions are nearly co-spatial with the outer shells of negative twist ($T_{w}$) as expected from the definition of $T_{w}$ (Figures~\ref{fig3}e and \ref{fig3}f). Regarding the evolution, the return current is initially much weaker than the direct current flowing through the flux rope, but becomes comparable by the final stage of the eruption as the direct current rapidly decreases. Figure~\ref{fig7}d presents the normal electric current ($J_{z}$) on the bottom plane. The modeled active region is strongly non-neutralised due to the pre-existing flux rope, with an initial $|DC/RC|$ ratio of 8.9 (Figure~\ref{fig8}a). The DC (RC) is computed from the integration of the positive/negative (negative/positive) current in positive/negative (negative/positive) magnetic-field polarity ($\int J_z dS$). Three typical features characterise the electric current evolution: (1) return currents form around the hooked ends of flare ribbons, corresponding to twisted field lines of opposite current helicity (Figure~\ref{fig3}f), but remain absent along the straight, separated ribbons \citep{Janvier2014}; (2) localised currents intensify along separated straight flare ribbons, consistent with \citet{Janvier2014}. This is due to the twist increase produced by magnetic reconnection, which converts sheared arcade fields into twisted flux-rope field lines, as such to increase the current density ($T_{w} = \int \frac{\mu_{0}J_{\parallel}}{4\pi B}dl$). (3) the current density at the flux-rope footpoints decreases, for two reasons. As shown in Figure~\ref{fig3}, the pre-existing flux rope rotates during the eruption, which decreases the twist number through the conversion of twist into writhe, thereby reducing the current density. In addition, the stretching of magnetic field lines can naturally reduce the twist density along a field line in ideal MHD condition, which further lowers the current density.

Figure~\ref{fig8}a shows the temporary evolution of |DC/RC|, which drops rapidly from 8.9 to 3.5 within 4 $\tau$ and then remains nearly constant thereafter. It is worth noting that, because our modeling is based on the TDm model incorporating a strongly current-carrying flux rope, the resulting |DC/RC| ratio is larger than the typical values reported in observations \citep{Liu2017}. To further examine the magnetic properties of return current, we cut two slits along $S_{1}$ (near poloidal polarity) and $S_{2}$ (near toroidal polarity) in Figures~\ref{fig7}(d3) and (d4), respectively. As illustrated in Figures~\ref{fig8}b and \ref{fig8}c, the normal and horizontal magnetic fields at the direct-return current (DC-RC) transition points differ significantly between these two regions. For the return current near the poloidal polarities, the DC-RC transition point corresponds to the maximum horizontal magnetic fields ($B_{h}$) and minimum normal magnetic field ($B_{z}$), closely resembling the conditions at flux-rope footpoints in a simple bipolar magnetic configuration \citep{Xing2024}. However, for the return current near the toroidal polarity, the DC–RC transition point corresponds instead to the minimum horizontal field $B_{h}$.

\section{Confined Mechanisms}\label{sec:res2}

As shown in Figure~\ref{fig5}, our numerical model successfully reproduced a confined flare. Here, we investigate the mechanisms responsible for this confinement, with particular emphasis on the roles of the downward Lorentz force and magnetic reconnection.

\subsection{Downward Lorentz Force Constraining the Solar Eruption}  \label{sec:LorentzForce} 

To investigate the origins and distributions of Lorentz force acting upon the flux rope, we decompose the total magnetic fields $B$ in the simulation into external toroidal field ($B_{\rm T}$), external poloidal field ($B_{\rm p}$) and flux-rope field ($B_{\rm FR}$):
\begin{align}
B=B_{\rm T}+B_{\rm p}+B_{\rm FR}
\end{align}
where $B_{\rm T}$ and $B_{\rm p}$ are directly determined from the prescribed analytical solutions of toroidal ($P_{T}$ and $N_{T}$) and poloidal ($P_{P}$ and $N_{P}$) magnetic charges, respectively. The flux-rope component is then obtained as $B_{\rm FR}=B-B_{\rm T}-B_{\rm p}$, enabling the calculation of the current density flowing through the flux rope via $J_{\rm FR}=\nabla \times B_{\rm FR}/\mu_{0}$. It should be emphasised that, because the magnetic field provided by the external poloidal and toroidal polarities satisfies the current-free potential field, the total electric current $J$ is entirely attributable to the flux rope, namely, $J = J_{\rm FR}$.

\begin{figure*}
  \includegraphics[width=15 cm,clip]{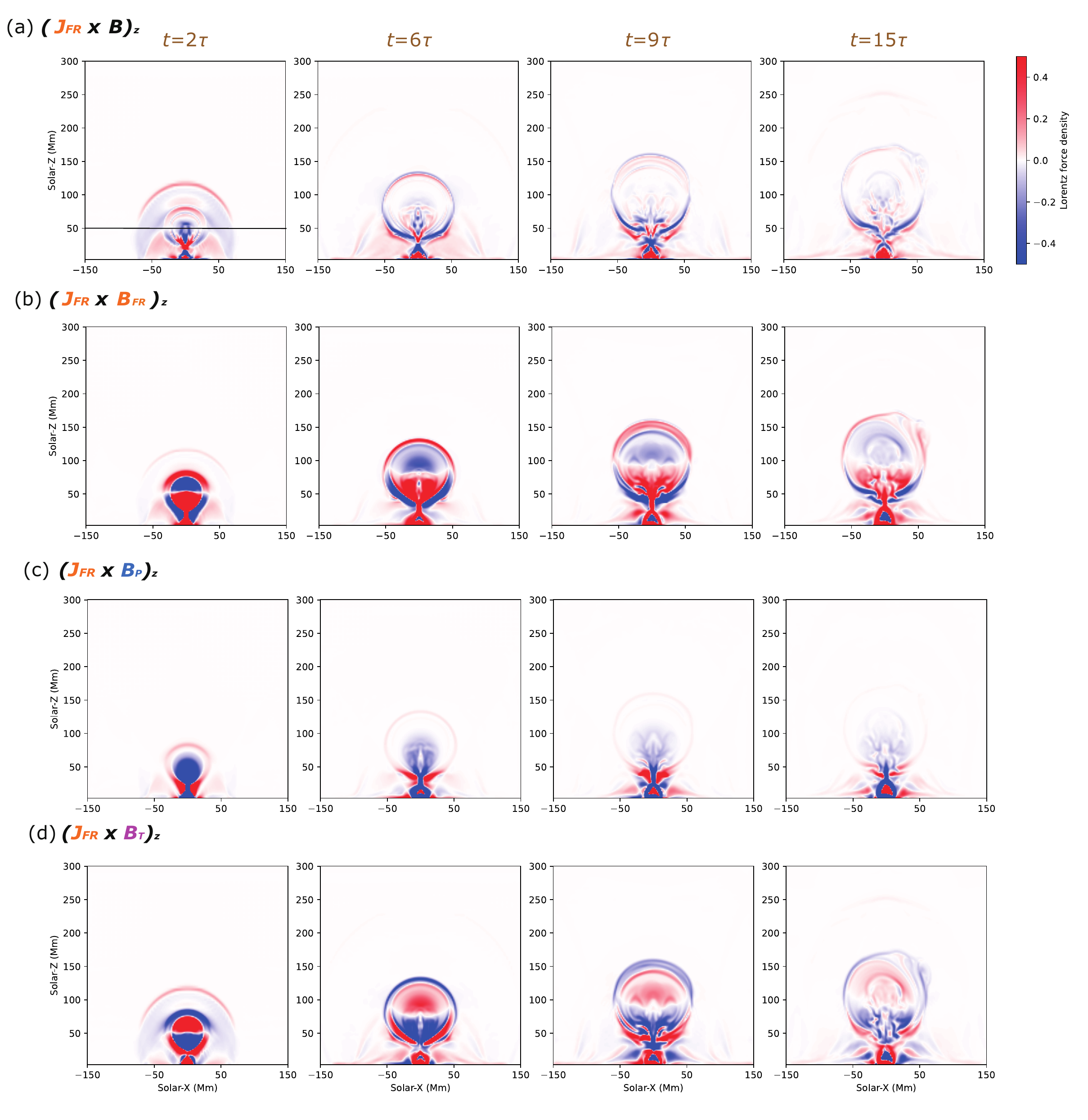}
  \centering
  \caption{Distribution of the vertical Lorentz force in the $x-z$ plane at the time of 2, 6, 9 and 15 $\tau$. Panels (a)--(d) exhibit the vertical component of Lorentz force induced from total magnetic fields, flux-rope fields, external poloidal fields and toroidal fields, respectively. 
  \label{fig9}}
\end{figure*}

Figure~\ref{fig9} displays the vertical Lorentz force acting on the flux rope as it activates in the $x-z$ plane. The temporal evolution of the net Lorentz force (Figure~\ref{fig9}a) shows that, initially, an upward force dominance in the lower part of the flux rope, driving its acceleration before $t=3\tau$ (Figure~\ref{fig4}c). Subsequently, the downward Lorentz force becomes dominant, leading to a deceleration and eventual suppression of the eruption, with the flux rope reaching a downward velocity of $-10$ km s$^{-1}$ at $t=12\tau$. 

To quantify the contributions from different magnetic components, we compute Lorentz force arising from the interaction of the current flowing through the flux rope ($J_{\rm FR}$) with (1) the flux rope itself ($B_{\rm FR}$, Figure~\ref{fig9}b), (2) the external poloidal fields ($B_{\rm p}$, Figure~\ref{fig9}c), and the external toroidal fields ($B_{\rm T}$, Figure~\ref{fig9}d). These are hereafter referred to as the self-force, $B_{\rm p}$-induced force, and $B_{\rm T}$-induced force, respectively. Following the terminology of \citet{Myers2015, Myers2016}, these correspond to the classical hoop force, strapping force, and tension force\footnote{It should be noted that these definitions are not rigid for two reasons. First, the orientations of the toroidal and poloidal components in the flux-rope coordinate system change as the rope rotates. Second, the self-force of the flux rope consists of both the hoop force ($J_{FR, T}\times B_{FR,P}$) and tension force ($J_{FR, P}\times B_{FR, T}$)}. In the following, we use this decomposition to assess how different magnetic field components govern the dynamics of the flux rope eruption.

Figure~\ref{fig9}a shows the distribution of the net Lorentz force, where the dominant downward Lorentz force arises mainly from the flux-rope core and its surrounding shell. Below, we investigate the origins of this downward component. Figure~\ref{fig9}b illustrates the self-force of the flux rope. A comparison with the net Lorentz force distribution indicates that the net downward Lorentz force in the flux-rope (panel a) is not due to this self-force. Figure~\ref{fig9}c demonstrates that the poloidal field-induced Lorentz force initially provides a downward strapping effect, covering the entire flux rope. However, as the flux rope rises, this restraining contribution weakens, approximately balancing the upward gradient of magnetic pressure. The expansion of the flux rope during the eruption further reduces $J_{\rm FR}$, while the simultaneous decrease in $B_{\rm p}$ diminishes the strapping effect, as reflected in the net Lorentz force distribution after $t > 8\tau$.

The toroidal field-induced Lorentz force, however, exhibits different behaviour. First, it acts in opposite directions on the upper and lower portions of the flux rope because the azimuthal current (Figure~\ref{fig7}a) interacts with the unit-directional $B_{\rm T}$ field (Figure~\ref{fig9}d). As a result, the distribution of the toroidal field-induced Lorentz force can be used to determine the occupied areas of the flux rope. Second, a concentric outer shell of negative-sign current develops around the leading part of the eruptive dome, generating strong downward forces as early as $t=3\tau$ and persisting thereafter. Because the flux rope is embedded in a strongly sheared, $B_{\rm T}$ dominated magnetic environment (Figure~\ref{fig3} a), the magnitude of the $B_{\rm T}$-induced force exceeds that of the $B_{\rm p}$-induced force by roughly an order of magnitude.  Consequently, the net Lorentz force globally remains directed downward (Figure~\ref{fig9}a), highlighting the toroidal field as the primary factor constraining the eruption. This conclusion is further supported by the decay index profiles derived from the different magnetic-field components (Figure~\ref{fig2}), which show that the toroidal field decreases more slowly with height than the poloidal field. 

\begin{figure*}
  \includegraphics[width=13 cm,clip]{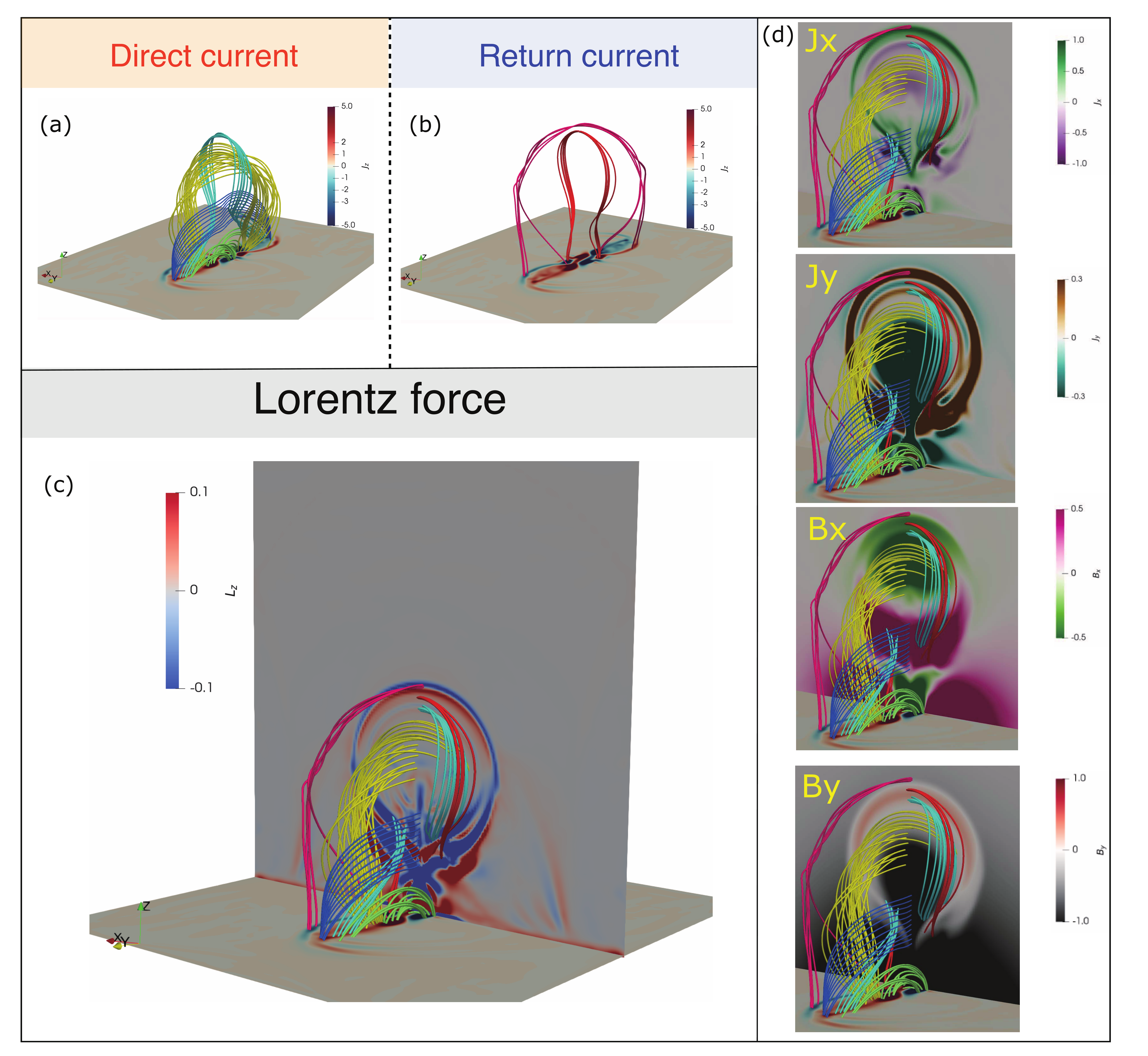}
  \centering
  \caption{The relationship between magnetic field lines and the electric current, Lorentz force and magnetic-field components. Panels (a) and (b) show the field lines traced from direct (yellow, green, cyan and blue lines) and return current (red and magenta lines). Panel (c) shows the distribution of the vertical Lorentz force on the $X$-$Z$ plane. Panel (d) displays the distribution of $J_{x}$, $J_{y}$, $B_{x}$ and $B_{y}$ on the $x$-$z$ plane.
  \label{fig10}}
\end{figure*}

Figure~\ref{fig10} illustrates the spatial relationships among the magnetic topology, electric-current systems (such as DC and RC) and Lorentz force. Figures~\ref{fig10}a and \ref{fig10}b display the field lines rooted in regions of direct currents (yellow, green, purple and pink lines) and return currents (cyan and red-wine lines), respectively. Figure~\ref{fig10}c presents the distribution of the vertical Lorentz force in the $x-z$ plane. It is found that the return current near the toroidal polarity (magenta lines) produces a downward Lorentz force, whereas the return current near the poloidal polarity (cyan lines) produces an upward Lorentz force. To further explore the origin of the downward Lorentz force, Figure~\ref{fig10}d shows the distributions of $J_{x}$, $J_{y}$, $B_{x}$, and $B_{y}$. The magenta and cyan field lines have the same $J_{x}$ but differ in $J_{y}$ in the direction due to rotation. In terms of magnetic fields, both exhibit negative $B_{x}$ components but differ in $B_{y}$. Consequently, the Lorentz force acting on the magenta and cyan field lines points downward and upward, respectively. In summary, these results suggest that the return currents near toroidal polarities are crucial in confining solar eruptions.

To further investigate the role of toroidal fields, we conducted a control group of an eruption without toroidal polarities, as described in Appendix~\ref{fig_ap1}. As shown in Figure~\ref{fig_ap1}, the control case demonstrates that a successful eruption occurs when only external poloidal fields are included (model FR-P), which suggests the critical role of external toroidal fields in producing failed eruptions. In the successful eruption case, flare ribbons exhibit the typical two-ribbon morphology and are associated with a higher degree of non-neutralised electric current compared to the confined eruption case.  

\subsection{Lateral Lorentz Force Driving the Flux-Rope Rotation}  \label{sec:Rota} 

As illustrated in Figure~\ref{fig3}b, the flux rope rotates clockwise (CW) during its eruption, thereby changing the distribution of the Lorentz force acting upon the flux rope. Consistent with the observations, we define a clockwise (CW) rotation as the rotation seen when viewing from above (top-down). Notably, the rotation angle in the successful eruption (model FR-P, Figure~\ref{fig_ap1}) is relatively smaller than in the confined eruption (model FR-PT). Hence, understanding the factors driving flux rope rotation is important for the dynamics of the flux rope.

To this end, similar to the analysis of the downward Lorentz force (Figure~\ref{fig9}), we decompose the lateral Lorentz force in the $x-y$ plane at a height of approximately 50 Mm into contributions from the flux rope self-force itself ($J_{\rm FR}\times B_{\rm FR}$), the external poloidal field-induced ($J_{\rm FR}\times B_{\rm p}$) and toroidal field-induced Lorentz forces ($J_{\rm FR}\times B_{\rm T}$). The flux rope rotation is primarily induced by the component orthogonal to the flux rope, mainly the $x$-component. All Lorentz force components have opposite signs in the two flux rope legs; this induces a torque that rotates the flux rope's upper part (as its footpoints are line-tied). 

\begin{figure*}
  \includegraphics[width=\textwidth,clip]{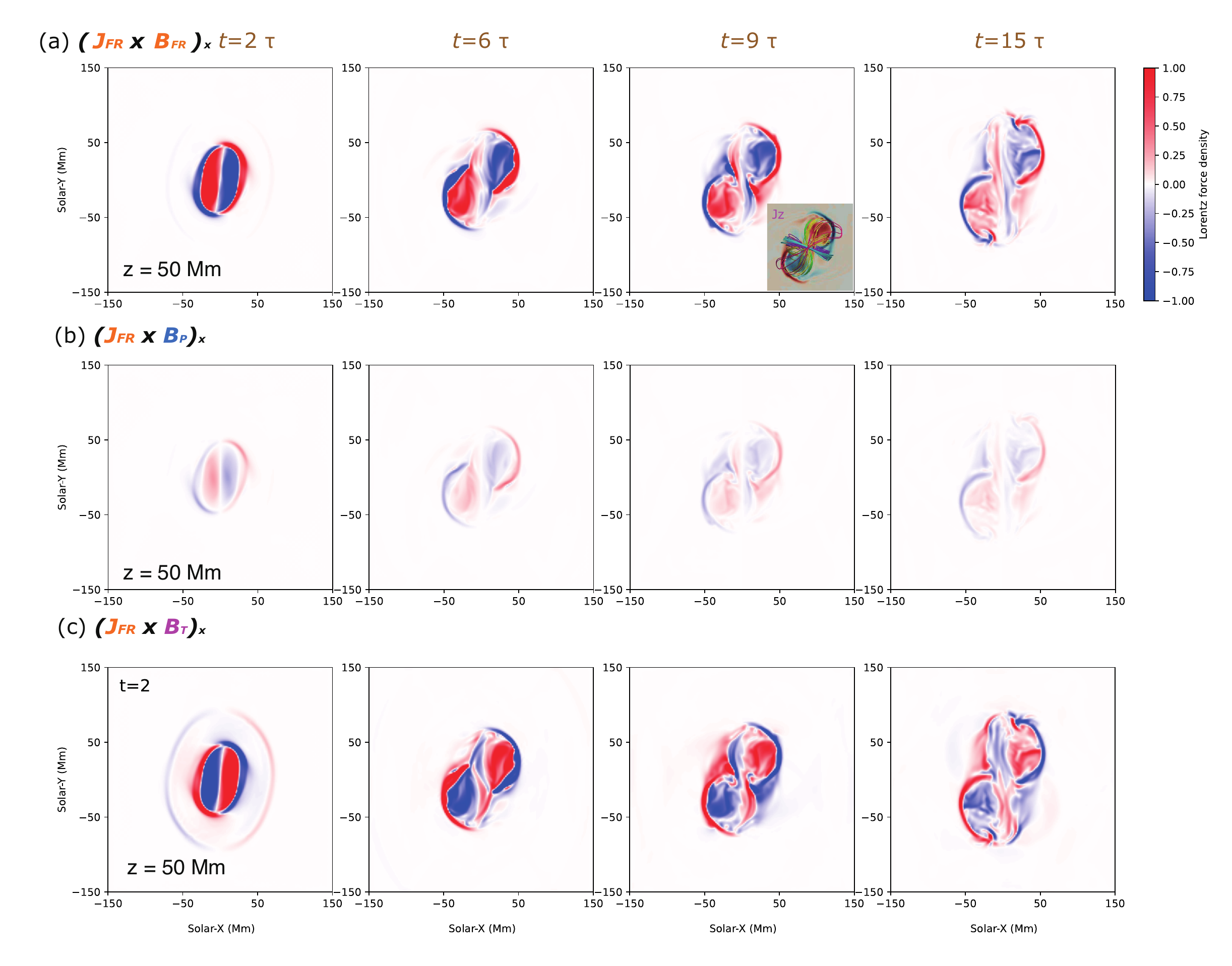}
  \centering
  \caption{Distribution of the lateral Lorentz force in the $x$-$z$ plane at the time of 2, 6, 9 and 15 $\tau$. Panels (a)--(c) exhibit the horizontal $x$ component of Lorentz force induced from the flux rope itself, external poloidal fields and toroidal fields, respectively. The plane is selected at $z=50\;$Mm shown in Figure~\ref{fig9}a. The inserted image in Panel (a) at $t=9$ shows the $J_{z}$ distribution and the corresponding field lines.
  \label{fig11}}
\end{figure*}

As shown in Figure~\ref{fig11}, the main forces are concentrated within the flux rope, with weaker, opposite-directed forces located at the flux rope boundaries within the return current region. The Lorentz force arising from the external poloidal fields is significantly smaller than the other two contributions. In contrast, the Lorentz forces from the flux rope itself and the external toroidal fields act in opposite directions, producing counter-clockwise and clockwise rotations, respectively. This implies that the external toroidal field dominates and determines the rotation direction of the flux rope. Consequently, the rotation does not necessarily reflect intrinsic magnetic properties of the flux rope, such as its twist, suggesting that rotation is not a reliable indicator of kink instability when the flux rope is embedded in strong external toroidal or sheared fields. 

To further assess the role of flux rope self-force itself, we remove both the poloidal and toroidal polarities and simulate the eruption of an isolated flux rope (Figure~\ref{fig_ap4}). This setup excludes other influences on the flux rope dynamics, such as magnetic reconnection at the leading edge and Lorentz forces from external fields. We find that the flux rope largely maintains a toroidal shape and exhibits only weak rotation during the eruption, indicating that its self-force is not the dominant factor controlling rotation.  

In summary, the Lorentz force analysis suggests that strong external toroidal fields can induce a return current at the leading front of the flux rope. This generates a substantial downward Lorentz force that suppresses the rise of the flux rope. Moreover, the external toroidal field itself plays the dominant role in driving flux rope rotation. This explains why so many failed eruptions are frequently accompanied by filament rotation \citep{Zhou2019}

\subsection{External Magnetic Reconnection destroying the Flux Rope}\label{sect:External_Magnetic_Reconnection}
A natural consequence of flux rope rotation is the formation of thinner QSLs in the interface between the flux rope and overlying field lines. This generates intense electric currents, which in turn drive magnetic reconnection involving the eruptive flux rope and finally destroys it. To examine the role of magnetic reconnection in modifying the flux-rope field lines, we traced two representative field lines located near the QSLs and followed their evolution in the eruption process. 

\begin{figure}
  \includegraphics[width=8.5 cm,clip]{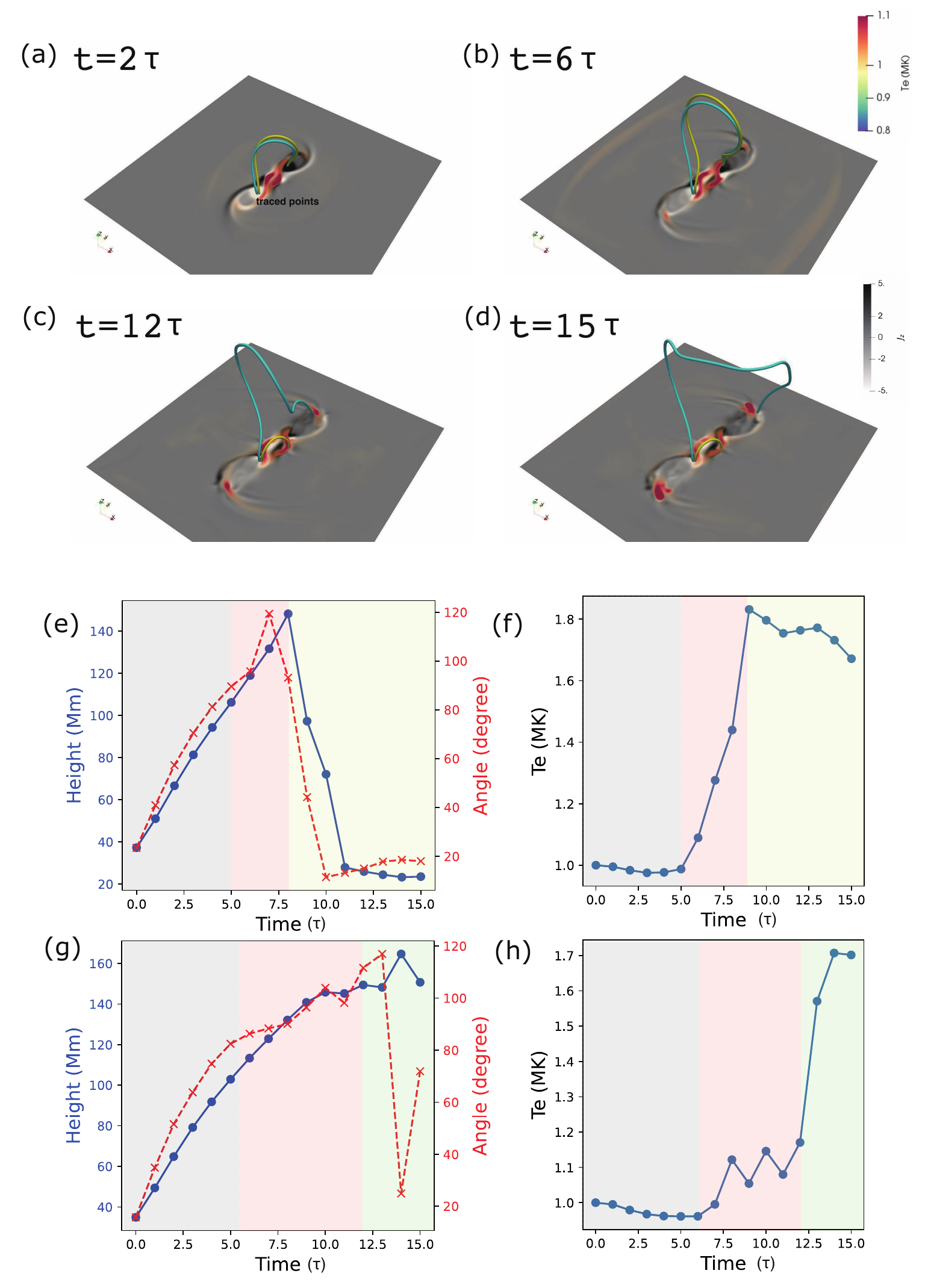}
  \centering
  \caption{Magnetic reconnection in the eruption process. Panels (a)--(d) show the evolution of two typical flux-rope field lines at the initial state. The bottom plane shows the distribution of $J_{z}$, which is overlaid by the temperature distribution with the translucent contours. Panels (e) and (g) show the temporal evolution of the apex height and rotation angle while panels (f) and (h) show the temporal evolution of the maximum temperature of the yellow/cyan field line, respectively. Different colour bands represent different stages (see Section \ref{sect:External_Magnetic_Reconnection}). 
  \label{fig12}}
\end{figure}

Figures~\ref{fig12}a--\ref{fig12}d illustrate the changes in connectivity of the traced field lines. Both field lines are rooted at fixed positions in the negative footpoint of the flux rope. Up to time $= 6\tau$, their field line footpoints nearly remain anchored at the same location while the flux rope is moving upward with a clockwise rotation. After several more $\tau$, however, the positive footpoints of both lines begin to significantly drift (Figure~\ref{fig12}c,d), consistent with slipping magnetic reconnection \citep{Aulanier2006, Aulanier2010, Janvier2013}. The yellow field line turns into a short flare-loop field line from a flux-rope field line. For the cyan field line, one of its footpoints migrates to the magnetic polarity $P_{T}$, forming a large-scale arcade located outside the twisted field region. This evolution is due to magnetic reconnection between the rotating flux rope and the overlying toroidal field.

To further investigate the effects of magnetic reconnection on the field-line connectivity, we compute the apex height, the angle with respect to the $y$--axis, and the maximum temperature along each field line to follow the thermodynamic response of the plasma. Figures~\ref{fig12}e and \ref{fig12}f show the evolution of the yellow field line, which can be divided into three stages. In the first stage (grey bands), the maximum temperature along the field line decreases slightly, corresponding to adiabatic expansion as the apex height gradually increases. During this phase, the flux rope rotates with an almost linear correlation between rotation angle and height. Following this (red bands), the maximum temperature increases sharping, accompanied by a rapid rotation and a sudden decrease in height. Specifically, the apex height drops from 150 Mm to 20 Mm within 4 $\tau$, indicating the transition from a flux-rope field line to a flare-loop field line. These provide clear evidence of magnetic reconnection involving the flux rope, specifically $ar-rf$ reconnection geometry as described by \citet{Aulanier2019}. This reconnection geometry occurs when an inclined arcade ($a$) field line reconnects with the leg of the erupting flux rope ($r$), creating a new flux rope (r) field line rooted far away from the original flux rope anchorage location and a flare (f) loop. This is summarised by the acronym $ar-rf$.

Similarly, Figures~\ref{fig12}g and \ref{fig12}h show that the cyan field line also undergoes three stages, with the first two stages resembling those of the yellow line. However, in the third stage, one of its footpoints migrates into the toroidal polarity ($P_{T}$), leading to a pronounced increase in temperature and rotation angle. Compared to low-lying flare loops, the flare loops formed by reconnection between the flux rope and overlying fields are longer, higher lying, and more twisted. The occurrence of such a reconnection indicates the breakup of the rising flux rope and the confinement of the eruption.

\section{Forward Modelings: thermal and non-thermal}\label{sec:res3}

\subsection{Manifestations on EUV Radiation Images}

To investigate the observational manifestations of solar eruptions in this magnetic configuration, we forward-model the simulation results using the temperature and density distributions to synthesise EUV images at 94 \AA\ as observed from the top, side and front views. In particular, the simulated temperature is scaled by an amplification factor in order to enhance the coronal emission response in high-temperature wavelength channels. Given that the coronal opacity is typically much less than unity, we adopt the optically thin approximation, considering only emission and neglecting absorption. The synthesis includes two steps. First, the emissivity in each cell is computed as $n_{e}^{2}G_{\lambda}(T)$. The synthesized image is then obtained by integrating this emissivity along the line of sight. The $G_{\lambda}(T)$ denotes the temperature response function of the EUV passband. For example, the $G_{94}(T)$ can be obtained from the SDO/AIA instrument response functions.

\begin{figure*}
  \includegraphics[width=16 cm,clip]{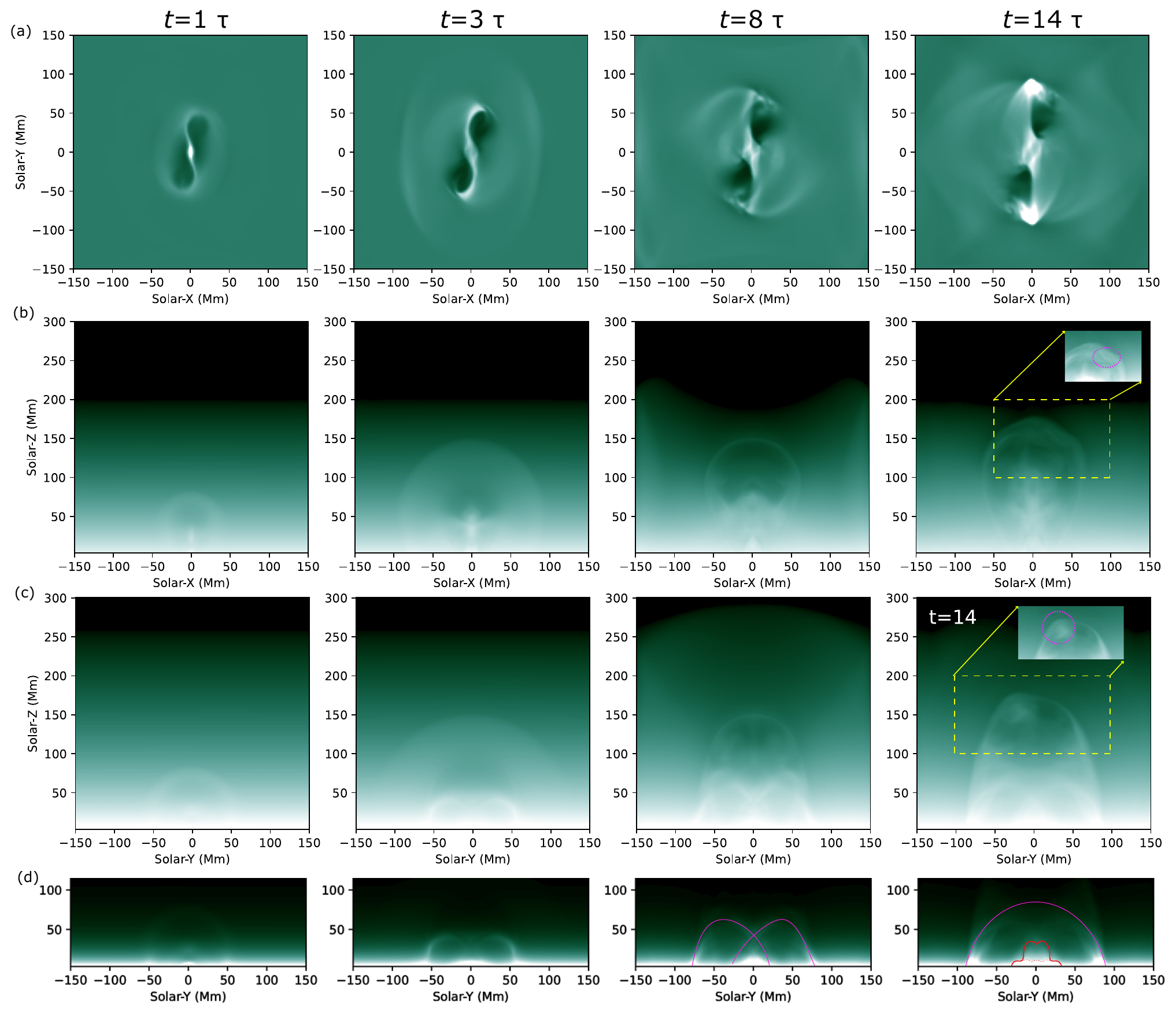}
  \centering
  \caption{Synthesised EUV 94 \AA\ radiation images viewed from the (a) top, (b) front and (c) side. The zoomed images in Panels (b) and (c) at $t=14$ indicate the plasma blobs due to magnetic reconnection between the flux rope and the overlying toroidal fields. Panel (d) displays the temporal evolution of post-flare loops. Pink and red lines at $t=14$ (bottom right panel) indicate the formed highly-sheared arcades and the underlying flare loops, which are formed due to magnetic reconnection from two group of sheared arcades (pink lines at $t=8$).}
  \label{fig13}
\end{figure*}

Figure~\ref{fig13}a illustrates the evolution of flare ribbons and their associated dimmings. The ribbon dynamics proceed in two stages. In the first stage, the flare ribbons elongate along the PIL, which is due to magnetic reconnection involving strong sheared fields \citep[as provided by the external toroidal magnetic fields, as in ][]{Qiu2014}. Then, the ribbons separate perpendicular to the PIL, indicating that increasingly potential magnetic fields, supplied by the poloidal polarities, participate in reconnection. The hooked structures at the ribbon ends, remain only partially closed, owing to the relatively low twist number of the flux rope \footnote{In principal, strong toroidal fields can decrease the twist number of the flux rope, while poloidal fields have opposite effects.}. In addition, two conjugated dark dimmings appear within the hooks and expand outward during the eruption, which is in accord with the evolution of QSLs and twist number shown in Figures~\ref{fig3} and \ref{fig6}. In the later stage of the eruption, two dot-like ribbons form at the far ends of the flux rope, situated at the borders of the toroidal polarities. These ribbons result from reconnection between the flux rope and the predominantly toroidal external magnetic field.

Figures~\ref{fig13}b--d present synthesised EUV images from the end and side perspectives, respectively. Although the eruption remains confined in our model, the end-view images (Figure~\ref{fig13}b) still exhibit the characteristic three-part structure of a CME: a bright leading front, a dark cavity, and a bright core, with flare loops visible beneath. By $t=14 \tau$ in Figure~\ref{fig13}b, a plasma blob (pink circle in the zoomed view) appears in the upper-right region of the dome-like structure. This feature serves as an observational signature of external reconnection involving the flux rope, consistent with previous observations \citep{Lilp2016, Lizf2023}.

From the side view (Figure~\ref{fig13}c), two downward-curved sheared arcades are visible beneath the dark cavity at $t=3 \tau$. These two sheared arcades subsequently reconnect, forming an extended hot arcade along with underlying flare loops. Figure~\ref{fig13}d provides a zoomed-in view below 50 Mm, where the flare-loop formation is traced with pink and red lines. The reconnection between the two groups of sheared arcades (pink lines) gives rise to high-lying flare loops that collectively display a transient “cowboy-hat” morphology, characterised by central loops that arch higher than those on either side. 

\begin{figure*}
  \includegraphics[width=13 cm,clip]{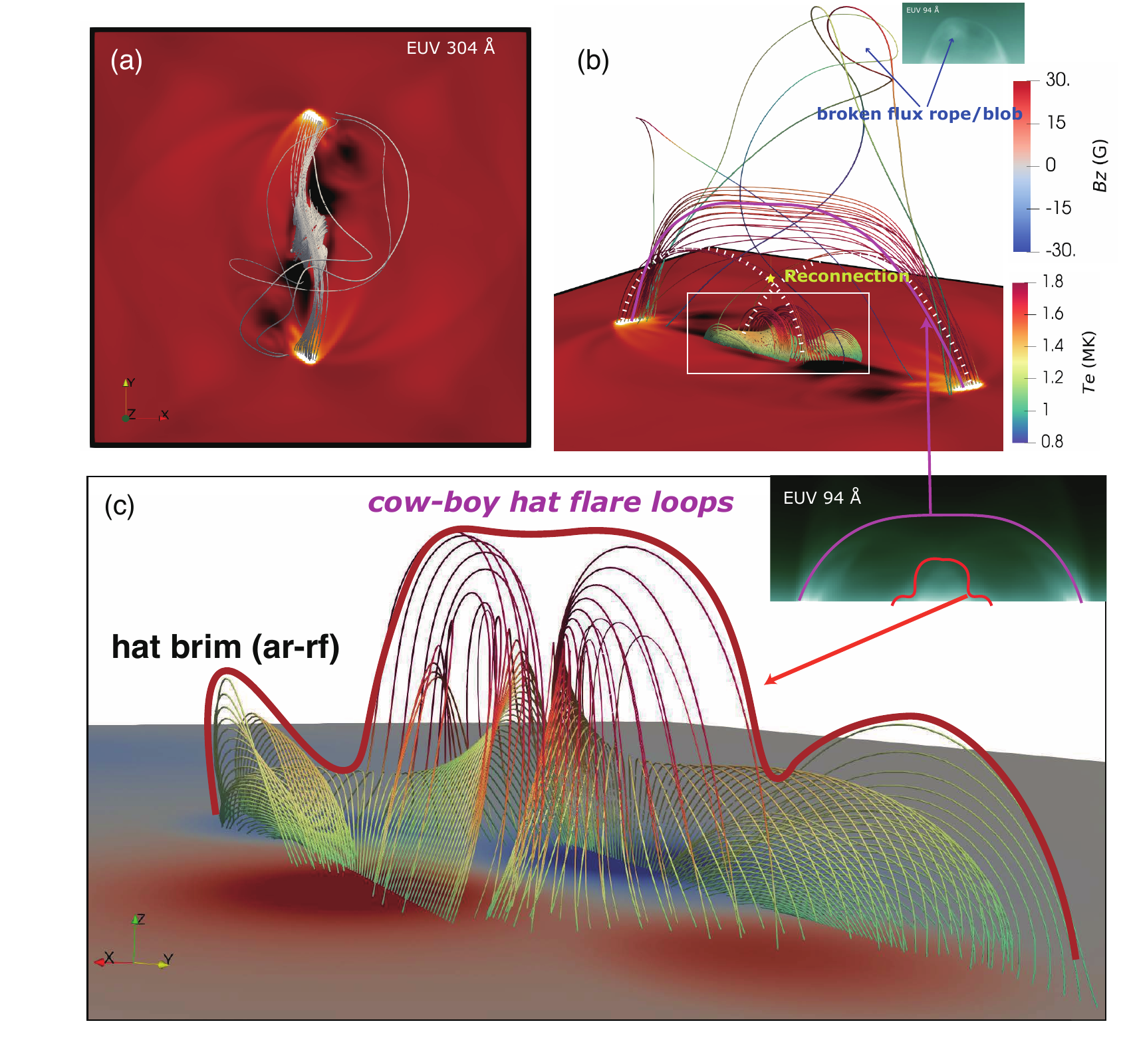}
  \centering
  \caption{Magnetic structure of flare loops and the synthesised radiation images. Panels (a) and (b) show the field lines traced from the bright regions in synthesised 304 \AA\ radiation images viewed from the top, in which the field lines are colour-coded by the temperature. Panel (c) shows the structure of post-flare loops and the comparison with synthesised 94 \AA\ images viewed from the side.
  \label{fig14}}
\end{figure*}

Figure~\ref{fig14} shows the magnetic structure of flare loops and the synthesised 304 \AA\ images viewed from the top. The highly sheared arcades connect the two bright regions, where the temperature is higher than the broken flux rope. One footpoint of the broken flux rope is anchored in dimming areas, whereas the other footpoint connects to the border of the dimming. The underlying flare loops exhibit a cowboy-hat morphology, in which the central portion (the hat crown) is elevated above both ends. Notably, the loop ends, corresponding to the hat brim, rise higher than the intermediate section between the brim and the crown. This configuration results from $ar-rf$ reconnection between the flux-rope field lines and the surrounding sheared arcades \citep{Lorincik2021}. As a result, the analysis based on synthesised EUV images and magnetic fields indicates that "cowboy-hat" morphology of the global set of flare loops is one manifestation of reconnection involved in the eruption located within the toroidal magnetic cage. It is worth noting that, the "cowboy-hat" morphology is an ensemble of flare loops formed with different reconnection geometries rather than an individual loop.

\subsection{Non-thermal Response of Electron Acceleration: Hard X-ray sources}

To investigate the non-thermal response during the eruption, we employ the test particle method based on the guiding-centre approximation (GCA) within the AMRVAC framework to study electron acceleration in the MHD environment. As described by \citet{Fabio2024} and \citet{Wu2025}, the GCA method is governed by six equations: the evolution of the guiding-centre position ($\mathbf{R}$), the parallel velocity component ($u_{||}$), and the conservation of the magnetic moment ($\mu$) (see Appendix~\ref{sec:GCA} for more details). Particle acceleration is primarily driven by direct electric fields, as well as curvature, polarisation, mirror, and relativistic drifts in solar environment. The direct electric field $\mathbf{E}$ is defined as $-\mathbf{v} \times \mathbf{B} + \eta \mathbf{J}$, where $\eta = 5 \times 10^{-6}$ in normalised units. For more details on the GCA method, see \citet{Fabio2024} and \citet{Wu2025}. In our simulation, $5 \times 10^5$ electrons following a Maxwellian velocity distribution with a temperature of $T = 1$ MK are injected into the magnetic field environment provided by the MHD model, uniformly distributed across all cells, which are not related to any MHD quantities. The hard X-ray sources are synthesised using the bremsstrahlung thick-target model \citep{Zhoux2016}.

\begin{figure*}
  \includegraphics[width=16cm,clip]{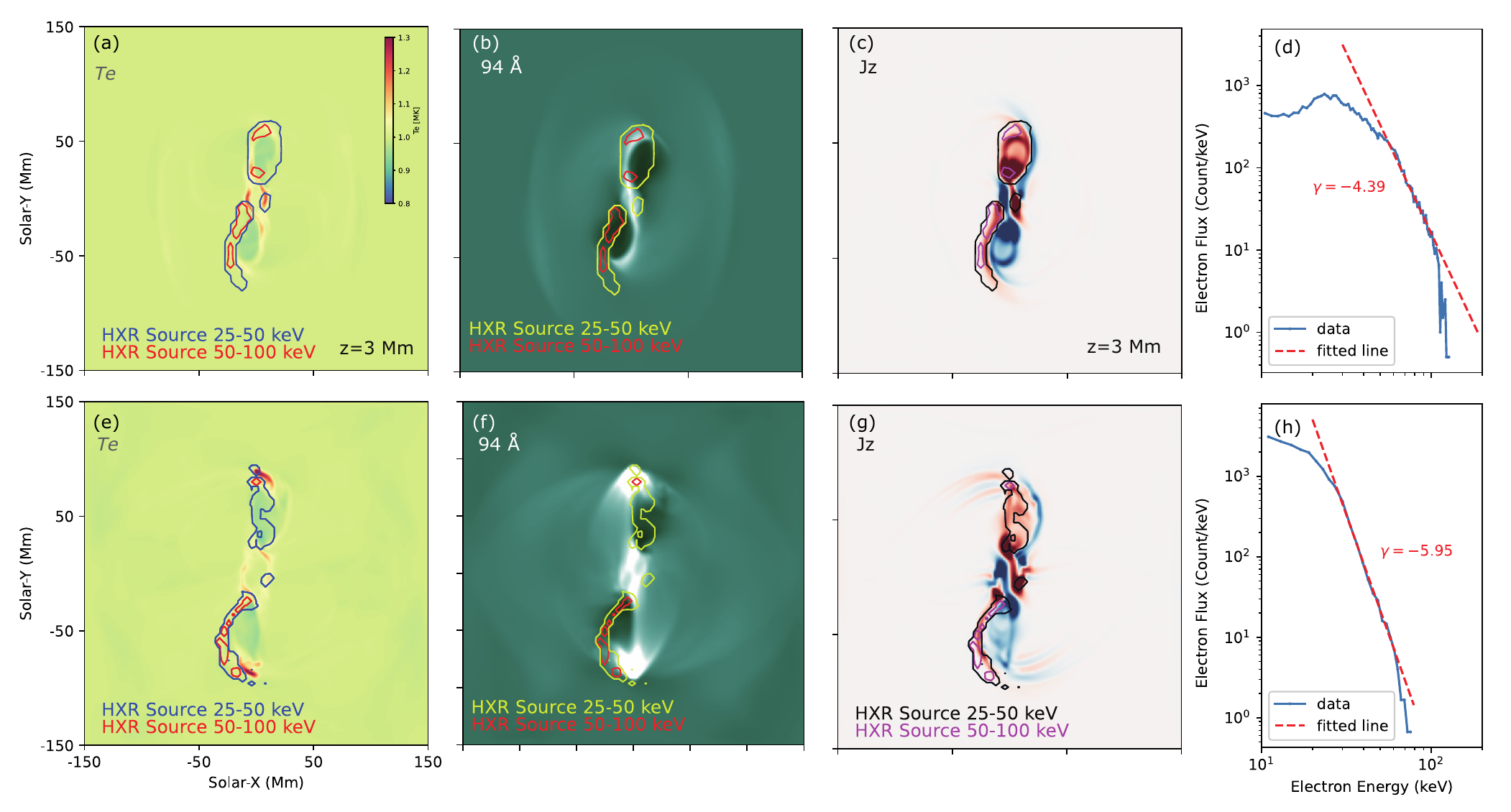}
  \centering
  \caption{Hard X-ray sources and electron spectra precipitating at the bottom boundary. The top and bottom rows correspond to $t=3$ and 14 $\tau$, respectively. Panels (a, e), (b, f) and (c, g) display the HXR sources (blue/yellow and red contours) overlaid on the bottom temperature distribution, synthesised EUV 94 \AA\ emissions and vertical electric current, respectively. Panels (d) and (h) present the spectrum of electrons precipitating at the bottom boundary at $t=3$ and 14 $\tau$, respectively. 
  \label{fig15}}
\end{figure*}

Figure~\ref{fig15} shows the hard X-ray sources overlaid on the bottom distribution of temperature, EUV 94 \AA\ images and normal electric currents. At $t = 3\tau$, hard X-ray sources with energies exceeding 50 keV are distributed in both the positive-polarity regions (associated with direct currents) and the negative-polarity regions (associated with return currents), with a spectral index of –4.39. Electrons in the 25–50 keV range are deposited in flux-rope footpoints, straight ribbons and return current. By contrast, at $t = 15\tau$, hard X-ray sources above 50 keV are primarily deposited in return-current regions, indicating that return-current regions have become the dominant sites of energetic particle (>50 keV) acceleration, and the spectral index becomes softer to –5.95. The softening of the energy spectrum between the two snapshots indicates a reduction in high-energy electrons, which may be associated with the decrease in the footpoint electric current. 

Furthermore, the spatial distributions of energetic electrons, heated regions and flare ribbons seen in synthesised EUV images are not co-spatial, revealing a divergence between thermal and non-thermal responses\footnote{It should be noted that, the feedback of energetic electrons on the MHD equation is neglected. Therefore, all heating arises solely from thermal mechanisms, including Joule heating, compression, and thermal conduction.}. The heated regions, corresponding to bright flare ribbons, are mainly co-spatial with the direct-current regions (see Figure~\ref{fig7}), whereas the $>50\;$keV electrons are primarily deposited in the return-current regions. This spatial discrepancy reflects the distinct roles of return currents in plasma heating (via Joule heating) versus electron acceleration. In particular, the thermal responses, such as EUV emissions and the temperature distribution at the bottom boundary, exhibit a symmetric pattern, whereas the hard X-ray sources display pronounced asymmetry, as reported by \citet{Jin2007}, \citet{Naus2022} and \citet{Shi2024}. 

Regarding the enhanced electron acceleration in the return-current region, it is found that the electrons are accelerated efficiently in return currents. In the toroidal magnetic cage configuration, a current sheet beneath the flux rope fails to develop efficiently during a failed eruption. Instead, the QSLs between the flux rope and the surrounding magnetic field become more pronounced, corresponding to regions of enhanced return current. The magnetic structures carrying the return current occupy a substantially larger area than those associated with the direct current, as shown in Figure \ref{fig10}. Consequently, a larger population of electrons is trapped and accelerated along the longer return-current field lines. In addition, the current intensity in the return-current region is comparable to that of the direct-current region, indicating that the acceleration efficiency in the return-current channels is likewise significant.

\section{Discussions} \label{sec:dis}

\subsection{The role and magnetic properties of return current in the dynamics of flux rope}

Previous studies have shown that electric current neutralisation (quantified as |DC/RC|) serves as a reliable indicator for determining the success or failure of a solar eruption \citep{Liu2017, Liu2024}. A positive correlation has been found between CME activity and the degree of current non-neutralisation in these studies, implying that return currents play a critical role in constraining CME production. Nevertheless, two key questions remain open: (1) Which magnetic configurations are prone to producing significant return currents? (2) How do return currents physically act to inhibit CME production?

Regarding the magnetic conditions that favour the generation of return currents, comparisons between the FR-PT and FR-P models indicate that external toroidal fields can induce strong return currents, reducing |DC/RC| ratio from about 30 to 10. As shown in Figures~\ref{fig7} and \ref{fig_ap2}, the presence of toroidal magnetic fields results in substantial return currents concentrated around the toroidal polarities. This is consistent with \citet{Duan2024}, who reported that the degree of current neutralization in an active region is positively correlated with its magnetic non-potentiality. In our modelling, the inclusion of toroidal magnetic fields naturally reduces the twist of the core fields while enhancing the background magnetic component, thereby leading to a more current-neutralised configuration. In addition, the return currents are mainly distributed around the hooked structure at the ribbon ends rather than along their straight parts, as shown in \citet{Janvier2014}. 

Our results show that the magnetic properties of return current differ between the toroidal and poloidal polarities (Figure~\ref{fig8}). The DC–RC transition point corresponds to a maximum in $B_{h}$ near poloidal polarities but to a minimum in $B_{h}$ around toroidal polarities. Magnetic field lines originating from poloidal polarities are more twisted (cyan lines in Figure~\ref{fig10}), corresponding to a larger $B_{h}$ field, compared to that around toroidal polarities (magenta lines in Figure~\ref{fig10}). It should be emphasised that the conclusions here are based on the TDm model, meaning that this work cannot capture how return currents are induced in a temporal evolution forming the magnetic configuration. As shown in \citet{Chintzoglou2019}, the collisional shearing motions of two emerging flux tubes may produce the toroidal magnetic cage configuration illustrated here. In future work, we intend to employ observational data-driven simulations to investigate mechanisms of return current generation in real active regions.

Regarding the role of return currents in confining eruptions, we show that return currents surrounding the toroidal polarities produce a downward Lorentz force (Figure~\ref{fig10}). This force arises from the interaction between the return currents and the overlying toroidal magnetic field (Figure~\ref{fig9}), which acts to restrain the rising of the flux rope and thereby leads to confined flares. In contrast, the Lorentz force from return currents around poloidal polarities is directed upward. This reveals the effects of the return current in governing the dynamics of the flux rope.

In summary, our simulations provide a self-consistent explanation for why active regions with more neutralised currents (lower |DC/RC| ratios) are more prone to form confined flares \citep{Liu2017}: the downward Lorentz force generated by return currents around toroidal polarities suppresses the ascent of the flux rope. These results suggest a potential link between magnetic structure (toroidal vs. poloidal fields), current neutralisation (|DC/RC|), and CME activity (success vs. failure eruptions), a relationship that could be further tested through future statistical studies. 

\subsection{Failed eruptions associated with filament rotation}

Intriguingly, many failed eruptions are closely associated with large-angle rotations of filaments or flux ropes \citep{Ji2003, Torok2005, Amari2018, Zhou2019, Jiang2023, Guo2024, Zhang2024}. \citet{Zhou2019} conducted a statistical analysis and found that all filaments associated with large-angle rotation in their sample ultimately halted at heights where decay index $n$ exceeded the commonly used torus-instability threshold of 1.5, indicating that the overlying poloidal field could not restrain the ascent of eruptive flux rope. These observations raise the two key questions: (1) Why do eruptions with large-angle filament rotation so often fail? (2) Why is the decay index at the halt height usually above 1.5 for such events? 

Our simulation results provide a self-consistent explanation for these issues. As shown in Figure~\ref{fig11}, it is that the lateral Lorentz force driving the flux-rope rotation is mainly contributed by external toroidal magnetic fields ($J \times B_{T}$), rather than the flux-rope itself ($J \times B_{FR}$) or the external poloidal field ($J \times B_{P}$). This finding is consistent with the parameter survey of \citet{Kliem2012} and the Lorentz-force torque analysis of \citet{Zhou2023}, which demonstrates that the external sheared/toroidal field is the primary driver of flux-rope rotation. \citet{Zhang2024} further demonstrated that the rotation angle of a flux rope is not strongly correlated with its twist number prior to the eruption. These indicate that kink instability, which arises from flux-rope self-force itself, is not the main driver of flux-rope rotation. In other words, kink instability can act as a trigger for the initiation of flux rope eruptions (such as to trigger slow-rise phase before impulsive phase), but not as the driver governing their rotation during eruption. We therefore suggest that large-angle rotation should not be regarded as an indicator that an eruption is driven by kink instability. Instead, external toroidal fields aligned with the flux-rope axis are the dominant factor controlling flux-rope rotation. The other effect of the external toroidal field is that it produces a downward Lorentz force, which constrains the eruption. Thus, external toroidal fields can both suppress the rising of the eruptive flux rope and drive its rotation, explaining why filament rotation and failed eruptions are frequently observed together. Additionally, we performed a set of benchmark runs with varying the toroidal magnetic-field strength to examine how solar eruptions depend on it in Appendix \ref{sec:ps}. It is found that the rotation angle increases as the external toroidal field becomes stronger. Moreover, when the toroidal field is sufficiently strong, the eruption is suppressed and the system transitions to a failed-eruption regime. Overall, these results suggest that a stronger external toroidal field tends to enhance flux-rope rotation while also making confinement, and hence failed eruptions, more likely.

Regarding the second issue, our simulations indicate that magnetic reconnection between the flux rope and overlying toroidal fields ultimately results in its destruction. In this scenario, reconnection-induced disruption is the primary mechanism restraining the eruption \citep{Jiang2023}. Furthermore, the constraining force is mainly due to the toroidal-field induced tension force \citep{Myers2015, Myers2016}, rather than the poloidal-field induced strapping force, which accounts for why the decay index at the stopping height often exceeds 1.5.

\subsection{Flare loops as a proxy for predicting CME activity}

Generally, confined and eruptive flares are classified based on the presence or absence of CMEs detected by coronagraphs. However, this criterion can be unreliable for eruptions occurring near the solar disk center or for slow CMEs. In such cases, the observational signatures in the source regions, such as flare ribbons and loops, become essential to distinguish between successful and failed eruptions.

\begin{figure*}
  \includegraphics[width=14 cm,clip]{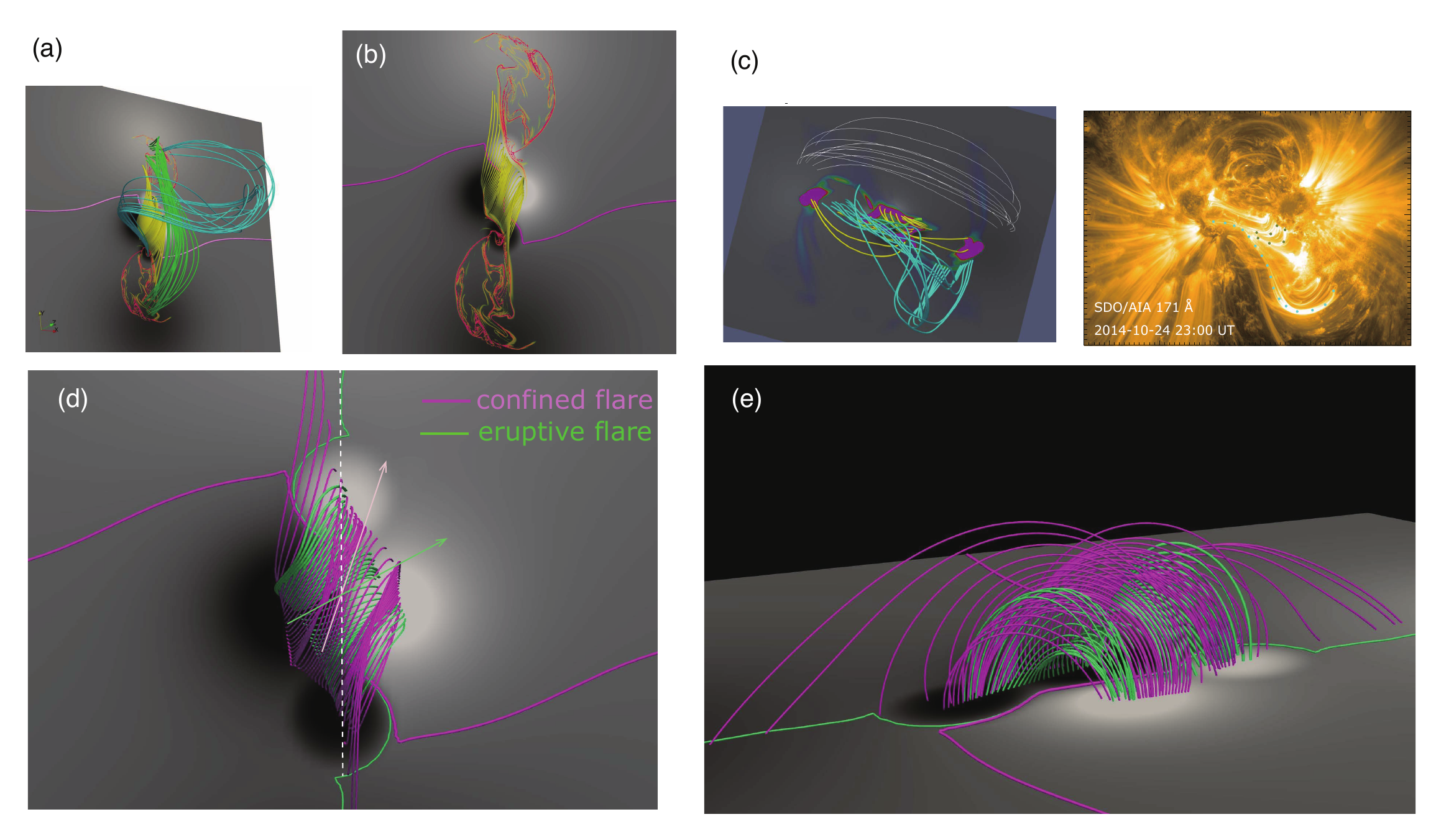}
  \centering
  \caption{Panels (a) and (v) show representative field lines traced from the bottom QSLs (red). Cyan and yellow lines denote the eruptive flux rope and the underlying flare loops, while green lines indicate the strongly sheared arcades connecting the two toroidal polarities. Pink curves mark the PILs on the bottom boundary. Panels (c) compare flare loops in simulations with corresponding observations. Panels (d) and (e) compare the underlying flare loops between the confined (pink lines) and eruptive cases (green lines).Pink and green curves denote the PILs on the bottom boundary for the confined and eruptive cases, respectively. \label{fig16}}
\end{figure*}

To this end, we compare the morphology of flare ribbons in confined (FE-PT) and eruptive (FE-P) flares in Figure~\ref{fig16} and Figure~\ref{fig_ap3}, as traced from the bottom QSLs. In the confined flare case, magnetic reconnection produces three distinct structures from bottom to top: (1) short underlying flare loops traced from the inner dynamic QSLs, (2) highly sheared loops anchored at both ends of the QSLs around the toroidal polarities, and (3) overlying broken twisted loops with one footpoint at the pre-eruptive flux-rope location and the other at the outer QSL boundary near toroidal polarities. 

To evaluate how well our simulation reproduces observations, we compare the modelled flare loops with those of a well-studied confined flare (Figure~\ref{fig16}c), 
also referred to as a “magnetic cage” \citep{Jiang2016, Amari2018}. It is seen that both the simulation and observations exhibit strongly sheared low-lying loops, and multiple ribbons (heated regions). Figures~\ref{fig16}d-f present the ribbons and loops of the eruptive flare. In contrast to the more complex structure in the confined flare case, the eruptive flare displays a typical two-ribbon pattern in the QSLs together with saddle-like flare loops \citep{Lorincik2021}, where the loop ends are elevated relative to the central. This geometry results from $ar$–$rf$ reconnection between flux-rope field lines rooted at the hook and the ambient arcade field lines \citep{Aulanier2019}. As shown in Figure~\ref{fig16}f, the modelled flare loops closely match the observations very well. These comparisons with observations suggest that our data-inspired models are capable of reflecting real-world observations.

The flare loops in confined and eruptive flares show different morphologies, making them a useful proxy to predict CME activity. In successful eruptions, flare loops typically display a saddle-like shape with a central dip. However, in confined flares they form a “cowboy-hat" shape, with the central loops elevated relative to the sides. Moreover, despite having nearly co-spatial central PILs (pink and green lines in Figure~\ref{fig16}d), the two cases differ significantly in shearing degree in flare loops: flare loops in confined flares are more strongly sheared (pink arrow) than those in the eruptive case (green arrow). These results suggest that the morphology and shearing degree of flare loops can serve as diagnostic indicators: strongly sheared, cowboy-hat-shaped loops are associated with confined flares, while weakly sheared, saddle-shaped loops are characteristic of successful eruptions. It should be noted that observations have shown that photospheric magnetic fields typically exhibit larger shear angles, with non-neutralised electric current in successful eruptions \citep{Liu2017, Liu2024}. In this work, however, the shear degree refers to the geometry of the flare loops rather than the photospheric magnetic field, which reflects the shearing degree of coronal magnetic fields rather than the photosphere. These results imply that accurately quantifying the degree of shear in coronal magnetic fields is crucial for distinguishing between eruptive and confined flares.

Moreover, the “cowboy-hat”–shaped flare loops are transient, relaxing into arch-like structures within a few Alfv\'en crossing times. This places stringent constraints on observations: to minimise relaxation effects, measurements should be obtained as close as possible to the flare peak, while the high temperatures of newly reconnected flare loops favor observations in hot channels such as 94 and 131~\AA .

\section{Summary} \label{sec:sum}

\begin{figure*}
  \includegraphics[width=17 cm,clip]{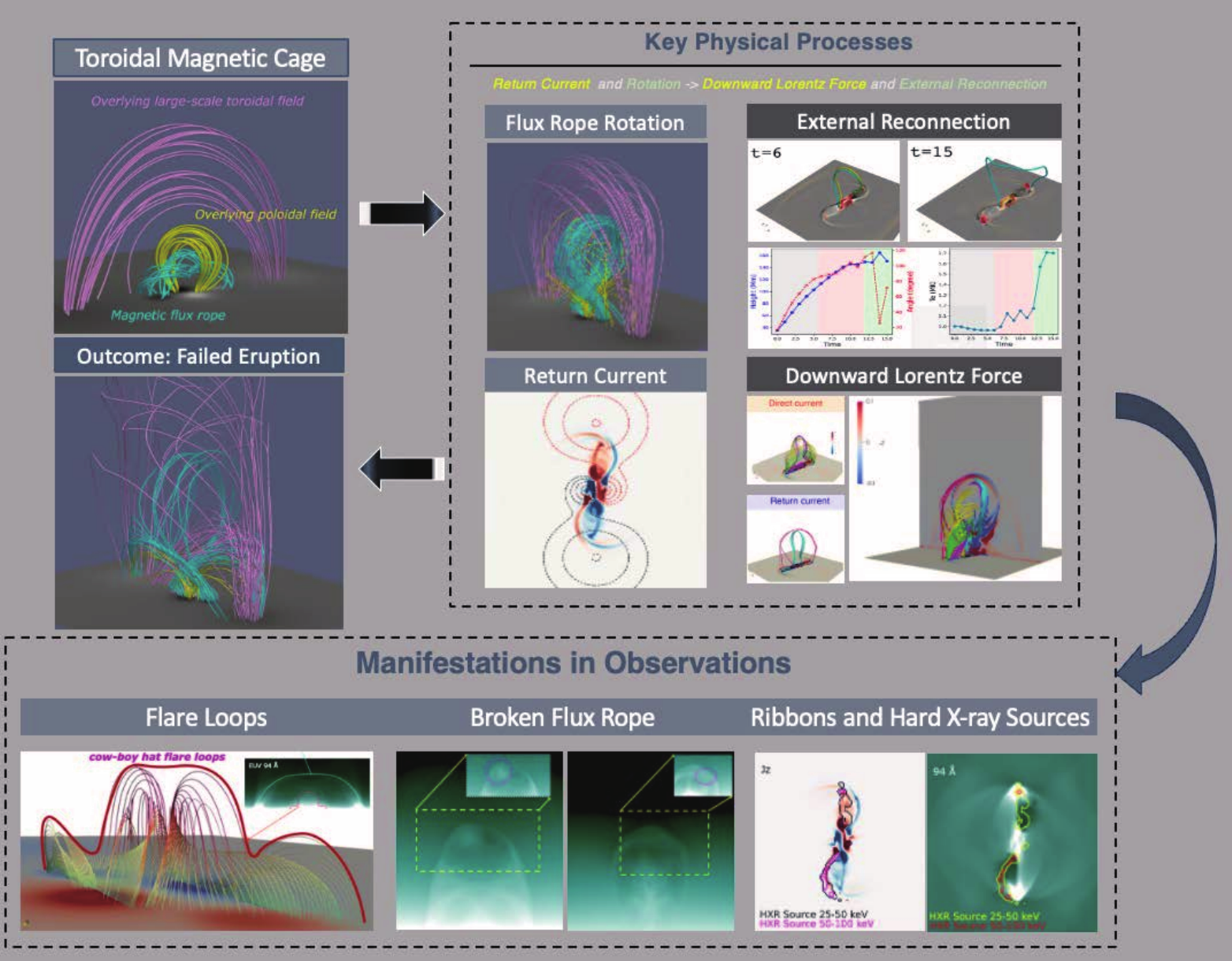}
  \centering
  \caption{Key physical processes and observational manifestations of failed eruptions in toroidal magnetic cage. \label{fig17}}
\end{figure*}

In this paper, we investigate the physical processes of solar eruptions in a configuration of a toroidal magnetic cage, with particular emphasis on the confined mechanisms and the observational manifestations. Figure~\ref{fig17} summarises the key physical processes of the toroidal magnetic cage, its role in driving failed eruptions, and the corresponding observational manifestations. Our simulations demonstrate that return currents play a critical role in restricting the rise of the eruptive flux rope, providing an explanation for the tendency of confined flares to occur preferentially in current-neutralised active regions. In addition, by comparing successful and failed eruption cases, we show that the shearing degree and the morphology of flare loops can serve as useful indicators to determine CME activity.

\begin{enumerate}

\item{\textbf{Confined mechanisms}: Our results reproduce the failure of a torus-unstable flux rope constrained by overlying toroidal magnetic fields, namely, the toroidal magnetic cage. A comparison with a control experiment excluding toroidal polarities reveals that external toroidal magnetic fields play a crucial role in producing confined flares. On the one hand, the return current around toroidal polarities induces a significant downward Lorentz force above the flux rope, which suppresses the flux rope's ascent, even as magnetic reconnection continues beneath it. On the other hand, the toroidal fields drive the rotation of the flux rope, which results in the reconnection between the rotating flux rope and the overlying toroidal magnetic fields. This ultimately causes the destruction of the eruptive flux rope. The rotation angle of the flux rope increases with the strength of the toroidal magnetic field, and the eruption becomes confined when the toroidal field is sufficiently strong. These two effects of toroidal fields provide a self-consistent explanation for why filament eruptions associated with large-angle rotation are often accompanied by failed eruptions.}\\

\item{\textbf{Observational manifestations in thermal and non-thermal emissions}: based on simulation results, we synthesise both thermal (EUV emissions from plasma temperature and density) and non-thermal (hard X-ray sources from energetic electrons) during the eruption. The flare ribbons exhibit a multi-ribbon shape with a clockwise rotation, and two bright kernels appear near the toroidal polarities in the final stage of the eruption. The flare loops consist of underlying “cowboy-hat-like” loops, strongly sheared arcades connecting toroidal polarities, and the overlying broken flux rope accompanied by plasma blobs above. The hard X-ray sources are mainly spatially displaced from one EUV flare ribbon, reflecting differences in plasma heating and electron acceleration sites. Moreover, in the late stage of the eruption, the majority of energetic electrons above 50 keV are deposited along the return current ribbon, where the plasma is only weakly heated.}\\

\item{\textbf{Proxy for distinguishing CME activity}: Our results suggest that the morphology and shearing angle of flare angle can serve as reliable diagnostics for distinguishing between confined and eruptive flares. In confined flares, flare loops display a relatively transient “cowboy-hat” shape, with the central part elevated above both ends. By contrast, eruptive flares show a “saddle-like” shape, where the loops at both ends rise higher than the centre. They are expected to observe near the flare peak time in high-temperature channel. Moreover, the shearing degree of flare loops, reflecting the shearing degree of coronal magnetic fields rather than the photosphere, in confined flares is considerably stronger than in eruptive flares. Consequently, the shearing degree of flare loops serves as a useful predictor of CME activity, with loops oriented nearly perpendicular to PIL being more likely to produce CMEs.}

\end{enumerate}

In summary, this study provides a systematic investigation of the magnetic cage in constraining solar eruptions, elucidating the coupled roles of the production of downward Lorentz force \citep{Zhong2021, Zhang2024, Guo2024}, flux-rope rotation \citep{Kliem2012, Zhou2023, Zhang2024} and external magnetic reconnection \citep{Lilp2016, Chen2023, Jiang2023, Lilp2025}. We show that external toroidal magnetic fields can induce return currents, whose associated downward Lorentz force acts to suppress the eruption. However, the present analysis is based on an idealised magnetic configuration and therefore requires further validation. Specifically, the TDm model employed here does not address the origin of the return current: whether it arises from flux emergence or from photospheric flows. Resolving this issue will require more realistic, observational data-driven simulations and comprehensive radiative MHD modelling. Furthermore, the conclusion that return currents can produce a substantial downward Lorentz force should also be tested in such realistic models. Finally, the comparative analysis of confined and eruptive flare cases suggests that the morphology and shearing degree of flare loops may serve as a diagnostic parameter for distinguishing between these two types. Determining the threshold shear angle of flare loops that distinguishes confined from eruptive flares should be a key objective of future dedicated-parameter surveys.

\begin{acknowledgements}
We are vert grateful to the discussions with Chen Xing and Xiaomeng Zhang at Nanjing University. J.H.G., P.F.C., and Y.G. are supported by the National Key R\&D Program of China (2020YFC2201200, 2022YFF0503004), NSFC (12503063, 12127901) and the China National Postdoctoral Program for Innovative Talents fellowship under Grant Number BX20240159. SP is funded by the European Union. However, the views and opinions expressed are those of the author(s) only and do not necessarily reflect those of the European Union or ERCEA. Neither the European Union nor the granting authority can be held responsible. His project (Open SESAME) has received funding under the Horizon Europe programme (ERC-AdG agreement No 101141362). These results were also obtained in the framework of the projects C16/24/010 C1 project Internal Funds KU Leuven), G0B5823N and G002523N (WEAVE) (FWO-Vlaanderen), and 4000145223 (SIDC Data Exploitation (SIDEX2), ESA Prodex). The numerical calculations in this paper were performed in the cluster system of the High Performance Computing Center (HPCC) of Nanjing University. 
\end{acknowledgements}
\bibliographystyle{aa}
\bibliography{ms}

@ARTICLE{Xu2020,
       author = {{Xu}, Yu and {Zhu}, Jiahao and {Guo}, Yang},
        title = "{Three-dimensional Reconstruction of Filament Axes and Their Writhe Numbers}",
      journal = {\apj},
         year = 2020,
        month = mar,
       volume = {892},
       number = {1},
          eid = {54},
        pages = {54},
          doi = {10.3847/1538-4357/ab791b},
       adsurl = {https://ui.adsabs.harvard.edu/abs/2020ApJ...892...54X}
}

@ARTICLE{Jin2007,
       author = {{Jin}, M. and {Ding}, M.~D.},
        title = "{Correlation and asymmetry between solar flare hard X-ray footpoints: a statistical study}",
      journal = {\aap},
         year = 2007,
        month = aug,
       volume = {471},
       number = {2},
        pages = {705-709},
          doi = {10.1051/0004-6361:20077202},
       adsurl = {https://ui.adsabs.harvard.edu/abs/2007A&A...471..705J}
}

@ARTICLE{Wu2025,
       author = {{Wu}, H. and {Guo}, Y. and {Keppens}, R. and {Xia}, C. and {Su}, Y. and {Kong}, X.~L. and {Ding}, M.~D.},
        title = "{Particle Acceleration and Transport in the Large-scale Current Sheet under an Erupting Magnetic Flux Rope}",
      journal = {\apj},
         year = 2025,
        month = oct,
       volume = {992},
       number = {1},
          eid = {81},
        pages = {81},
          doi = {10.3847/1538-4357/adfed4},
archivePrefix = {arXiv},
       eprint = {2509.22265},
 primaryClass = {astro-ph.SR},
       adsurl = {https://ui.adsabs.harvard.edu/abs/2025ApJ...992...81W}
}

@ARTICLE{Fan2007,
       author = {{Fan}, Y. and {Gibson}, S.~E.},
        title = "{Onset of Coronal Mass Ejections Due to Loss of Confinement of Coronal Flux Ropes}",
      journal = {\apj},
         year = 2007,
        month = oct,
       volume = {668},
       number = {2},
        pages = {1232-1245},
          doi = {10.1086/521335},
       adsurl = {https://ui.adsabs.harvard.edu/abs/2007ApJ...668.1232F}
}

@ARTICLE{Duan2024,
       author = {{Duan}, Aiying and {Xing}, Yaoyu and {Jiang}, Chaowei},
        title = "{Are Solar Active Regions Born with Neutralized Currents?}",
      journal = {Research in Astronomy and Astrophysics},
         year = 2024,
        month = jul,
       volume = {24},
       number = {7},
          eid = {075005},
        pages = {075005},
          doi = {10.1088/1674-4527/ad50b6},
       adsurl = {https://ui.adsabs.harvard.edu/abs/2024RAA....24g5005D}
}

@ARTICLE{Lit2021,
       author = {{Li}, Ting and {Chen}, Anqin and {Hou}, Yijun and {Veronig}, Astrid M. and {Yang}, Shuhong and {Zhang}, Jun},
        title = "{Magnetic Flux and Magnetic Nonpotentiality of Active Regions in Eruptive and Confined Solar Flares}",
      journal = {\apjl},
         year = 2021,
        month = aug,
       volume = {917},
       number = {2},
          eid = {L29},
        pages = {L29},
          doi = {10.3847/2041-8213/ac1a15},
archivePrefix = {arXiv},
       eprint = {2108.01299},
 primaryClass = {astro-ph.SR},
       adsurl = {https://ui.adsabs.harvard.edu/abs/2021ApJ...917L..29L}
}

@ARTICLE{Fabio2024,
       author = {{Bacchini}, Fabio and {Ruan}, Wenzhi and {Keppens}, Rony},
        title = "{Particle trapping and acceleration in turbulent post-flare coronal loops}",
      journal = {\mnras},
         year = 2024,
        month = apr,
       volume = {529},
       number = {3},
        pages = {2399-2412},
          doi = {10.1093/mnras/stae723},
archivePrefix = {arXiv},
       eprint = {2403.07107},
 primaryClass = {astro-ph.SR},
       adsurl = {https://ui.adsabs.harvard.edu/abs/2024MNRAS.529.2399B}
}

@ARTICLE{Torok2010,
       author = {{T{\"o}r{\"o}k}, T. and {Berger}, M.~A. and {Kliem}, B.},
        title = "{The writhe of helical structures in the solar corona}",
      journal = {\aap},
         year = 2010,
        month = jun,
       volume = {516},
          eid = {A49},
        pages = {A49},
          doi = {10.1051/0004-6361/200913578},
archivePrefix = {arXiv},
       eprint = {1004.3918},
 primaryClass = {astro-ph.SR},
       adsurl = {https://ui.adsabs.harvard.edu/abs/2010A&A...516A..49T}
}

@ARTICLE{Meng2014,
       author = {{Meng}, Y. and {Lin}, J. and {Zhang}, L. and {Reeves}, K.~K. and {Zhang}, Q.~S. and {Yuan}, F.},
        title = "{An MHD Model for Magnetar Giant Flares}",
      journal = {\apj},
         year = 2014,
        month = apr,
       volume = {785},
       number = {1},
          eid = {62},
        pages = {62},
          doi = {10.1088/0004-637X/785/1/62},
archivePrefix = {arXiv},
       eprint = {1402.5824},
 primaryClass = {astro-ph.HE},
       adsurl = {https://ui.adsabs.harvard.edu/abs/2014ApJ...785...62M}
}

@ARTICLE{Liu2017,
       author = {{Liu}, Yang and {Sun}, Xudong and {T{\"o}r{\"o}k}, Tibor and {Titov}, Viacheslav S. and {Leake}, James E.},
        title = "{Electric-current Neutralization, Magnetic Shear, and Eruptive Activity in Solar Active Regions}",
      journal = {\apjl},
         year = 2017,
        month = sep,
       volume = {846},
       number = {1},
          eid = {L6},
        pages = {L6},
          doi = {10.3847/2041-8213/aa861e},
archivePrefix = {arXiv},
       eprint = {1708.04411},
 primaryClass = {astro-ph.SR},
       adsurl = {https://ui.adsabs.harvard.edu/abs/2017ApJ...846L...6L}
}

@ARTICLE{Liu2024,
       author = {{Liu}, Y. and {T{\"o}r{\"o}k}, T. and {Titov}, V.~S. and {Leake}, J.~E. and {Sun}, X. and {Jin}, M.},
        title = "{Nonneutralized Electric Currents as a Proxy for Eruptive Activity in Solar Active Regions}",
      journal = {\apj},
         year = 2024,
        month = feb,
       volume = {961},
       number = {2},
          eid = {148},
        pages = {148},
          doi = {10.3847/1538-4357/ad11da},
       adsurl = {https://ui.adsabs.harvard.edu/abs/2024ApJ...961..148L}
}

@ARTICLE{Aulanier2012,
       author = {{Aulanier}, G. and {Janvier}, M. and {Schmieder}, B.},
        title = "{The standard flare model in three dimensions. I. Strong-to-weak shear transition in post-flare loops}",
      journal = {\aap},
         year = 2012,
        month = jul,
       volume = {543},
          eid = {A110},
        pages = {A110},
          doi = {10.1051/0004-6361/201219311},
       adsurl = {https://ui.adsabs.harvard.edu/abs/2012A&A...543A.110A}
}

@ARTICLE{Zuccarello2017,
       author = {{Zuccarello}, F.~P. and {Chandra}, R. and {Schmieder}, B. and {Aulanier}, G. and {Joshi}, R.},
        title = "{Transition from eruptive to confined flares in the same active region}",
      journal = {\aap},
         year = 2017,
        month = may,
       volume = {601},
          eid = {A26},
        pages = {A26},
          doi = {10.1051/0004-6361/201629836},
archivePrefix = {arXiv},
       eprint = {1702.02477},
 primaryClass = {astro-ph.SR},
       adsurl = {https://ui.adsabs.harvard.edu/abs/2017A&A...601A..26Z}
}

@ARTICLE{Kilpua2021,
       author = {{Kilpua}, Emilia K.~J. and {Pomoell}, Jens and {Price}, Daniel and {Sarkar}, Ranadeep and {Asvestari}, Eleanna},
        title = "{Estimating the magnetic structure of an erupting CME flux rope from AR12158 using data-driven modelling}",
      journal = {Frontiers in Astronomy and Space Sciences},
         year = 2021,
        month = mar,
       volume = {8},
          eid = {35},
        pages = {35},
          doi = {10.3389/fspas.2021.631582},
       adsurl = {https://ui.adsabs.harvard.edu/abs/2021FrASS...8...35K}
}

@ARTICLE{Jiang2023,
       author = {{Jiang}, Chaowei and {Duan}, Aiying and {Zou}, Peng and {Zhou}, Zhenjun and {Bian}, Xinkai and {Feng}, Xueshang and {Zuo}, Pingbing and {Wang}, Yi},
        title = "{A model of failed solar eruption initiated and destructed by magnetic reconnection}",
      journal = {\mnras},
         year = 2023,
        month = nov,
       volume = {525},
       number = {4},
        pages = {5857-5867},
          doi = {10.1093/mnras/stad2658},
archivePrefix = {arXiv},
       eprint = {2307.15847},
 primaryClass = {astro-ph.SR},
       adsurl = {https://ui.adsabs.harvard.edu/abs/2023MNRAS.525.5857J}
}

@ARTICLE{Dai2006,
       author = {{Dai}, Z.~G. and {Wang}, X.~Y. and {Wu}, X.~F. and {Zhang}, B.},
        title = "{X-ray Flares from Postmerger Millisecond Pulsars}",
      journal = {Science},
         year = 2006,
        month = feb,
       volume = {311},
       number = {5764},
        pages = {1127-1129},
          doi = {10.1126/science.1123606},
archivePrefix = {arXiv},
       eprint = {astro-ph/0602525},
 primaryClass = {astro-ph},
       adsurl = {https://ui.adsabs.harvard.edu/abs/2006Sci...311.1127D}
}

@ARTICLE{Chen2023,
       author = {{Chen}, Jun and {Cheng}, Xin and {Kliem}, Bernhard and {Ding}, MingDe},
        title = "{A Model for Confined Solar Eruptions Including External Reconnection}",
      journal = {\apjl},
         year = 2023,
        month = jul,
       volume = {951},
       number = {2},
          eid = {L35},
        pages = {L35},
          doi = {10.3847/2041-8213/acdef5},
archivePrefix = {arXiv},
       eprint = {2306.04993},
 primaryClass = {astro-ph.SR},
       adsurl = {https://ui.adsabs.harvard.edu/abs/2023ApJ...951L..35C}
}

@ARTICLE{Myers2015,
       author = {{Myers}, Clayton E. and {Yamada}, Masaaki and {Ji}, Hantao and {Yoo}, Jongsoo and {Fox}, William and {Jara-Almonte}, Jonathan and {Savcheva}, Antonia and {Deluca}, Edward E.},
        title = "{A dynamic magnetic tension force as the cause of failed solar eruptions}",
      journal = {\nat},
         year = 2015,
        month = dec,
       volume = {528},
       number = {7583},
        pages = {526-529},
          doi = {10.1038/nature16188},
       adsurl = {https://ui.adsabs.harvard.edu/abs/2015Natur.528..526M}
}

@ARTICLE{Myers2016,
       author = {{Myers}, C.~E. and {Yamada}, M. and {Ji}, H. and {Yoo}, J. and {Jara-Almonte}, J. and {Fox}, W.},
        title = "{Laboratory study of low-{\ensuremath{\beta}} forces in arched, line-tied magnetic flux ropes}",
      journal = {Physics of Plasmas},
         year = 2016,
        month = nov,
       volume = {23},
       number = {11},
          eid = {112102},
        pages = {112102},
          doi = {10.1063/1.4966691},
       adsurl = {https://ui.adsabs.harvard.edu/abs/2016PhPl...23k2102M}
}

@ARTICLE{Zhong2021,
       author = {{Zhong}, Ze and {Guo}, Yang and {Ding}, M.~D.},
        title = "{The role of non-axisymmetry of magnetic flux rope in constraining solar eruptions}",
      journal = {Nature Communications},
         year = 2021,
        month = jan,
       volume = {12},
          eid = {2734},
        pages = {2734},
          doi = {10.1038/s41467-021-23037-8},
archivePrefix = {arXiv},
       eprint = {2105.07339},
 primaryClass = {astro-ph.SR},
       adsurl = {https://ui.adsabs.harvard.edu/abs/2021NatCo..12.2734Z}
}

@ARTICLE{Kliem2012,
       author = {{Kliem}, B. and {T{\"o}r{\"o}k}, T. and {Thompson}, W.~T.},
        title = "{A Parametric Study of Erupting Flux Rope Rotation. Modeling the ``Cartwheel CME'' on 9 April 2008}",
      journal = {\solphys},
         year = 2012,
        month = nov,
       volume = {281},
       number = {1},
        pages = {137-166},
          doi = {10.1007/s11207-012-9990-z},
archivePrefix = {arXiv},
       eprint = {1112.3389},
 primaryClass = {astro-ph.SR},
       adsurl = {https://ui.adsabs.harvard.edu/abs/2012SoPh..281..137K}
}

@ARTICLE{Zhou2023,
       author = {{Zhou}, Zhenjun and {Jiang}, Chaowei and {Yu}, Xiaoyu and {Wang}, Yuming and {Hao}, Yongqiang and {Cui}, Jun},
        title = "{The mechanism of magnetic flux rope rotation during solar eruption}",
      journal = {Frontiers in Physics},
         year = 2023,
        month = feb,
       volume = {11},
          eid = {1119637},
        pages = {1119637},
          doi = {10.3389/fphy.2023.1119637},
archivePrefix = {arXiv},
       eprint = {2302.11103},
 primaryClass = {astro-ph.SR},
       adsurl = {https://ui.adsabs.harvard.edu/abs/2023FrP....1119637Z}
}

@ARTICLE{Green2007,
       author = {{Green}, L.~M. and {Kliem}, B. and {T{\"o}r{\"o}k}, T. and {van Driel-Gesztelyi}, L. and {Attrill}, G.~D.~R.},
        title = "{Transient Coronal Sigmoids and Rotating Erupting Flux Ropes}",
      journal = {\solphys},
         year = 2007,
        month = dec,
       volume = {246},
       number = {2},
        pages = {365-391},
          doi = {10.1007/s11207-007-9061-z},
       adsurl = {https://ui.adsabs.harvard.edu/abs/2007SoPh..246..365G}
}

@ARTICLE{Liu2016,
       author = {{Liu}, Rui and {Kliem}, Bernhard and {Titov}, Viacheslav S. and {Chen}, Jun and {Wang}, Yuming and {Wang}, Haimin and {Liu}, Chang and {Xu}, Yan and {Wiegelmann}, Thomas},
        title = "{Structure, Stability, and Evolution of Magnetic Flux Ropes from the Perspective of Magnetic Twist}",
      journal = {\apj},
         year = 2016,
        month = feb,
       volume = {818},
       number = {2},
          eid = {148},
        pages = {148},
          doi = {10.3847/0004-637X/818/2/148},
archivePrefix = {arXiv},
       eprint = {1512.02338},
 primaryClass = {astro-ph.SR},
       adsurl = {https://ui.adsabs.harvard.edu/abs/2016ApJ...818..148L}
}

@ARTICLE{Aulanier2006,
       author = {{Aulanier}, G. and {Pariat}, E. and {D{\'e}moulin}, P. and {Devore}, C.~R.},
        title = "{Slip-Running Reconnection in Quasi-Separatrix Layers}",
      journal = {\solphys},
         year = 2006,
        month = nov,
       volume = {238},
       number = {2},
        pages = {347-376},
          doi = {10.1007/s11207-006-0230-2},
       adsurl = {https://ui.adsabs.harvard.edu/abs/2006SoPh..238..347A}
}

@ARTICLE{Aulanier2019,
       author = {{Aulanier}, Guillaume and {Dud{\'\i}k}, Jaroslav},
        title = "{Drifting of the line-tied footpoints of CME flux-ropes}",
      journal = {\aap},
         year = 2019,
        month = jan,
       volume = {621},
          eid = {A72},
        pages = {A72},
          doi = {10.1051/0004-6361/201834221},
archivePrefix = {arXiv},
       eprint = {1811.04253},
 primaryClass = {astro-ph.SR},
       adsurl = {https://ui.adsabs.harvard.edu/abs/2019A&A...621A..72A}
}

@article{Berger&Prior2006,
  title={The writhe of open and closed curves},
  author={Mitchell A. Berger and Chris Prior},
  journal={Journal of Physics A},
  year={2006},
  volume={39},
  pages={8321-8348}
}

@ARTICLE{Jiang2021,
       author = {{Jiang}, Chaowei and {Feng}, Xueshang and {Liu}, Rui and {Yan}, XiaoLi and {Hu}, Qiang and {Moore}, Ronald L. and {Duan}, Aiying and {Cui}, Jun and {Zuo}, Pingbing and {Wang}, Yi and {Wei}, Fengsi},
        title = "{A fundamental mechanism of solar eruption initiation}",
      journal = {Nature Astronomy},
         year = 2021,
        month = jul,
       volume = {5},
        pages = {1126-1138},
          doi = {10.1038/s41550-021-01414-z},
archivePrefix = {arXiv},
       eprint = {2107.08204},
 primaryClass = {astro-ph.SR},
       adsurl = {https://ui.adsabs.harvard.edu/abs/2021NatAs...5.1126J}
}

@ARTICLE{chen12,
       author = {{Chen}, P.~F. and {Su}, J.~T. and {Guo}, Y. and {Deng}, Y.~Y.},
        title = "{Where do flare ribbons stop?}",
      journal = {Chinese Science Bulletin},
         year = 2012,
        month = jan,
       volume = {57},
        pages = {1393-1396},
          doi = {10.1007/s11434-011-4829-9},
archivePrefix = {arXiv},
       eprint = {1109.0381},
 primaryClass = {astro-ph.SR},
       adsurl = {https://ui.adsabs.harvard.edu/abs/2012ChSBu..57.1393C}
}

@ARTICLE{Tsurutani2023,
       author = {{Tsurutani}, Bruce T. and {Zank}, Gary P. and {Sterken}, Veerle J. and {Shibata}, Kazunari and {Nagai}, Tsugunobu and {Mannucci}, Anthony J. and {Malaspina}, David M. and {Lakhina}, Gurbax S. and {Kanekal}, Shrikanth G. and {Hosokawa}, Keisuke and {Horne}, Richard B. and {Hajra}, Rajkumar and {Glassmeier}, Karl-Heinz and {Gaunt}, C. Trevor and {Chen}, Peng-Fei and {Akasofu}, Syun-Ichi},
        title = "{Space Plasma Physics: A Review}",
      journal = {IEEE Transactions on Plasma Science},
         year = 2023,
        month = jul,
       volume = {51},
       number = {7},
        pages = {1595-1655},
          doi = {10.1109/TPS.2022.3208906},
archivePrefix = {arXiv},
       eprint = {2209.14545},
 primaryClass = {physics.space-ph},
       adsurl = {https://ui.adsabs.harvard.edu/abs/2023ITPS...51.1595T}
}

@ARTICLE{Aulanier2010,
       author = {{Aulanier}, G. and {T{\"o}r{\"o}k}, T. and {D{\'e}moulin}, P. and {DeLuca}, E.~E.},
        title = "{Formation of Torus-Unstable Flux Ropes and Electric Currents in Erupting Sigmoids}",
      journal = {\apj},
         year = 2010,
        month = jan,
       volume = {708},
       number = {1},
        pages = {314-333},
          doi = {10.1088/0004-637X/708/1/314},
       adsurl = {https://ui.adsabs.harvard.edu/abs/2010ApJ...708..314A}
}

@ARTICLE{Nindos2015,
       author = {{Nindos}, A. and {Patsourakos}, S. and {Vourlidas}, A. and {Tagikas}, C.},
        title = "{How Common Are Hot Magnetic Flux Ropes in the Low Solar Corona? A Statistical Study of EUV Observations}",
      journal = {\apj},
         year = 2015,
        month = aug,
       volume = {808},
       number = {2},
          eid = {117},
        pages = {117},
          doi = {10.1088/0004-637X/808/2/117},
archivePrefix = {arXiv},
       eprint = {1507.03766},
 primaryClass = {astro-ph.SR},
       adsurl = {https://ui.adsabs.harvard.edu/abs/2015ApJ...808..117N}
}

@ARTICLE{Jiang2016,
       author = {{Jiang}, Chaowei and {Wu}, S.~T. and {Yurchyshyn}, Vasyl and {Wang}, Haiming and {Feng}, Xueshang and {Hu}, Qiang},
        title = "{How Did a Major Confined Flare Occur in Super Solar Active Region 12192?}",
      journal = {\apj},
         year = 2016,
        month = sep,
       volume = {828},
       number = {1},
          eid = {62},
        pages = {62},
          doi = {10.3847/0004-637X/828/1/62},
archivePrefix = {arXiv},
       eprint = {1606.09334},
 primaryClass = {astro-ph.SR},
       adsurl = {https://ui.adsabs.harvard.edu/abs/2016ApJ...828...62J}
}

@ARTICLE{Ji2003,
       author = {{Ji}, Haisheng and {Wang}, Haimin and {Schmahl}, Edward J. and {Moon}, Y. -J. and {Jiang}, Yunchun},
        title = "{Observations of the Failed Eruption of a Filament}",
      journal = {\apjl},
         year = 2003,
        month = oct,
       volume = {595},
       number = {2},
        pages = {L135-L138},
          doi = {10.1086/378178},
       adsurl = {https://ui.adsabs.harvard.edu/abs/2003ApJ...595L.135J}
}

@ARTICLE{Torok2005,
       author = {{T{\"o}r{\"o}k}, T. and {Kliem}, B.},
        title = "{Confined and Ejective Eruptions of Kink-unstable Flux Ropes}",
      journal = {\apjl},
         year = 2005,
        month = sep,
       volume = {630},
       number = {1},
        pages = {L97-L100},
          doi = {10.1086/462412},
archivePrefix = {arXiv},
       eprint = {astro-ph/0507662},
 primaryClass = {astro-ph},
       adsurl = {https://ui.adsabs.harvard.edu/abs/2005ApJ...630L..97T}
}

@ARTICLE{Zhou2019,
       author = {{Zhou}, Zhenjun and {Cheng}, Xin and {Zhang}, Jie and {Wang}, Yuming and {Wang}, Dong and {Liu}, Lijuan and {Zhuang}, Bin and {Cui}, Jun},
        title = "{Why Do Torus-unstable Solar Filaments Experience Failed Eruptions?}",
      journal = {\apjl},
         year = 2019,
        month = jun,
       volume = {877},
       number = {2},
          eid = {L28},
        pages = {L28},
          doi = {10.3847/2041-8213/ab21cb},
archivePrefix = {arXiv},
       eprint = {1905.00224},
 primaryClass = {astro-ph.SR},
       adsurl = {https://ui.adsabs.harvard.edu/abs/2019ApJ...877L..28Z}
}

@ARTICLE{Guo2024tor,
       author = {{Guo}, J.~H. and {Linan}, L. and {Poedts}, S. and {Guo}, Y. and {Schmieder}, B. and {Lani}, A. and {Ni}, Y.~W. and {Brchnelova}, M. and {Perri}, B. and {Baratashvili}, T. and {Li}, S.~T. and {Chen}, P.~F.},
        title = "{Dependence of coronal mass ejections on the morphology and toroidal flux of their source magnetic flux ropes}",
      journal = {\aap},
         year = 2024,
        month = oct,
       volume = {690},
          eid = {A189},
        pages = {A189},
          doi = {10.1051/0004-6361/202449731},
archivePrefix = {arXiv},
       eprint = {2407.09457},
 primaryClass = {astro-ph.SR},
       adsurl = {https://ui.adsabs.harvard.edu/abs/2024A&A...690A.189G}
}

@ARTICLE{Schmieder2024,
       author = {{Schmieder}, Brigitte and {Guo}, Jinhan and {Poedts}, Stefaan},
        title = "{Recent advances in solar data-driven MHD simulations of the formation and evolution of CME flux ropes}",
      journal = {Reviews of Modern Plasma Physics},
         year = 2024,
        month = aug,
       volume = {8},
       number = {1},
          eid = {27},
        pages = {27},
          doi = {10.1007/s41614-024-00166-3},
archivePrefix = {arXiv},
       eprint = {2408.06595},
 primaryClass = {astro-ph.SR},
       adsurl = {https://ui.adsabs.harvard.edu/abs/2024RvMPP...8...27S}
}

@ARTICLE{Guo2023b,
       author = {{Guo}, J.~H. and {Qiu}, Y. and {Ni}, Y.~W. and {Guo}, Y. and {Li}, C. and {Gao}, Y.~H. and {Schmieder}, B. and {Poedts}, S. and {Chen}, P.~F.},
        title = "{Understanding the Lateral Drifting of an Erupting Filament with a Data-constrained Magnetohydrodynamic Simulation}",
      journal = {\apj},
         year = 2023,
        month = oct,
       volume = {956},
       number = {2},
          eid = {119},
        pages = {119},
          doi = {10.3847/1538-4357/acf198},
archivePrefix = {arXiv},
       eprint = {2308.08831},
 primaryClass = {astro-ph.SR},
       adsurl = {https://ui.adsabs.harvard.edu/abs/2023ApJ...956..119G}
}

@ARTICLE{Thalmann2015,
       author = {{Thalmann}, J.~K. and {Su}, Y. and {Temmer}, M. and {Veronig}, A.~M.},
        title = "{The Confined X-class Flares of Solar Active Region 2192}",
      journal = {\apjl},
         year = 2015,
        month = mar,
       volume = {801},
       number = {2},
          eid = {L23},
        pages = {L23},
          doi = {10.1088/2041-8205/801/2/L23},
archivePrefix = {arXiv},
       eprint = {1502.05157},
 primaryClass = {astro-ph.SR},
       adsurl = {https://ui.adsabs.harvard.edu/abs/2015ApJ...801L..23T}
}

@ARTICLE{Chen2011,
       author = {{Chen}, P.~F.},
        title = "{Coronal Mass Ejections: Models and Their Observational Basis}",
      journal = {Living Reviews in Solar Physics},
         year = 2011,
        month = apr,
       volume = {8},
       number = {1},
          eid = {1},
        pages = {1},
          doi = {10.12942/lrsp-2011-1},
       adsurl = {https://ui.adsabs.harvard.edu/abs/2011LRSP....8....1C}
}

@ARTICLE{Titov1999,
       author = {{Titov}, V.~S. and {D{\'e}moulin}, P.},
        title = "{Basic topology of twisted magnetic configurations in solar flares}",
      journal = {\aap},
         year = 1999,
        month = nov,
       volume = {351},
        pages = {707-720},
       adsurl = {https://ui.adsabs.harvard.edu/abs/1999A&A...351..707T}
}

@ARTICLE{Titov2014,
       author = {{Titov}, V.~S. and {T{\"o}r{\"o}k}, T. and {Mikic}, Z. and
         {Linker}, J.~A.},
        title = "{A Method for Embedding Circular Force-free Flux Ropes in Potential Magnetic Fields}",
      journal = {\apj},
         year = 2014,
        month = aug,
       volume = {790},
       number = {2},
          eid = {163},
        pages = {163},
          doi = {10.1088/0004-637X/790/2/163},
       adsurl = {https://ui.adsabs.harvard.edu/abs/2014ApJ...790..163T}
}

@ARTICLE{Titov2018,
       author = {{Titov}, Viacheslav S. and {Downs}, Cooper and {Miki{\'c}}, Zoran and
         {T{\"o}r{\"o}k}, Tibor and {Linker}, Jon A. and {Caplan}, Ronald M.},
        title = "{Regularized Biot-Savart Laws for Modeling Magnetic Flux Ropes}",
      journal = {\apjl},
         year = 2018,
        month = jan,
       volume = {852},
       number = {2},
          eid = {L21},
        pages = {L21},
          doi = {10.3847/2041-8213/aaa3da},
archivePrefix = {arXiv},
       eprint = {1712.06708},
 primaryClass = {astro-ph.SR},
       adsurl = {https://ui.adsabs.harvard.edu/abs/2018ApJ...852L..21T}
}

@ARTICLE{Gardiner2005,
       author = {{Gardiner}, Thomas A. and {Stone}, James M.},
        title = "{An unsplit Godunov method for ideal MHD via constrained transport}",
      journal = {Journal of Computational Physics},
         year = 2005,
        month = may,
       volume = {205},
       number = {2},
        pages = {509-539},
          doi = {10.1016/j.jcp.2004.11.016},
archivePrefix = {arXiv},
       eprint = {astro-ph/0501557},
 primaryClass = {astro-ph},
       adsurl = {https://ui.adsabs.harvard.edu/abs/2005JCoPh.205..509G}
}

@ARTICLE{Xia2018,
       author = {{Xia}, C. and {Teunissen}, J. and {El Mellah}, I. and {Chan{\'e}}, E. and
         {Keppens}, R.},
        title = "{MPI-AMRVAC 2.0 for Solar and Astrophysical Applications}",
      journal = {\apjs},
         year = 2018,
        month = feb,
       volume = {234},
       number = {2},
          eid = {30},
        pages = {30},
          doi = {10.3847/1538-4365/aaa6c8},
archivePrefix = {arXiv},
       eprint = {1710.06140},
 primaryClass = {astro-ph.SR},
       adsurl = {https://ui.adsabs.harvard.edu/abs/2018ApJS..234...30X}
}

@ARTICLE{Amari2018,
       author = {{Amari}, Tahar and {Canou}, Aur{\'e}lien and {Aly}, Jean-Jacques and {Delyon}, Francois and {Alauzet}, Fr{\'e}deric},
        title = "{Magnetic cage and rope as the key for solar eruptions}",
      journal = {\nat},
         year = 2018,
        month = feb,
       volume = {554},
       number = {7691},
        pages = {211-215},
          doi = {10.1038/nature24671},
       adsurl = {https://ui.adsabs.harvard.edu/abs/2018Natur.554..211A}
}

@ARTICLE{Guo2016b,
       author = {{Guo}, Y. and {Xia}, C. and {Keppens}, R. and {Valori}, G.},
        title = "{Magneto-frictional Modeling of Coronal Nonlinear Force-free Fields. I. Testing with Analytic Solutions}",
      journal = {\apj},
         year = 2016,
        month = sep,
       volume = {828},
       number = {2},
          eid = {82},
        pages = {82},
          doi = {10.3847/0004-637X/828/2/82},
       adsurl = {https://ui.adsabs.harvard.edu/abs/2016ApJ...828...82G}
}

@ARTICLE{Guo2024,
       author = {{Guo}, J.~H. and {Ni}, Y.~W. and {Guo}, Y. and {Xia}, C. and {Schmieder}, B. and {Poedts}, S. and {Zhong}, Z. and {Zhou}, Y.~H. and {Yu}, F. and {Chen}, P.~F.},
        title = "{Data-driven Modeling of a Coronal Magnetic Flux Rope: From Birth to Death}",
      journal = {\apj},
         year = 2024,
        month = jan,
       volume = {961},
       number = {1},
          eid = {140},
        pages = {140},
          doi = {10.3847/1538-4357/ad088d},
archivePrefix = {arXiv},
       eprint = {2310.19617},
 primaryClass = {astro-ph.SR},
       adsurl = {https://ui.adsabs.harvard.edu/abs/2024ApJ...961..140G}
}

@ARTICLE{Keppens2023,
       author = {{Keppens}, R. and {Popescu Braileanu}, B. and {Zhou}, Y. and {Ruan}, W. and {Xia}, C. and {Guo}, Y. and {Claes}, N. and {Bacchini}, F.},
        title = "{MPI-AMRVAC 3.0: Updates to an open-source simulation framework}",
      journal = {\aap},
         year = 2023,
        month = may,
       volume = {673},
          eid = {A66},
        pages = {A66},
          doi = {10.1051/0004-6361/202245359},
archivePrefix = {arXiv},
       eprint = {2303.03026},
 primaryClass = {astro-ph.IM},
       adsurl = {https://ui.adsabs.harvard.edu/abs/2023A&A...673A..66K}
}

@ARTICLE{Zhang2024,
       author = {{Zhang}, X.~M. and {Guo}, J.~H. and {Guo}, Y. and {Ding}, M.~D. and {Keppens}, Rony},
        title = "{Rotation and Confined Eruption of a Double Flux-rope System}",
      journal = {\apj},
         year = 2024,
        month = feb,
       volume = {961},
       number = {2},
          eid = {145},
        pages = {145},
          doi = {10.3847/1538-4357/ad1521},
archivePrefix = {arXiv},
       eprint = {2312.07406},
 primaryClass = {astro-ph.SR},
       adsurl = {https://ui.adsabs.harvard.edu/abs/2024ApJ...961..145Z}
}

@ARTICLE{Lilp2016,
       author = {{Li}, Leping and {Zhang}, Jun and {Peter}, Hardi and {Priest}, Eric and {Chen}, Huadong and {Guo}, Lijia and {Chen}, Feng and {Mackay}, Duncan},
        title = "{Magnetic reconnection between a solar filament and nearby coronal loops}",
      journal = {Nature Physics},
         year = 2016,
        month = sep,
       volume = {12},
       number = {9},
        pages = {847-851},
          doi = {10.1038/nphys3768},
archivePrefix = {arXiv},
       eprint = {1605.03320},
 primaryClass = {astro-ph.SR},
       adsurl = {https://ui.adsabs.harvard.edu/abs/2016NatPh..12..847L}
}

@ARTICLE{Lilp2025,
       author = {{Li}, Leping and {Song}, Hongqiang and {Hou}, Yijun and {Zhou}, Guiping and {Tan}, Baolin and {Ji}, Kaifan and {Xiang}, Yongyuan and {Hou}, Zhenyong and {Guo}, Yang and {Qiu}, Ye and {Su}, Yingna and {Ji}, Haisheng and {Zhang}, Qingmin and {Ou}, Yudi},
        title = "{Failure of a Solar Filament Eruption Caused by Magnetic Reconnection with Overlying Coronal Loops}",
      journal = {\apj},
         year = 2025,
        month = feb,
       volume = {979},
       number = {2},
          eid = {113},
        pages = {113},
          doi = {10.3847/1538-4357/ad9a56},
archivePrefix = {arXiv},
       eprint = {2412.01126},
 primaryClass = {astro-ph.SR},
       adsurl = {https://ui.adsabs.harvard.edu/abs/2025ApJ...979..113L}
}

@ARTICLE{Wang2007,
       author = {{Wang}, Yuming and {Zhang}, Jie},
        title = "{A Comparative Study between Eruptive X-Class Flares Associated with Coronal Mass Ejections and Confined X-Class Flares}",
      journal = {\apj},
         year = 2007,
        month = aug,
       volume = {665},
       number = {2},
        pages = {1428-1438},
          doi = {10.1086/519765},
archivePrefix = {arXiv},
       eprint = {0808.2976},
 primaryClass = {astro-ph},
       adsurl = {https://ui.adsabs.harvard.edu/abs/2007ApJ...665.1428W}
}

@ARTICLE{Lit2022,
       author = {{Li}, Ting and {Sun}, Xudong and {Hou}, Yijun and {Chen}, Anqin and {Yang}, Shuhong and {Zhang}, Jun},
        title = "{A New Magnetic Parameter of Active Regions Distinguishing Large Eruptive and Confined Solar Flares}",
      journal = {\apjl},
         year = 2022,
        month = feb,
       volume = {926},
       number = {2},
          eid = {L14},
        pages = {L14},
          doi = {10.3847/2041-8213/ac5251},
archivePrefix = {arXiv},
       eprint = {2202.09966},
 primaryClass = {astro-ph.SR},
       adsurl = {https://ui.adsabs.harvard.edu/abs/2022ApJ...926L..14L}
}

@ARTICLE{Wang2017,
       author = {{Wang}, Dong and {Liu}, Rui and {Wang}, Yuming and {Liu}, Kai and {Chen}, Jun and {Liu}, Jiajia and {Zhou}, Zhenjun and {Zhang}, Min},
        title = "{Critical Height of the Torus Instability in Two-ribbon Solar Flares}",
      journal = {\apjl},
         year = 2017,
        month = jul,
       volume = {843},
       number = {1},
          eid = {L9},
        pages = {L9},
          doi = {10.3847/2041-8213/aa79f0},
archivePrefix = {arXiv},
       eprint = {1706.03169},
 primaryClass = {astro-ph.SR},
       adsurl = {https://ui.adsabs.harvard.edu/abs/2017ApJ...843L...9W}
}

@ARTICLE{Guo2010,
       author = {{Guo}, Y. and {Ding}, M.~D. and {Schmieder}, B. and {Li}, H. and {T{\"o}r{\"o}k}, T. and {Wiegelmann}, T.},
        title = "{Driving Mechanism and Onset Condition of a Confined Eruption}",
      journal = {\apjl},
         year = 2010,
        month = dec,
       volume = {725},
       number = {1},
        pages = {L38-L42},
          doi = {10.1088/2041-8205/725/1/L38},
       adsurl = {https://ui.adsabs.harvard.edu/abs/2010ApJ...725L..38G}
}

@ARTICLE{Lizf2023,
       author = {{Li}, Z.~F. and {Cheng}, X. and {Ding}, M.~D. and {Chitta}, L.~P. and {Peter}, H. and {Berghmans}, D. and {Smith}, P.~J. and {Auch{\`e}re}, F. and {Parenti}, S. and {Barczynski}, K. and {Harra}, L. and {Sch{\"u}hle}, U. and {Buchlin}, {\'E}. and {Verbeeck}, C. and {Aznar Cuadrado}, R. and {Zhukov}, A.~N. and {Long}, D.~M. and {Teriaca}, L. and {Rodriguez}, L.},
        title = "{Evidence of external reconnection between an erupting mini-filament and ambient loops observed by Solar Orbiter/EUI}",
      journal = {\aap},
         year = 2023,
        month = may,
       volume = {673},
          eid = {A83},
        pages = {A83},
          doi = {10.1051/0004-6361/202245814},
archivePrefix = {arXiv},
       eprint = {2303.16046},
 primaryClass = {astro-ph.SR},
       adsurl = {https://ui.adsabs.harvard.edu/abs/2023A&A...673A..83L}
}

@ARTICLE{Zhoux2016,
       author = {{Zhou}, X. and {B{\"u}chner}, J. and {B{\'a}rta}, M. and {Gan}, W. and {Liu}, S.},
        title = "{Electron Acceleration by Cascading Reconnection in the Solar Corona. II. Resistive Electric Field Effects}",
      journal = {\apj},
         year = 2016,
        month = aug,
       volume = {827},
       number = {2},
          eid = {94},
        pages = {94},
          doi = {10.3847/0004-637X/827/2/94},
       adsurl = {https://ui.adsabs.harvard.edu/abs/2016ApJ...827...94Z}
}

@ARTICLE{Naus2022,
       author = {{Naus}, S.~J. and {Qiu}, J. and {DeVore}, C.~R. and {Antiochos}, S.~K. and {Dahlin}, J.~T. and {Drake}, J.~F. and {Swisdak}, M.},
        title = "{Correlated Spatio-temporal Evolution of Extreme-Ultraviolet Ribbons and Hard X-Rays in a Solar Flare}",
      journal = {\apj},
         year = 2022,
        month = feb,
       volume = {926},
       number = {2},
          eid = {218},
        pages = {218},
          doi = {10.3847/1538-4357/ac4028},
archivePrefix = {arXiv},
       eprint = {2109.15314},
 primaryClass = {astro-ph.SR},
       adsurl = {https://ui.adsabs.harvard.edu/abs/2022ApJ...926..218N}
}

@ARTICLE{Shi2024,
       author = {{Shi}, Guanglu and {Feng}, Li and {Chen}, Jun and {Ying}, Beili and {Li}, Shuting and {Li}, Qiao and {Li}, Hui and {Li}, Ying and {Ji}, Kaifan and {Huang}, Yu and {Li}, Youping and {Li}, Jingwei and {Zhao}, Jie and {Lu}, Lei and {Xue}, Jianchao and {Zhang}, Ping and {Song}, Dechao and {Tian}, Zhengyuan and {Su}, Yingna and {Zhang}, Qingmin and {Ge}, Yunyi and {Shan}, Jiahui and {Zhou}, Yue and {Tian}, Jun and {Li}, Gen and {Liu}, Xiaofeng and {Jing}, Zhichen and {Lei}, Shijun and {Gan}, Weiqun},
        title = "{Asymmetric Hard X-ray Radiation of Two Ribbons in a Thermal-Dominated C-Class Flare}",
      journal = {\solphys},
         year = 2024,
        month = jul,
       volume = {299},
       number = {7},
          eid = {104},
        pages = {104},
          doi = {10.1007/s11207-024-02349-0},
archivePrefix = {arXiv},
       eprint = {2407.13099},
 primaryClass = {astro-ph.SR},
       adsurl = {https://ui.adsabs.harvard.edu/abs/2024SoPh..299..104S}
}

@ARTICLE{Titov2002,
       author = {{Titov}, Vyacheslav S. and {Hornig}, Gunnar and {D{\'e}moulin}, Pascal},
        title = "{Theory of magnetic connectivity in the solar corona}",
      journal = {Journal of Geophysical Research (Space Physics)},
         year = 2002,
        month = aug,
       volume = {107},
       number = {A8},
          eid = {1164},
        pages = {1164},
          doi = {10.1029/2001JA000278},
       adsurl = {https://ui.adsabs.harvard.edu/abs/2002JGRA..107.1164T}
}

@ARTICLE{Chintzoglou2019,
       author = {{Chintzoglou}, Georgios and {Zhang}, Jie and {Cheung}, Mark C.~M. and {Kazachenko}, Maria},
        title = "{The Origin of Major Solar Activity: Collisional Shearing between Nonconjugated Polarities of Multiple Bipoles Emerging within Active Regions}",
      journal = {\apj},
         year = 2019,
        month = jan,
       volume = {871},
       number = {1},
          eid = {67},
        pages = {67},
          doi = {10.3847/1538-4357/aaef30},
archivePrefix = {arXiv},
       eprint = {1811.02186},
 primaryClass = {astro-ph.SR},
       adsurl = {https://ui.adsabs.harvard.edu/abs/2019ApJ...871...67C}
}

@ARTICLE{Janvier2013,
       author = {{Janvier}, M. and {Aulanier}, G. and {Pariat}, E. and {D{\'e}moulin}, P.},
        title = "{The standard flare model in three dimensions. III. Slip-running reconnection properties}",
      journal = {\aap},
         year = 2013,
        month = jul,
       volume = {555},
          eid = {A77},
        pages = {A77},
          doi = {10.1051/0004-6361/201321164},
archivePrefix = {arXiv},
       eprint = {1305.4053},
 primaryClass = {astro-ph.SR},
       adsurl = {https://ui.adsabs.harvard.edu/abs/2013A&A...555A..77J}
}

@ARTICLE{Janvier2014,
       author = {{Janvier}, M. and {Aulanier}, G. and {Bommier}, V. and {Schmieder}, B. and {D{\'e}moulin}, P. and {Pariat}, E.},
        title = "{Electric Currents in Flare Ribbons: Observations and Three-dimensional Standard Model}",
      journal = {\apj},
         year = 2014,
        month = jun,
       volume = {788},
       number = {1},
          eid = {60},
        pages = {60},
          doi = {10.1088/0004-637X/788/1/60},
archivePrefix = {arXiv},
       eprint = {1402.2010},
 primaryClass = {astro-ph.SR},
       adsurl = {https://ui.adsabs.harvard.edu/abs/2014ApJ...788...60J}
}

@ARTICLE{Zhao2016,
       author = {{Zhao}, Jie and {Gilchrist}, Stuart A. and {Aulanier}, Guillaume and {Schmieder}, Brigitte and {Pariat}, Etienne and {Li}, Hui},
        title = "{Hooked Flare Ribbons and Flux-rope-related QSL Footprints}",
      journal = {\apj},
         year = 2016,
        month = may,
       volume = {823},
       number = {1},
          eid = {62},
        pages = {62},
          doi = {10.3847/0004-637X/823/1/62},
archivePrefix = {arXiv},
       eprint = {1603.07563},
 primaryClass = {astro-ph.SR},
       adsurl = {https://ui.adsabs.harvard.edu/abs/2016ApJ...823...62Z}
}

@ARTICLE{Xing2024,
       author = {{Xing}, Chen and {Aulanier}, Guillaume and {Schmieder}, Brigitte and {Cheng}, Xin and {Ding}, Mingde},
        title = "{Identifying footpoints of pre-eruptive and coronal mass ejection flux ropes with sunspot scars}",
      journal = {\aap},
         year = 2024,
        month = feb,
       volume = {682},
          eid = {A3},
        pages = {A3},
          doi = {10.1051/0004-6361/202347053},
archivePrefix = {arXiv},
       eprint = {2310.13532},
 primaryClass = {astro-ph.SR},
       adsurl = {https://ui.adsabs.harvard.edu/abs/2024A&A...682A...3X}
}

@ARTICLE{Qiu2014,
       author = {{Qiu}, Jiong and {Longcope}, Dana W. and {Cassak}, Paul A. and {Priest}, Eric R.},
        title = "{Elongation of Flare Ribbons}",
      journal = {\apj},
         year = 2017,
        month = mar,
       volume = {838},
       number = {1},
          eid = {17},
        pages = {17},
          doi = {10.3847/1538-4357/aa6341},
archivePrefix = {arXiv},
       eprint = {1707.02478},
 primaryClass = {astro-ph.SR},
       adsurl = {https://ui.adsabs.harvard.edu/abs/2017ApJ...838...17Q}
}

@ARTICLE{Teraoka2025,
       author = {{Teraoka}, Kouhei and {Yamasaki}, Daiki and {Kawabata}, Yusuke and {Imada}, Shinsuke and {Shimizu}, Toshifumi},
        title = "{Observational Comparison between Confined and Eruptive Flares: Magnetohydrodynamics Instability Parameters in a Similar Magnetic Configuration}",
      journal = {\apj},
         year = 2025,
        month = apr,
       volume = {983},
       number = {2},
          eid = {126},
        pages = {126},
          doi = {10.3847/1538-4357/adc12d},
archivePrefix = {arXiv},
       eprint = {2503.16857},
 primaryClass = {astro-ph.SR},
       adsurl = {https://ui.adsabs.harvard.edu/abs/2025ApJ...983..126T}
}

@ARTICLE{Duan2019,
       author = {{Duan}, Aiying and {Jiang}, Chaowei and {He}, Wen and {Feng}, Xueshang and {Zou}, Peng and {Cui}, Jun},
        title = "{A Study of Pre-flare Solar Coronal Magnetic Fields: Magnetic Flux Ropes}",
      journal = {\apj},
         year = 2019,
        month = oct,
       volume = {884},
       number = {1},
          eid = {73},
        pages = {73},
          doi = {10.3847/1538-4357/ab3e33},
archivePrefix = {arXiv},
       eprint = {1908.08643},
 primaryClass = {astro-ph.SR},
       adsurl = {https://ui.adsabs.harvard.edu/abs/2019ApJ...884...73D}
}

@ARTICLE{Jing2018,
       author = {{Jing}, Ju and {Liu}, Chang and {Lee}, Jeongwoo and {Ji}, Hantao and {Liu}, Nian and {Xu}, Yan and {Wang}, Haimin},
        title = "{Statistical Analysis of Torus and Kink Instabilities in Solar Eruptions}",
      journal = {\apj},
         year = 2018,
        month = sep,
       volume = {864},
       number = {2},
          eid = {138},
        pages = {138},
          doi = {10.3847/1538-4357/aad6e4},
archivePrefix = {arXiv},
       eprint = {1808.08924},
 primaryClass = {astro-ph.SR},
       adsurl = {https://ui.adsabs.harvard.edu/abs/2018ApJ...864..138J}
}

@ARTICLE{Priest1995,
       author = {{Priest}, E.~R. and {D{\'e}moulin}, P.},
        title = "{Three-dimensional magnetic reconnection without null points. 1. Basic theory of magnetic flipping}",
      journal = {\jgr},
         year = 1995,
        month = dec,
       volume = {100},
       number = {A12},
        pages = {23443-23464},
          doi = {10.1029/95JA02740},
       adsurl = {https://ui.adsabs.harvard.edu/abs/1995JGR...10023443P}
}

@ARTICLE{Demoulin1996FR,
       author = {{D{\'e}moulin}, P. and {Priest}, E.~R. and {Lonie}, D.~P.},
        title = "{Three-dimensional magnetic reconnection without null points 2. Application to twisted flux tubes}",
      journal = {\jgr},
         year = 1996,
        month = apr,
       volume = {101},
       number = {A4},
        pages = {7631-7646},
          doi = {10.1029/95JA03558},
       adsurl = {https://ui.adsabs.harvard.edu/abs/1996JGR...101.7631D}
}

@ARTICLE{Demoulin1996,
       author = {{Demoulin}, P. and {Henoux}, J.~C. and {Priest}, E.~R. and {Mandrini}, C.~H.},
        title = "{Quasi-Separatrix layers in solar flares. I. Method.}",
      journal = {\aap},
         year = 1996,
        month = apr,
       volume = {308},
        pages = {643-655},
       adsurl = {https://ui.adsabs.harvard.edu/abs/1996A&A...308..643D}
}

@ARTICLE{Lorincik2021,
       author = {{L{\"o}rin{\v{c}}{\'\i}k}, Juraj and {Dud{\'\i}k}, Jaroslav and {Aulanier}, Guillaume},
        title = "{Saddle-shaped Solar Flare Arcades}",
      journal = {\apjl},
         year = 2021,
        month = mar,
       volume = {909},
       number = {1},
          eid = {L4},
        pages = {L4},
          doi = {10.3847/2041-8213/abe7f7},
archivePrefix = {arXiv},
       eprint = {2102.10858},
 primaryClass = {astro-ph.SR},
       adsurl = {https://ui.adsabs.harvard.edu/abs/2021ApJ...909L...4L}
}

@ARTICLE{Jiang2021b,
       author = {{Jiang}, Chaowei and {Chen}, Jun and {Duan}, Aiying and {Bian}, Xinkai and {Wang}, Xinyi and {Li}, Jiaying and {Zou}, Peng and {Feng}, Xueshang},
        title = "{Formation of Magnetic Flux Rope during Solar Eruption. I. Evolution of Toroidal Flux and Reconnection Flux}",
      journal = {Frontiers in Physics},
         year = 2021,
        month = oct,
       volume = {9},
          eid = {575},
        pages = {575},
          doi = {10.3389/fphy.2021.746576},
archivePrefix = {arXiv},
       eprint = {2109.08422},
 primaryClass = {astro-ph.SR},
       adsurl = {https://ui.adsabs.harvard.edu/abs/2021FrP.....9..575J}
}

\newpage
\begin{appendix}

\section{Control Experiment 1: Successful Eruption in a Magnetic Configuration Dominated by Poloidal Fields}  \label{sec:Experiment1}

\begin{figure*}
  \includegraphics[width=16 cm,clip]{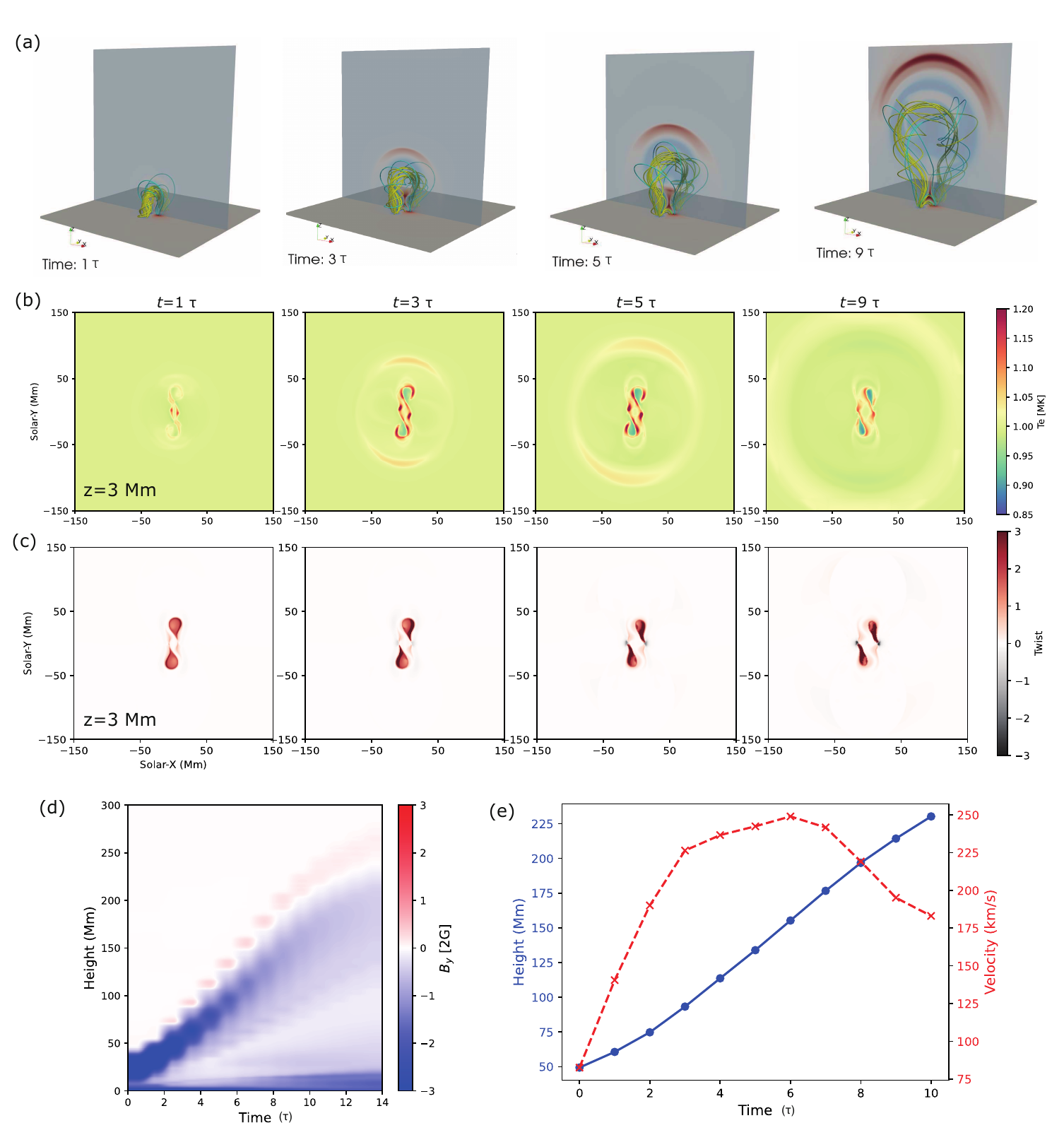}
  \centering
  \caption{Evolution of the model FR-P in the eruption process. Panel (a) displays 3D magnetic structures during the eruption. The yellow and cyan lines represent the magnetic fields traced from the pre-eruptive flux-rope footpoints and the poloidal polarities, respectively. Panels (b) and (c) show the distribution of the temperature and twist on the bottom plane, respectively. Panel (d) shows the time-distance diagram of the $B_{y}$ along the $z$-axis. Panel (e) shows the kinetics of the flux rope by measuring the $+B_{y}$ front in Panel (d), wherein the blue and red lines represent the evolution of height and speed. 
  \label{fig_ap1}}
\end{figure*}

\begin{figure*}
  \includegraphics[width=16 cm,clip]{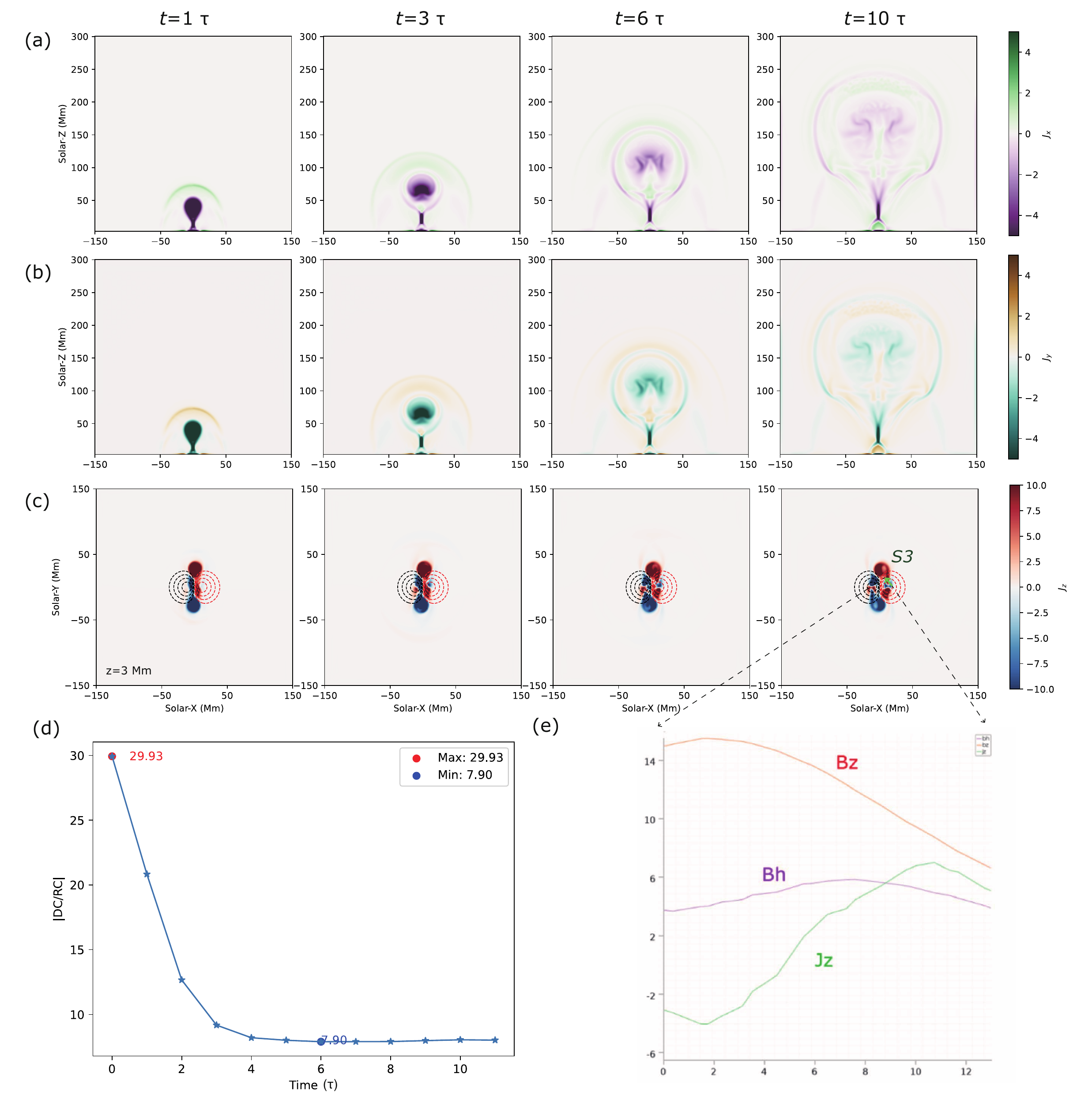}
  \centering
  \caption{Evolution of the electric current during the eruption. Panels (a)--(b) display the distribution of $J_{x}$, $J_{y}$ on the $x-y$ plane, respectively. Panel (c) shows the distribution of $J_{z}$ on the bottom plane. Panel (d) illustrates the temporal evolution of |DC/RC| during the eruption process. Panel (e) shows the magnetic properties ($B_{z}$, $J_{z}$ and $B_{h}$) along the slit $S3$ in Panel (c) at $t=14$ . 
  \label{fig_ap2}}
\end{figure*}


To investigate the role of external toroidal magnetic fields on flux rope eruption, we conduct a control experiment referred to as model FR-P. In this model, we eliminate the external toroidal magnetic fields generated by $N_{\rm _{T}}$ and $P_{\rm _{T}}$ polarities while maintaining consistency with the other parameters in the benchmark case described in Section~\ref{sec:met} (referred to as model FR-PT). In contrast to the failed eruption with significant rotation in model FR-PT, the flux rope in model FR-P ascends to a height of 300 Mm within 12 $\tau$ (Figure~\ref{fig_ap1}d). Moreover, magnetic reconnection primarily occurs beneath the flux rope in model FR-P, converting overlying loops into part of the flux rope and thereby reducing their downward magnetic tension, which opposes the flux rope rise. In addition, for model FR-PT, magnetic reconnection also takes place between the rotated flux rope and the overlying toroidal magnetic fields, ultimately disrupting the flux rope and leading to its failed eruption. These differences highlight the critical role of overlying large-scale toroidal magnetic fields in triggering failed solar eruptions with significant rotation.

Figure~\ref{fig_ap1}b shows the morphology of flare ribbons, outlined by the heated regions. The modelled flare ribbons display a separated double J-shaped structure, which aligns well with the 3D standard flare model \citep{Janvier2014}. Additionally, the hooked structures at both ends of flare ribbons, typically surrounding the flux rope footpoints, become increasingly closed. This reflects an increase in the twist at the flux rope border during the eruption process due to the magnetic reconnection of the dominantly poloidal magnetic field below the erupting flux rope \citep{Demoulin1996FR,Zhao2016}. In contrast, the hooked flare ribbons in model FR-PT remain relatively open, indicating a lower twist for three reasons. First, the presence of external toroidal fields reduces the ratio of poloidal to toroidal flux within the flux rope, thereby lowering its twist. Second, the magnetic field lines in the current sheet beneath the flux rope are more strongly sheared in model FR-PT. As a result, the amount of twisted flux injected into the flux rope via magnetic reconnection is also diminished. Third, 3D magnetic reconnection involving the flux rope itself can ruin its structure, further decreasing its twist. 

Figure~\ref{fig_ap1}c shows the distribution of the twist number on the bottom plane, indicating the footpoints of the flux rope. The areas of the flux rope footpoints increase first and then decrease,  but the twist number increases during the eruption, which is significantly different from the confined eruption in model FR-PT. The increase in the flux-rope footpoint area corresponds to reconnection occurring in the arcade field lines beneath it, which converts sheared arcades into the flux rope ($aa$–$rf$),. Hereafter, a decrease in the flux-rope footpoint area reflects reconnection involving the flux rope itself, e.g., reconnection within the flux-rope field lines ($rr$–$rf$) or the reconnection between the eruptive flux rope and the overlying field lines ($ar$–$rf$), as also reported by \citet{Jiang2021b}. Figures~\ref{fig_ap1}d and \ref{fig_ap1}e show the kinetics of the flux rope by measuring the $+B_{y}$ leading front of the flux rope. Its velocity increases to about 250 km s$^{-1}$ at $t=6\tau$ but subsequently starts to decrease due to the boundary issues in the finite computation domain. 

\begin{figure*}
  \includegraphics[width=19 cm,clip]{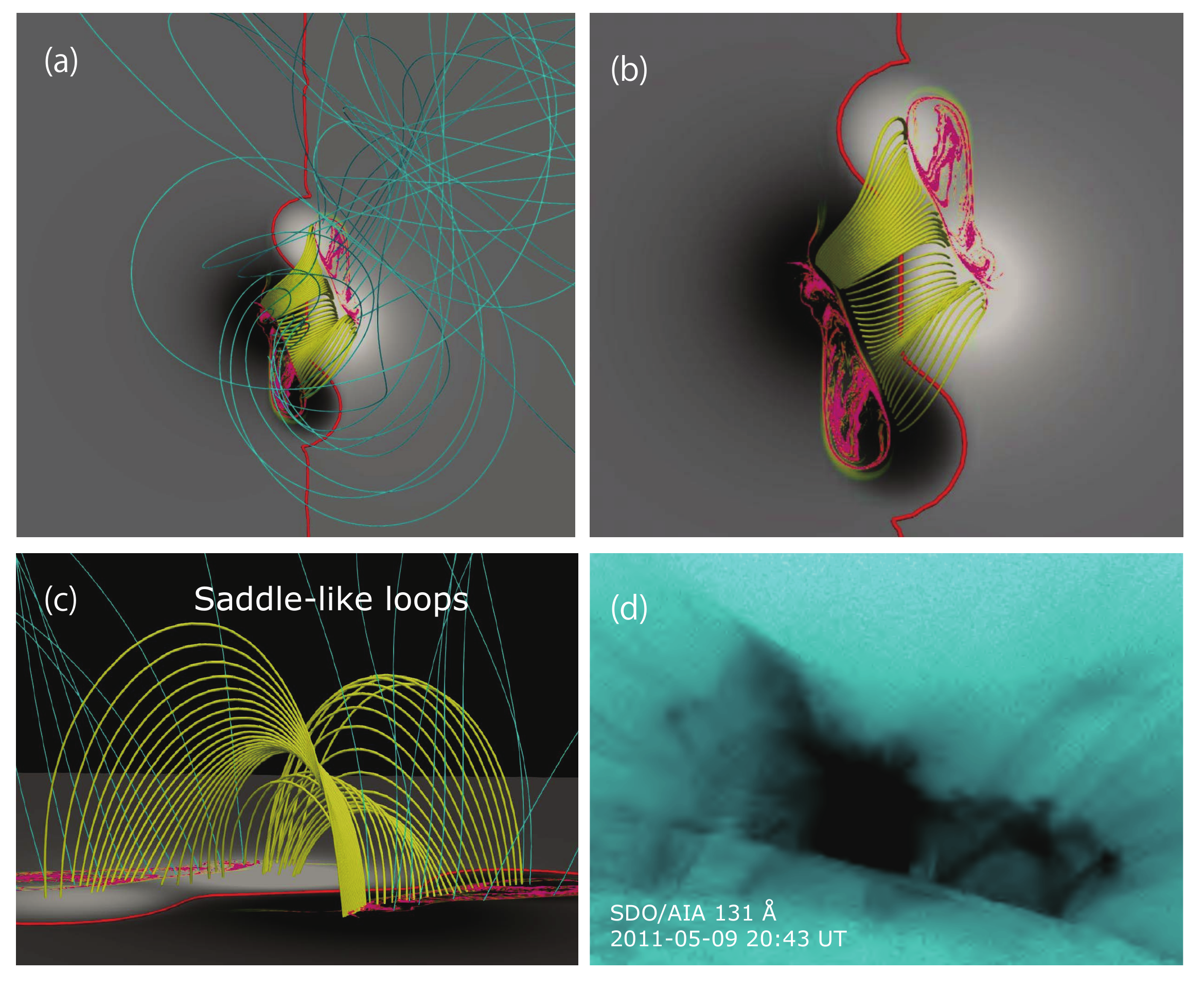}
  \centering
  \caption{Flare loops, QSLs, and comparisons with observations in the successful-eruption case. The yellow lines represent the flare loops, while the cyan lines correspond to the eruptive flux rope. The red line marks the PIL on the bottom plane, and the red contours denote the QSLs.}
  \label{fig_ap3}
\end{figure*}

The evolution of electric currents is illustrated in Figure~\ref{fig_ap2}. The return current around the flux rope in the FR-P model is significantly lower compared to that in the FR-PT model. This suggests that the toroidal magnetic fields are crucial to induce the return current. Moreover, the degree of current neutralisation, represented by $|DC/RC|$, is significantly higher in model FR-P ($\sim$30), so that the currents are less neutralised, compared to model FR-PT ($\sim$9; Figure~\ref{fig9}a). This is consistent with the statistical findings of \citet{Liu2017},\citet{Liu2024} and \citet{Duan2024}, who found that the electric currents of eruptive core fields are closer to unity in confined eruptions compared to eruptive events. However, it is essential to note that the flux rope in our model is based on the TDm model, which only considers the direct current flowing through the flux rope. As a result, the neutralisation degree measured by $|DC/RC|$ is higher than the typically observed values. Figure~\ref{fig_ap3} shows the flare loops and their comparison with observations. The flare loops in the FR-P case exhibit a pronounced saddle-like morphology, with the loops on both sides rising higher than those in the central section, consistent with the observations reported by \citet{Lorincik2021}.

\section{Control Experiment 2: Evolution of an Individual Flux Rope}
\label{sec:Experiment2}

\begin{figure*}
  \includegraphics[width=19 cm,clip]{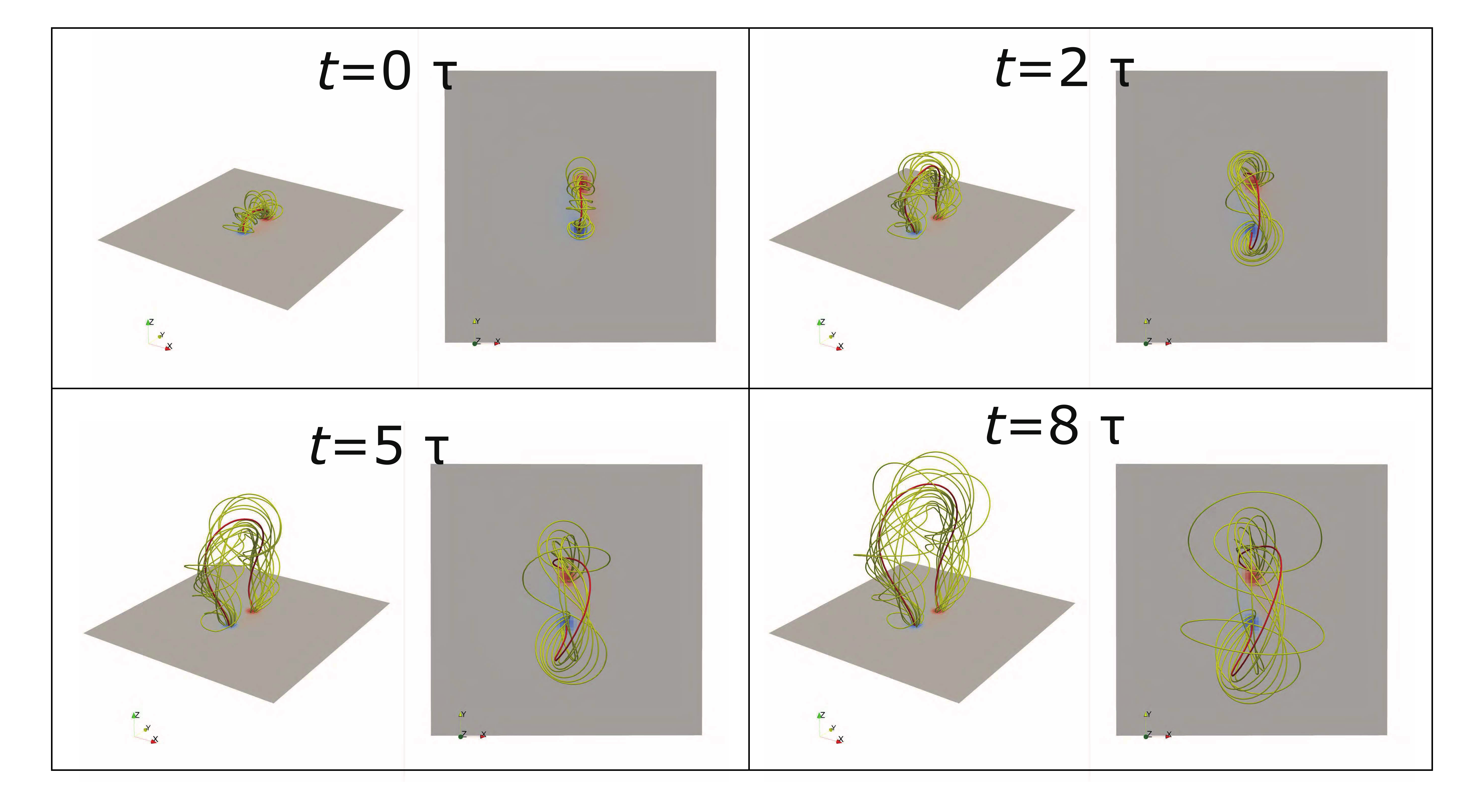}
  \centering
  \caption{3D magnetic fields of an individual flux rope during the eruption. Yellow and red lines represent the twisted field lines and the flux-rope axis, respectively. 
  \label{fig_ap4}}
\end{figure*}

To isolate the effects of the Lorentz force from external magnetic fields and magnetic reconnection, we remove both the external toroidal and poloidal magnetic fields while retaining the flux rope in the initial magnetic configuration, referred to as model FR. Figure~\ref{fig_ap4} presents the evolution of an individual flux rope. Initially, the flux rope undergoes slight deformation accompanied by counter-clockwise rotation, consistent with the distribution of its self-induced Lorentz force, as shown in Figure~\ref{fig11}c, which may be related to the kink instability. Subsequently, the flux rope almost maintains a fixed morphology as it rises. Notably, the writhe number of an S-shaped curve anchored to the surface is positive when it is low-lying, but becomes negative once its apex reaches a critical height \citep{Torok2010, Xu2020}. The simulation between models FR, FR-PT and FR-P indicates that the external toroidal magnetic fields are the key to inducing the rotation of the flux rope.

\section{Test-Particle Method for Solving Particle Motion with GCA}  \label{sec:GCA}

Here, we introduce the test-particle method implemented in the MPI-AMRVAC framework based on the GCA. This approach relies on two assumptions. First, the electron lifetime is much shorter than the evolution timescale of the coronal MHD system. Second, the electron gyro-radius and gyro-period are much smaller than the characteristic spatial and temporal scales of the large-scale MHD evolution. As such, the gyromotion of particles perpendicular to the magnetic field can be averaged out, and the particle dynamics can be described in terms of the guiding-center position ($\boldsymbol{R}$), the motion parallel to the magnetic field ($u_\parallel$), and the magnetic moment ($\mu = mu_{\perp}^2/2B\kappa=mu_{\perp}^2/2B (1-v_E^2/c^2)^{-1/2}$), as follows \citep{Fabio2024, Wu2025}:

\begin{equation}
\frac{d\boldsymbol{R}}{dt} = \frac{u_\parallel}{\gamma} \boldsymbol{b} + \boldsymbol{v}_E + \boldsymbol{v}_{\text{curv}} + \boldsymbol{v}_{\text{pol}} + \boldsymbol{v}_{\nabla B} + \boldsymbol{v}_{\text{rel}},
\end{equation}

\begin{equation}
\frac{du_\parallel}{dt} = \frac{q}{m} E_\parallel + a_{\text{curv}} + a_{\nabla B},
\end{equation}

\begin{equation}
\frac{d\mu}{dt} = 0,
\end{equation}
where electric field is computed from $\boldsymbol{E}=-\mathbf{v} \times \mathbf{B} + \eta \mathbf{J}$, and the parallel electric field ($E_{\parallel}$) is defined as $\boldsymbol{E} \cdot \boldsymbol{B}$. The guiding-centre position of electrons is dominated by parallel motion ($u_\parallel$), $\boldsymbol{E}\times \boldsymbol{B}$ drift ($\boldsymbol{v}_E$), curvature drift ($\boldsymbol{v}_{\text{curv}}$), gradient drift ($\boldsymbol{v}_{\nabla B}$), polarization drift $\boldsymbol{v}_{\text{pol}}$, and the relativistic correction ($\boldsymbol{v}_{\text{rel}}$). They are described as follows:

\begin{equation}
\boldsymbol{v}_{\text{E}} = \boldsymbol{E} \times \boldsymbol{B} / B^2,
\end{equation}

\begin{equation}
\boldsymbol{v}_{\text{curv}} = \frac{mc\kappa^2}{qB} \boldsymbol{b} \times \left[ \frac{u_\parallel^2}{\gamma} (\boldsymbol{b} \cdot \nabla) \boldsymbol{b} + u_\parallel (\boldsymbol{v}_E \cdot \nabla) \boldsymbol{b} \right],
\end{equation}

\begin{equation}
\boldsymbol{v}_{\text{pol}} = \frac{mc\kappa^2}{qB} \boldsymbol{b} \times \left[ u_\parallel (\boldsymbol{b} \cdot \nabla) \boldsymbol{v}_E + \gamma (\boldsymbol{v}_E \cdot \nabla) \boldsymbol{v}_E \right],
\end{equation}

\begin{equation}
\boldsymbol{v}_{\nabla B} = \frac{\mu c \kappa^2}{\gamma q B} \boldsymbol{b} \times \nabla \left( \frac{B}{\kappa} \right),
\end{equation}

\begin{equation}
\boldsymbol{v}_{\text{rel}} = \frac{u_\parallel E_\parallel \kappa^2}{c \gamma B} \boldsymbol{b} \times \boldsymbol{v}_E,
\end{equation}
where coefficient $\kappa$ is derived from $1/\sqrt{1-v_E^2/c^2}$,  $\gamma$ is the Lorentz factor, $\boldsymbol{b}$ is the unit vector along the magnetic field direction, $q$ is the electric charge of the particle, and $c$ is the light speed. Regarding the parallel motion along magnetic fields, the acceleration items resulting from the curvature and gradient of magnetic fields are described by:

\begin{equation}
a_{\text{curv}} = \boldsymbol{v}_E \cdot \left[ u_\parallel (\boldsymbol{b} \cdot \nabla) \boldsymbol{b} + \gamma (\boldsymbol{v}_E \cdot \nabla) \boldsymbol{b} \right], 
\end{equation}

\begin{equation}
a_{\nabla B} = -\frac{\mu}{m} \boldsymbol{b} \cdot \nabla \left( \frac{B}{\kappa} \right).
\end{equation}

\section{Parameter Survey for the Strength of Toroidal Magnetic Fields}  \label{sec:ps}

\begin{figure*}
  \includegraphics[width=19 cm,clip]{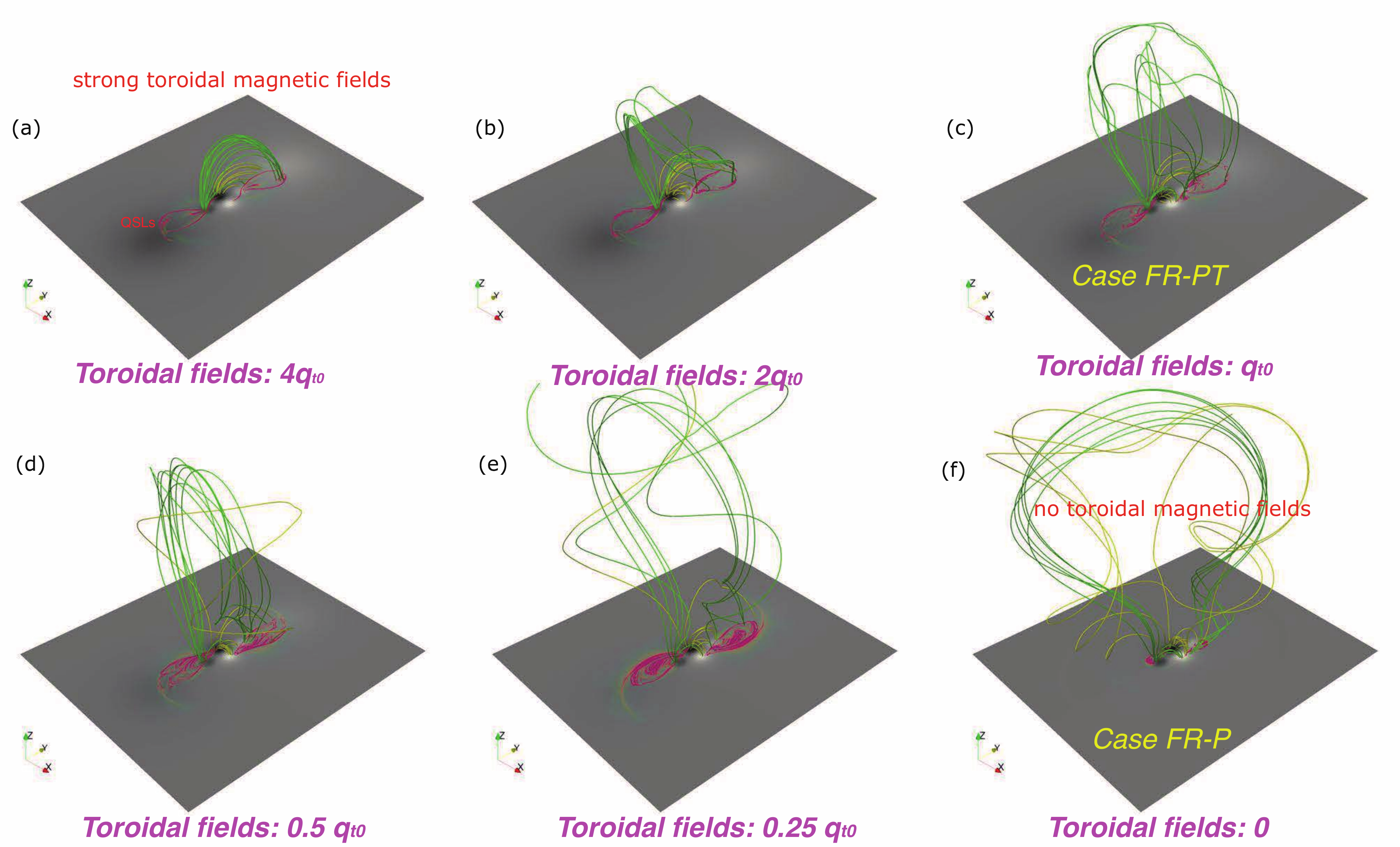}
  \centering
  \caption{Parameter survey of the toroidal magnetic-field strength. The red contours at the bottom indicate QSLs, used here as proxies for flare ribbons. The toroidal field strength decreases from panel (a) to panel (f).
  \label{fig_ap5}}
\end{figure*}

To investigate the role of the toroidal magnetic field strength, we perform a parameter survey by varying \(q_t\) from \(4q_{t0}\), \(2q_{t0}\), and \(q_{t0}\) to \(0.5q_{t0}\), \(0.25q_{t0}\), and \(0\). The case with \(q_t=q_{t0}\) corresponds to Case FR-PT, whereas the case with \(q_t=0\) corresponds to Case FR-P. The three-dimensional magnetic field structures at the same time are shown in Figure~\ref{fig_ap5}. It is found that the rotation angle of the flux rope increases with increasing toroidal magnetic field strength. In addition, when the toroidal magnetic field becomes sufficiently strong, the eruption turns into a failed eruption. These results confirm that the toroidal magnetic field promotes the rotation of the flux rope and can suppress the eruption when its strength is sufficiently large.

\end{appendix}

\end{document}